\documentclass[12pt]{article}
\pdfoutput=1
\usepackage{bbm}
\usepackage{color}
\usepackage{diagbox}
\usepackage{authblk}
\usepackage{latexsym}
\usepackage{epsfig,amssymb,euscript, mathrsfs,cite}
\usepackage{amsmath}
\usepackage{mathrsfs} 
\usepackage{enumerate}
\usepackage{slashed}
\definecolor{MyDarkBlue}{rgb}{0.15,0.15,0.45}
\usepackage[linktocpage=true]{hyperref}
\hypersetup{
colorlinks=true,
citecolor=MyDarkBlue,
linkcolor=MyDarkBlue,
urlcolor=MyDarkBlue,
pdfauthor={},
pdftitle={},
pdfsubject={hep-th}
}
\newcommand{\vev}[1]{\left< #1 \right>} 

\newcounter{savefootnote}
\newcounter{symfootnote}
\newcommand{\symfootnote}[1]{%
   \setcounter{savefootnote}{\value{footnote}}%
   \setcounter{footnote}{\value{symfootnote}}%
   \ifnum\value{footnote}>8\setcounter{footnote}{0}\fi%
   \let\oldthefootnote=\thefootnote%
   \renewcommand{\thefootnote}{\fnsymbol{footnote}}%
   \footnote{#1}%
   \let\thefootnote=\oldthefootnote%
   \setcounter{symfootnote}{\value{footnote}}%
   \setcounter{footnote}{\value{savefootnote}}%
}

\usepackage{mathtools}

\newsavebox{\ns}
\newsavebox{\dbrane}
\newsavebox{\dbshort}

\def\be{\begin{equation}}
\def\ee{\end{equation}}
\def\bea{\begin{eqnarray}}
\def\eea{\end{eqnarray}}

\newcommand{\nn}{\notag \\}

\def\eq#1 { \begin{equation} #1 \end{equation} }

\newcommand{\vol}{\mathrm{vol}}

\newlength{\sswidth}

\numberwithin{equation}{section}       

\begin{document}

\begin{titlepage}

\vfill

\begin{flushright}
APCTP Pre2024-004\\
CCTP-2024-10\\   
ITCP-2024/10
\end{flushright}

\vfill

\begin{center}
   \baselineskip=16pt
   {\Large\bf Superconformal Monodromy Defects\\ in ABJM and mABJM
   Theory}
  \vskip 1cm
Igal Arav$^1$,  Jerome P. Gauntlett$^2$, Yusheng Jiao$^2$\\
Matthew M. Roberts$^{3,4}$ and Christopher Rosen$^5$\\
     \vskip 1cm     
                                                    \begin{small}
                                \textit{$^1$KU Leuven, Instituut voor Theoretische Fysica,\\
                                Celestijnenlaan 200D, B-3001 Leuven, Belgium}
        \end{small}\\
        \begin{small}\vskip .3cm
      \textit{$^2$Blackett Laboratory, 
  Imperial College\\ Prince Consort Rd., London, SW7 2AZ, U.K.}
        \end{small}\\
                \begin{small}\vskip .3cm
      \textit{$^3$Asia Pacific Center for Theoretical Physics,\\
      Pohang, 37673, Korea}
        \end{small}\\
                \begin{small}\vskip .3cm
      \textit{$^4$Department of Physics, Pohang University of Science and Technology,\\
      Pohang 37673, Korea}
        \end{small}\\
             \begin{small}\vskip .3cm
      \textit{$^5$Crete Center for Theoretical Physics, Department of Physics, University of Crete,\\
71003 Heraklion, Greece}
        \end{small}\\
                       \end{center}
\vfill

\begin{center}
\textbf{Abstract}
\end{center}
\begin{quote}
We study $D=11$ supergravity solutions which are dual to 
one-dimensional superconformal defects in $d=3$ SCFTs. We consider defects in ABJM theory with 
monodromy for $U(1)^4\subset SO(8)$ global symmetry, as well as in
$\mathcal{N}=2$ mABJM SCFT, which
arises from the RG flow of a mass deformation of ABJM theory, with monodromy for 
$U(1)^3\subset SU(3)\times U(1)$ global symmetry. 
We show that the defects of the two SCFTs are connected by a line of bulk marginal mass deformations and argue that  they are also related by bulk RG flow. In all cases we allow for the possibility of conical singularities at the location of the defect.
Various physical observables of the defects are computed including the defects conformal weight and the partition function, as well as associated  supersymmetric Renyi entropies.

\end{quote}

\vfill

\end{titlepage}

\tableofcontents

\newpage

\section{Introduction}\label{sec:intro}

The study of defects in conformal field theories is an important ongoing endeavour (for some review see\cite{Andrei:2018die,Billo:2016cpy}).
In this paper we study one-dimensional (line) defects in $d=3$ SCFTs, 
that preserve a residual conformal symmetry $SO(1,2)\times SO(2)\subset SO(3,2)$ as well as
two Poincar\'e and two superconformal supersymmetries. The defects are monodromy defects, with a non-trivial monodromy for an abelian subgroup of the global symmetry as one circles the
point defect on a constant time slice. Our analysis will closely parallel a similar investigation of two-dimensional monodromy defects
in $d=4$ SCFTs that was carried out in \cite{Arav:2024exg}.

We study these defects using holography by analysing supersymmetric solutions of $D=11$ supergravity.
We consider ABJM theory, dual to $AdS_4\times S^7$, with non-trivial monodromy for a $U(1)^4$ subgroup
of the $SO(8)$ global symmetry, extending the investigations\footnote{Co-dimension two monodromy defects have been studied holographically in various spacetime dimensions, including \cite{Gutperle:2018fea,Chen:2020mtv,Gutperle:2020rty,Gutperle:2022pgw,Gutperle:2023yrd,Capuozzo:2023fll}. We also refer to \cite{Penati:2021tfj} for a review of line defects in ABJM theory and mABJM theory from a complementary viewpoint.}  initiated in \cite{Gutperle:2018fea,Chen:2020mtv}. Viewing the CFT as living in flat spacetime, the defect lies at the origin
of the spatial $\mathbb{R}^2$ and we also allow for the possibility of a conical singularity at the origin.
Via a Weyl transformation we can map the flat spacetime to $AdS_2\times S^1$ with the conical singularity then pushed off to infinity
and a non-trivial holonomy for the global symmetry around the $S^1$. The data associated with the conical singularity
is replaced with the ratio of the radii of the $AdS_2$ and $S^1$ factors.

We also consider such defects in ABJM theory which have, in addition, supersymmetric mass deformations that depend on the two flat spatial directions transverse to the defect and preserving the superconformal
invariance of the defect. This generalises the mass deformations of ABJM theory that depend on one spatial dimension transverse to
an interface that were studied in \cite{Kim:2018qle,Arav:2018njv,Arav:2020asu,Kim:2019kns}. In the context of the
$AdS_2\times S^1$ Weyl frame, these additional mass deformations are constant deformations.

In addition, we also study line defects in mABJM theory, which we recall is a $d=3$, $\mathcal{N}=2$ SCFT that arises as the IR limit of an RG flow from ABJM theory with a Poincar\'e invariant mass deformation \cite{Klebanov:2008vq,Benna:2008zy}. mABJM theory has $SU(3)\times U(1)_R$ global symmetry, where
the $U(1)_R$ is the R-symmetry, and is dual to an $AdS_4\times S^7$ solution, with a non-round metric on the $S^7$ \cite{Corrado:2001nv}. For mABJM theory we study line defects with non-trivial monodromy for a $U(1)^3$
subgroup of $SU(3)\times U(1)_R$. 

For all cases, we utilise a sub-truncation of $D=4$ maximal $SO(8)$ gauged supergravity, whose solutions can be uplifted on $S^7$ to obtain solutions
of $D=11$ supergravity.
We compute various observables including the one-point functions of both the stress tensor and
the flavour and R-symmetry currents and relate these to the monodromy sources. 
We compute the on-shell action
and define a ``defect free energy'' $I_D$ via
\begin{align}
I_D=I-nI_0\,,
\end{align}
where $I$ is the free energy of the CFT with a defect, $I_0$ is the free energy of the CFT with no defect and
the conical deficit angle is $2\pi(1-\frac{1}{n})$. Writing $\log g= -I_D$, in the case of no conical singularity with $n=1$, we can identify $g$ 
with the defect $g$-function
of \cite{Cuomo:2021rkm}. Here we are able to express $I_D$ in 
terms of the monodromy sources
for line defects in both ABJM and mABJM theory and show that it shares some similar properties to
the defect central charge $b$ that arises in the context of co-dimension two monodromy defects in $d=4$ SCFTs (e.g. see the discussion in section 1 of \cite{Arav:2024exg}).
We show that the conformal weight, $h_D$, of the monodromy defects is determined by the one-point function of
the R-symmetry current. For $\mathcal{N}=(0,2)$ co-dimension two defects in $d=4$ SCFTs this is expected \cite{Bianchi:2019sxz}, but we are unaware of a general argument for defects in $d=3$ SCFTs.

When $0<n<1$ there is a conical deficit angle and when $n>1$ there is a conical excess angle.
We will see that there are two possible branches of solutions with branch 1, the ``main branch", existing for $n>0$ which, in particular,
is continuously connected to the solutions with no conical singularity when $n=1$. 
By contrast the branch 2 solutions 
can only exist (at most) for $0<n<1$. 
We show that all solutions on branch 2 have $h_D>0$ (and $-I_D\ge 0$).
For solutions on the main branch with $n\ge 1$ we also always have $h_D\ge 0$ (as well as $-I_D\ge 0$), but for $n<1$ there are solutions with $h_D<0$ (and $-I_D< 0$). That $h_D<0$ for $n<1$ has also been observed in 
in the context of holography in \cite{Bianchi:2016xvf,Baiguera:2022sao} and also for free theories \cite{Dowker:1987mn,Dowker:2015qta}. It has been shown that $h_D>0$ when 
there is no conical singularity (i.e. $n=1$) and when the ANEC is satisfied \cite{Jensen:2018rxu}; it would be interesting to understand why this result does not apply when $n<1$.

For the case of defects 
in ABJM theory with no spatially dependent mass deformations, we can utilise known analytic solutions 
of the STU model. However, as in \cite{Arav:2024exg}, we will again see that 
many results can be obtained without using the explicit solution, but just by examining the BPS equations with the relevant boundary conditions.
Furthermore, our analysis allows us to obtain results for defects in ABJM theory that have spatially dependent mass deformations
as well as for defects in mABJM theory, for which explicit analytic solutions are not known.
It is particularly interesting that the bulk core behaviour of both the uncharged scalar fields and the metric warp function
are precisely the same in the ABJM case with spatially modulated mass sources as in the mABJM case. This reveals
that a kind of attractor mechanism is at work. Furthermore, the expectation values for the conserved currents
and the on-shell action, expressed in terms of the monodromy parameters, are also exactly the same in these two cases. 

\begin{figure}[ht!]
\begin{center}
\includegraphics[scale=.3]{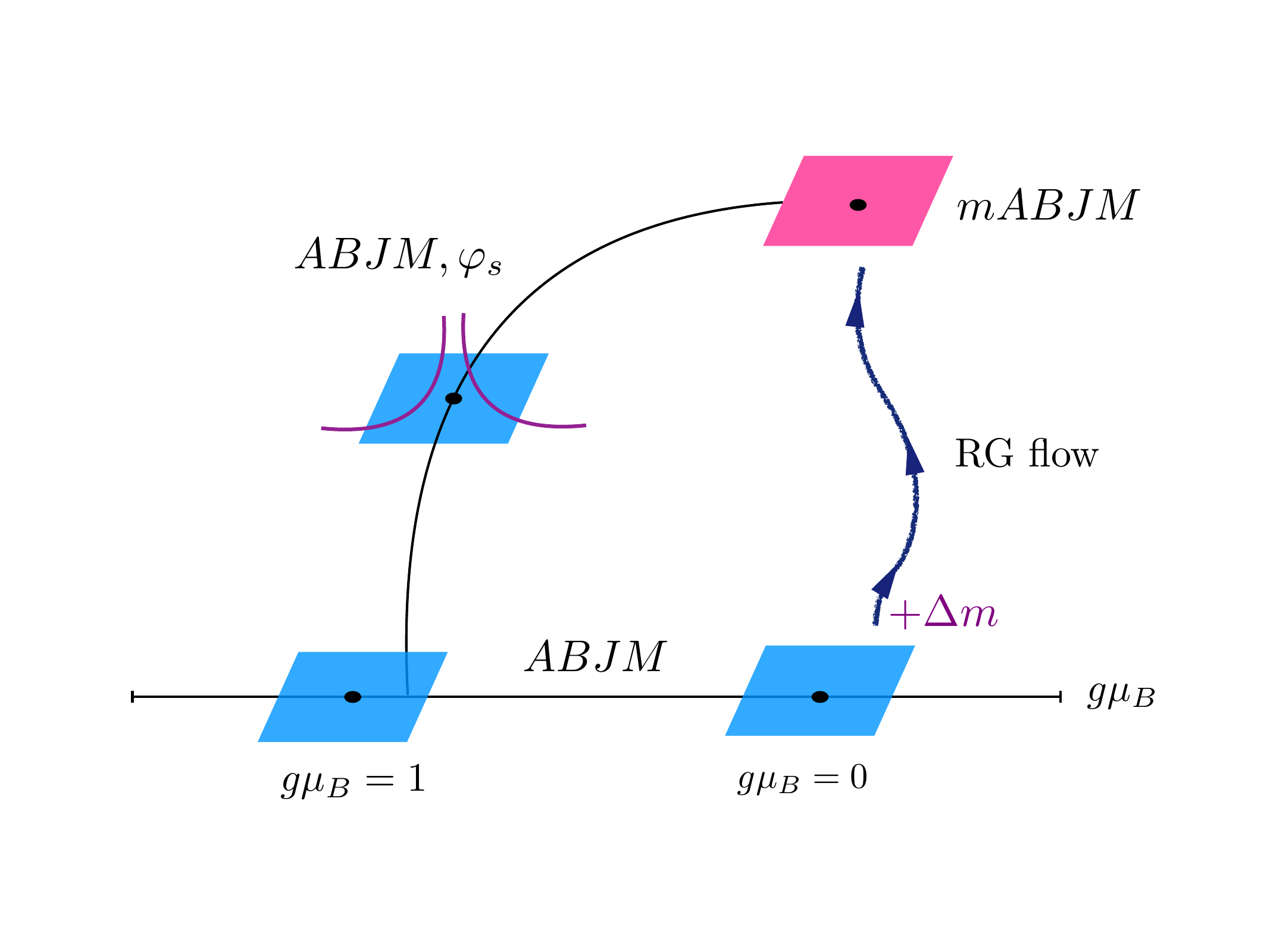}~
\caption{Part of the solution space for the monodromy defects with no conical singularity, $n=1$.
We have suppressed the flavour monodromy fluxes $g\mu_{F_i}$ (and also set $\kappa=1$).
}
\label{fig:sumsolutions}
\end{center}
\end{figure}

We will also construct some new solutions numerically and this reveals an interesting solution space as summarised 
in figure \ref{fig:sumsolutions} (and analogous to what was seen in \cite{Arav:2024exg}). 
The figure illustrates defects in flat space and with no conical singularities (for simplicity), 
which implies, as we later show, a vanishing monodromy for the R-symmetry, $g\mu_R=0$.
The horizontal line corresponds to ABJM defects of the STU model, parametrised by monodromy parameters $g\mu_B$ and $g\mu_{F_1}, g\mu_{F_2}$, for the $U(1)^3$ flavour symmetry,
with $g\mu_{F_i}$ suppressed in the figure. For $g\mu_B=1$ another branch of solutions emerges, parametrised by
$\varphi_s$ and is associated with spatially dependent mass deformations preserving the superconformal invariance of the defect and breaking one of the $U(1)$ flavour symmetries; this is a line of ``bulk marginal mass deformations" in the sense of \cite{Herzog:2019bom}.
Interestingly, for large values of $\varphi_s$ the solutions describing defects in ABJM theory closely approximate
the defect solutions of mABJM theory, before sharply turning back to the ABJM vacuum at the boundary. In the limit
that $\varphi_s\to\infty$ we recover the solutions describing defects in mABJM theory, which have $g\mu_B=0$ and 
parametrised by $g\mu_{F_i}$. Remarkably, this feature combined with the fact that the on-shell action is independent of $\varphi_s$, 
allows us to determine the partition function of mABJM defects, as a function of 
the monodromy parameters, just  from a knowledge of the partition function of defects in ABJM theory (i.e. with $\varphi_s=0$). On figure \ref{fig:sumsolutions} we have also indicated
homogeneous mass deformations ($\Delta m$) that break bulk conformal invariance, which induce an RG flow\footnote{These RG flows should be contrasted with those driven by deformations localised on the defect itself studied in e.g. \cite{Cuomo:2021rkm}. We also point out that bulk RG flows 
have been studied in a different setting in \cite{Green:2007wr} (see also \cite{Shachar:2024ubf}).} 
from ABJM defects in the UV to mABJM defects in the IR.

The computation of the on-shell action also allows us to compute supersymmetric Renyi entropies for the two $d=3$ SCFTs that are
associated with a circular entangling surface. 
Recall that Renyi entropy is a one parameter generalisation of entanglement entropy that utilises an $n$-fold covering space of the entangling surface, with a corresponding conical singularity. In a supersymmetric context this will typically break supersymmetry,
but by turning on suitable background $R$-symmetry sources it is possible to preserve supersymmetry giving rise to the notion of supersymmetric Renyi entropy (SRE)
\cite{Nishioka:2013haa,Nishioka:2014mwa}. 
We emphasise that there are various ways to define a supersymmetric Renyi entropy (SRE) depending on which monodromy sources one keeps fixed as one adjusts the conical deficit; this can be viewed as different supersymmetric loci of charged Renyi entropies \cite{Belin:2013uta}.
For the case of ABJM theory without spatially dependent mass deformations we make contact with previous computations \cite{Nishioka:2014mwa,Hosseini:2019and} as well as present some new results. We also 
compute various SREs for mABJM theory.

The plan of the rest of the paper is as follows. In section \ref{sec:sugramodel} we present the supergravity model that we use in the rest of the paper.
In section \ref{sec:ads3} we consider the BPS equations for solutions associated with defects in ABJM theory with spatially dependent mass deformations (in flat space)
and defects in mABJM theory; these cases have bulk gravitational solutions with a non-vanishing charged scalar field and are naturally treated together.
In section \ref{sectstu} we discuss solutions dual to defects in ABJM theory without such mass deformations; while there are known analytic solutions of the STU model to describe these configurations, we will not use them in our analysis. Section \ref{sec:partfn} discusses the partition function, defect free energy and supersymmetric Renyi entropies for ABJM and mABJM theory.  
We conclude in section \ref{sec:discussion} with some discussion. We have summarised some features of the BPS equations in appendix \ref{sec:appa} and presented some details of holographic renormalisation 
in appendices \ref{appa} and \ref{secappcmabjm}, for ABJM and mABJM theory, respectively.
In appendix \ref{minsugraapp} we cast the known solutions of minimal gauged supergravity into the conventions of this paper and show, in particular, 
that solutions on both the main branch and on branch 2 exist. In appendix
\ref{positivehdmId} we demonstrate some positivity results for $h_D$ and $-I_D$.
 
\section{The supergravity model}\label{sec:sugramodel}

We use the $U(1)^2 \subset SU(3)\subset SO(6)\subset SO(8)$ invariant consistent truncation of $D=4$ maximal $SO(8)$ gauged supergravity of \cite{Bobev:2018uxk}. This is an $\mathcal{N}=2$ gauged supergravity theory with three vector multiplets and one hypermultiplet.
As in \cite{Bobev:2018uxk} we can consistently truncate one of the two complex scalar fields in the hypermultiplet to zero.
Furthermore, as in \cite{Suh:2022pkg}, considering solutions with $F\wedge F=0$,
we can take the three complex scalars in the vector multiplets to be real. This then leads to a $D=4$ theory with 
a metric, four gauge fields $A^{\alpha}\equiv (A^{0},A^{1},A^{2},A^{3})$, three real and neutral scalars $\lambda_i$
and a single complex scalar field $\zeta\equiv \varphi e^{i\theta}$ which is charged with respect to a specific linear combination of the four $U(1)$'s. The model can be viewed as an extension of the STU model by an additional complex scalar field. 
We will largely follow the notation of the published version of \cite{Suh:2022pkg}.

The bosonic part of the Lagrangian, in a \emph{mostly plus} signature, is given by 
\begin{align} \label{mlag}
\mathcal{L}=&\frac{1}{2}R-\partial_\mu\varphi\partial^\mu\varphi-\frac{1}{4}\sinh^2\left(2\varphi\right)D_\mu\theta{D}^\mu\theta-\sum_{i=1}^3\partial_\mu\lambda_i\partial^\mu\lambda_i-g^2\mathcal{P} \notag \\
&-\frac{1}{4}\Big[e^{-2\left(\lambda_1+\lambda_2+\lambda_3\right)}F_{\mu\nu}^0F^{0\mu\nu}
+e^{-2\left(\lambda_1-\lambda_2-\lambda_3\right)}F_{\mu\nu}^1F^{1\mu\nu}\nn
&\qquad+e^{-2\left(-\lambda_1+\lambda_2-\lambda_3\right)}F_{\mu\nu}^2F^{2\mu\nu}
+e^{-2\left(-\lambda_1-\lambda_2+\lambda_3\right)}F_{\mu\nu}^3F^{3\mu\nu}\Big]\,.
\end{align}
The bulk action is normalized as 
\begin{align}\label{bulkactandnorm}
S_{bulk} =\frac{1}{8\pi G}\int \sqrt{-g} \mathcal{L}, \qquad \frac{1}{G} =\frac{4\sqrt{2}g^2}{3} N^{3/2}\,,
\end{align}
where $N$ is the rank of the gauge groups in ABJM theory which is dual to the vacuum $AdS_4$ solution.
We have
\begin{align}\label{dthetastext}
 D\theta\equiv\,d\theta- g\left(A^0-A^1-A^2-A^3\right)\equiv d\theta- gA_B\,,
\end{align}
and the scalar potential $\mathcal{P}$ is given by
\begin{equation}
\mathcal{P}\,=\,\frac{1}{2}\left(\frac{\partial{W}}{\partial\varphi}\right)^2+\frac{1}{2}\sum_{i=1}^3\left(\frac{\partial{W}}{\partial\lambda_i}\right)^2-\frac{3}{2}W^2\,,
\end{equation}
where $W$ is defined by
\begin{equation}
\label{superpottext}
W={e^{\lambda_1+\lambda_2+\lambda_3}\sinh^2\varphi-\frac{1}{2}\left(e^{\lambda_1+\lambda_2+\lambda_3}+e^{\lambda_1-\lambda_2-\lambda_3}+e^{-\lambda_1+\lambda_2-\lambda_3}+e^{-\lambda_1-\lambda_2+\lambda_3}\right)\cosh^2\varphi}\,.
\end{equation}

In order that a solution preserves some of the supersymmetry of the maximal gauged supergravity theory
we require
\begin{equation}\label{susy1}
\left[\nabla_\mu-\frac{i}{2}Q_\mu-\frac{g}{2\sqrt{2}}W\gamma_\mu-i\frac{1}{4\sqrt{2}}H_{\nu\rho}\gamma^{\nu\rho}\gamma_\mu\right]\epsilon\,=\,0\,,
\end{equation}
where $\epsilon$ is a complex $D=4$ Dirac spinor and we define
\begin{align} \label{hhbbdef}
H_{\mu\nu}&\equiv \overline{F}^{78}=\,\frac{1}{2}\left(e^{-\lambda_1-\lambda_2-\lambda_3}F_{\mu\nu}^{0}+e^{-\lambda_1+\lambda_2+\lambda_3}F_{\mu\nu}^{1}+e^{\lambda_1-\lambda_2+\lambda_3}F_{\mu\nu}^{2}+e^{\lambda_1+\lambda_2-\lambda_3}F_{\mu\nu}^{3}\right)\,, \notag \\
Q_\mu&\equiv\,-g\left(A^0_\mu+A^1_\mu+A^2_\mu+A^3_\mu\right)-\frac{1}{2}\left(\cosh2\varphi-1\right)D_\mu\theta\,.
\end{align}
Notice that the supersymmetry parameters are only charged with respect to the 
gauge field $g\left(A^0_\mu+A^1_\mu+A^2_\mu+A^3_\mu\right)$
and have charge $-1/2$. 
In addition, we also require
\begin{align}\label{susy2}
\left[\gamma^\mu\partial_\mu\lambda_1+\frac{g}{\sqrt{2}}\partial_{\lambda_1}W+i\frac{1}{2\sqrt{2}}\gamma^{\mu\nu}\overline{F}_{\mu\nu}^{12}\right]\epsilon\,=&\,0\,, \notag \\
\left[\gamma^\mu\partial_\mu\lambda_2+\frac{g}{\sqrt{2}}\partial_{\lambda_2}W+i\frac{1}{2\sqrt{2}}\gamma^{\mu\nu}\overline{F}_{\mu\nu}^{34}\right]\epsilon\,=&\,0\,, \notag \\
\left[\gamma^\mu\partial_\mu\lambda_3+\frac{g}{\sqrt{2}}\partial_{\lambda_3}W+i\frac{1}{2\sqrt{2}}\gamma^{\mu\nu}\overline{F}_{\mu\nu}^{56}\right]\epsilon\,=&\,0\,, \notag \\
\left[\gamma^\mu\partial_\mu\varphi+\frac{g}{\sqrt{2}}\partial_\varphi{W}+i\frac{1}{2}\partial_\varphi{Q}_\mu\gamma^\mu\right]\epsilon\,=&\,0\,,
\end{align}
where 
\begin{align}\label{effrels}
\overline{F}^{12}_{\mu\nu}=\,\frac{1}{2}\left(e^{-\lambda_1-\lambda_2-\lambda_3}F_{\mu\nu}^{0}+e^{-\lambda_1+\lambda_2+\lambda_3}F_{\mu\nu}^{1}-e^{\lambda_1-\lambda_2+\lambda_3}F_{\mu\nu}^{2}-e^{\lambda_1+\lambda_2-\lambda_3}F_{\mu\nu}^{3}\right)\,, \notag \\
\overline{F}^{34}_{\mu\nu}=\,\frac{1}{2}\left(e^{-\lambda_1-\lambda_2-\lambda_3}F_{\mu\nu}^{0}-e^{-\lambda_1+\lambda_2+\lambda_3}F_{\mu\nu}^{1}+e^{\lambda_1-\lambda_2+\lambda_3}F_{\mu\nu}^{2}-e^{\lambda_1+\lambda_2-\lambda_3}F_{\mu\nu}^{3}\right)\,, \notag \\
\overline{F}^{56}_{\mu\nu}=\,\frac{1}{2}\left(e^{-\lambda_1-\lambda_2-\lambda_3}F_{\mu\nu}^{0}-e^{-\lambda_1+\lambda_2+\lambda_3}F_{\mu\nu}^{1}-e^{\lambda_1-\lambda_2+\lambda_3}F_{\mu\nu}^{2}+e^{\lambda_1+\lambda_2-\lambda_3}F_{\mu\nu}^{3}\right)\,, \notag \\
\overline{F}^{78}_{\mu\nu}=\,\frac{1}{2}\left(e^{-\lambda_1-\lambda_2-\lambda_3}F_{\mu\nu}^{0}+e^{-\lambda_1+\lambda_2+\lambda_3}F_{\mu\nu}^{1}+e^{\lambda_1-\lambda_2+\lambda_3}F_{\mu\nu}^{2}+e^{\lambda_1+\lambda_2-\lambda_3}F_{\mu\nu}^{3}\right)\,, 
\end{align}
and recall that in \eqref{susy1} $H_{\mu\nu}\equiv \overline{F}^{78}_{\mu\nu}$.

\subsection{The ABJM $AdS_4$ vacuum}
This model admits the maximally supersymmetric $AdS_4$ vacuum solution with vanishing matter fields and the $AdS_4$ metric having radius squared equal to
\begin{align}\label{abjmelldef}
L^2\equiv \frac{1}{2g^2}\,.
\end{align}
This solution uplifts to the $AdS_4\times S^7$ solution dual to ABJM theory.
Within ABJM theory we can identify the scalar fields $\lambda_i$ with 
bosonic mass operators, of conformal dimension $\Delta=1$, while $\zeta$ is dual to a fermionic mass 
operator, with scaling dimension $\Delta=2$.

The four gauge fields $(A^0,A^1,A^2,A^3)$ are dual to $U(1)^4\subset SO(8)$ R-symmetry currents $J_\alpha$.
Associated with the decomposition $SO(8)\to SO(6)\times SO(2)$ and then $SO(6)\to SU(3)\times U(1)\to U(1)^3$ 
it is also natural to consider
the gauge fields
\begin{align}\label{abjmdefsa}
A_R&=A^0+A^1+A^2+A^3\,,\qquad
A_{F_1}=A^1-A^2\,,\nn
A_{F_2}&=A^2-A^3\,,\qquad\qquad\qquad
A_{F'}=\frac{1}{2}(3A^0 -A^1-A^2-A^3)\,,
\end{align}
with the associated currents 
\begin{align}\label{abjmdefsacurr}
J_R^{ABJM}&=\frac{1}{4}(J_0+J_1+J_2+J_3)
\,,\qquad
J_{F_1}=\frac{1}{3}(2J_1-J_2-J_3)\,,\nn
J_{F_2}&=\frac{1}{3}(J_1+J_2-2J_3)\,,\qquad\qquad
J_{F'}=\frac{1}{6}(3J_0-J_1-J_2-J_3)\,,
\end{align}
and note $A^\alpha J_\alpha=A_RJ_R^{ABJM}+A_{F_1}J_{F_1}+A_{F_2}J_{F_2}+A_{F'}J_{F'}$.

For ABJM solutions with $\varphi\ne 0$ we can instead consider
\begin{align}\label{abjmdefsaphinz}
A_R&=A^0+A^1+A^2+A^3\,,\qquad
A_{F_1}=A^1-A^2\,,\nn
A_{F_2}&=A^2-A^3\,,\qquad\qquad\qquad
A_{B}=A^0 -A^1-A^2-A^3\,,
\end{align}
with associated currents 
\begin{align}\label{abjmfnzjs}
J_R^{\varphi}&=\frac{1}{6}(3J_0+J_1+J_2+J_3)\,,\qquad
J_{F_1}=\frac{1}{3}(2J_1-J_2-J_3)\,,\nn
J_{F_2}&=\frac{1}{3}(J_1+J_2-2J_3)\,,\qquad\qquad
J_B=\frac{1}{6}(3J_0-J_1-J_2-J_3)\,,
\end{align}
and note $A^\alpha J_\alpha=A_RJ_R^{\varphi}+A_{F_1}J_{F_1}+A_{F_2}J_{F_2}+A_{B}J_B$.

When $\varphi=0$ all currents in \eqref{abjmdefsacurr} and \eqref{abjmfnzjs}
are conserved, but when $\varphi\ne 0$ the conserved currents are
$J_R^{\varphi}$, $J_{F_1}$ and $J_{F_2}$. It will be important later to note
that $J_R^{ABJM}=J_R^{\varphi}-\frac{1}{2}J_{B}$. We also note that $J_B=J_{F'}$.

\subsection{The mABJM $AdS_4$ vacuum}
The model also admits a supersymmetric $AdS_4$ solution \cite{Warner:1983vz} with 
\begin{align}\label{lsfpsc}
\tilde L^2 \equiv \frac{2}{3\sqrt{3}g^2},\qquad 
e^{\lambda_i}=3^{1/4},\qquad \tanh\varphi\,=\,\frac{1}{\sqrt{3}}\,,
\end{align}
where $\tilde L$ is the radius of the $AdS_4$,
and vanishing gauge fields.
This solution preserves $SU(3)\times U(1)_R$ global symmetry and is dual
to the $d=3$, $\mathcal{N}=2$ mABJM SCFT that arises as the IR fixed point in the RG flow of mass-deformed
ABJM theory \cite{Ahn:2000aq,Corrado:2001nv,Benna:2008zy,Klebanov:2008vq}. 
In this background $\lambda_i,\varphi$ mix and are dual
to relevant operators with scaling dimension
 $\Delta =1, 1, \frac{1}{2}+\frac{1}{2}\sqrt{17}$  as well as an irrelevant operator of dimension $\frac{5}{2}+\frac{1}{2}\sqrt{17}$. 
In more detail, the combinations $\lambda_1-\lambda_2$ and $\lambda_2-\lambda_3$ do not mix while
$\lambda_1+\lambda_2+\lambda_3$ does mix with $\varphi$. Since $\lambda_1-\lambda_2$ and $\lambda_2-\lambda_3$ do not participate in the RG flow from ABJM to mABJM, we conclude that they remain dual to $\Delta=1$ operators and are each a component of the multiplets containing the
$U(1)^2\subset SU(3)$ flavour symmetry currents that we discuss below.

There is also a massive vector dual to an irrelevant vector operator with scaling dimension $\Delta=\frac{3}{2}+\frac{1}{2}\sqrt{17}$,
and three conserved currents consisting of an R-symmetry current and two flavour symmetry currents.
For the mABJM analysis it will be convenient to define
\begin{align}\label{mabjmbasis}
A_R^{mABJM}&=\frac{1}{2}(A^0+3A^1+3A^2+3A^3)\,,\qquad
A_{F_1}=A^1-A^2\,,\nn
A_{F_2}&=A^2-A^3\,,\qquad\qquad\qquad
A_{B}=A^0 -A^1-A^2-A^3\,,
\end{align}
where $A_B$ is the massive vector, dual to an irrelevant vector operator, while $A_R^{mABJM}$ and $A_{F_i}$ are massless vectors dual to the R-symmetry and the flavour symmetries, respectively. 
Notice that if we set $A_B=0$ then $A_R^{mABJM}=A_R$, with $A_R$ appearing in
\eqref{abjmdefsa}, \eqref{abjmdefsaphinz}. 
The associated currents in mABJM theory, dual to $A_R^{mABJM}$ and $A_{F_i}$, are again denoted
$J_R^{\varphi}$ and $J_{F_i}$, respectively.

The various bosonic operators can be arranged in terms of (partial) $\mathcal{N}=2$ SCFT multiplets.
In the notation of \cite{Cordova:2016emh}, the scalars with irrational dimension and the massive vector are part of
a massive vector multiplet labelled $L\bar L$ multiplet on page 67 of  \cite{Cordova:2016emh};
the R-symmetry current and the stress tensor are in 
the graviton multiplet as in (5.45) of  \cite{Cordova:2016emh}; the scalars with $\Delta=1$ and the flavour symmetry currents
are partial current multiplets as in (5.44) of  \cite{Cordova:2016emh}. Further comments on
the spectrum can be found in \cite{Klebanov:2008vq,Bobev:2010ib}.

\subsection{Further subtruncations}\label{secfurthertruncs}
 There are additional consistent truncations one can consider. We can set $\zeta=0$ to recover the STU model \cite{Cvetic:1999xp}; and if we further set 
 $\lambda_i=0$ and $A^{0}=A^{1}=A^{2}=A^{3}$ we get minimal gauged supergravity with an $AdS_4$ vacuum that uplifts to 
the  $AdS_4\times S^7$ solution dual of ABJM theory.

Alternatively, with $\zeta\ne 0$ we can set $A^{1}=A^{2}=A^{3}$ and $\lambda_1=\lambda_2=\lambda_3$ to get a theory with $\zeta\ne 0$
 containing both 
 $AdS_4$ solutions. It is a truncation of the $SU(3)$ invariant theory constructed in \cite{Bobev:2010ib}. 
In particular,
 solutions of this theory will preserve the $SU(3)$ flavour symmetry of the mABJM SCFT fixed point.
  Finally, we can set $\frac{1}{3}A^{0}=A^{1}=A^{2}=A^{3}$ and the set the scalars to the constant values
 in \eqref{lsfpsc}; we again get minimal gauged supergravity but now with an $AdS_4$ vacuum that uplifts to the 
$AdS_4\times S^7$ solution dual to the mABJM SCFT.

When $\zeta =0$, there is also a larger ``2+2" truncation of the STU model with
$A^{0}=A^{1}$ and $A^2=A^3$ and $\lambda_2=\lambda_3=0$, which is a truncation of
$\mathcal{N}=4$ gauged supergravity (e.g. \cite{deWit:1984wbb,Cvetic:1999au,Ferrero:2021ovq}).
For this theory, generically the defect solutions that we will consider do not get an enhancement of
supersymmetry. However, for defect solutions of the truncated theory with\footnote{Similar comments apply to
the truncation with $\zeta=0$, $A^0=A^1=0$ and $\lambda^2=-\lambda^3$.}
 $\zeta=0$, 
$A^2=A^3=0$ and $\lambda^2=\lambda^3$ we do get an enhancement of supersymmetry to four Poincar\'e supersymmetries;
in particular the subset of these solutions which further have $A^{0}=A^{1}$ and 
$\lambda^2=\lambda^3=0$ also lie with the $2+2$ truncation just mentioned.
For defect solutions of the truncated theory with $\zeta=0$, $A^1=A^2=A^3=0$ and $\lambda^1=\lambda^2=\lambda^3$ there is a further enhancement of supersymmetry
to eight Poincar\'e supersymmetries (these solutions necessarily have $n\ne1 $ i.e. conical singularities on the boundary).
 
\section{$AdS_2$ ansatz}\label{sec:ads3}
We are interested in constructing supersymmetric solutions using the ansatz
\begin{align}\label{ads2ans}
ds^2&=e^{2V}ds^2(AdS_2)+f^2dy^2+h^2 dz^2\,,\nn
A^{\alpha}&=a^{\alpha}dz\,.
\end{align}
Here $ds^2(AdS_2)$ is a unit radius metric on $AdS_2$ and $V,f,h,a^{\alpha}$ are functions of $y$ only. 
The scalar fields $\lambda_i,\varphi$ are functions of $y$ only. The phase of the complex scalar field, $\theta$, 
is linear in $z$, $\theta=\bar\theta z$, with $\bar\theta$ a constant and hence 
we can write $Q_\mu dx^\mu\equiv Q_z(y)dz$. 
As we will discuss, regularity of the solutions with non-vanishing $\varphi$ requires we work in a gauge with $\bar\theta=0$.

We will take $z$ to be a periodic coordinate with period $\Delta z=2\pi$. We also take $y\in [y_{core},\infty)$, with $y=y_{core}$ 
the core of the solution (in the bulk) and $y\to \infty$ associated with an $AdS_4$ vacuum, dual to either ABJM or the mABJM
fixed point. We will be more precise about the boundary conditions at the $AdS_4$ boundary later, but here we note that we will have
$e^{2V}\to e^{2V_0}R^2 y^2+\dots$,  $h^2\to e^{2V_0} R^2 n^2y^2+\dots$ and
$f^2\to R^2y^{-2}$, so that the metric approaches $AdS_4$ in the form
\begin{align}\label{ads3ansbdy}
ds^2&=R^2\left(\frac{dy^2}{y^2}+e^{2V_0}y^2[ds^2(AdS_2)+n^2 dz^2]\right)+\dots \,,
\end{align}
where $R=L$ or $\tilde L$ is the $AdS_4$ radius for the ABJM and mABJM vacua, respectively,
and $n>0$ is a constant.
The associated four-dimensional boundary is a regular $AdS_2\times S^1$ with metric given by
\begin{align}\label{flatbdy}
e^{2V_0}(ds^2(AdS_2)+n^2 dz^2)=e^{2V_0}\left(\frac{1}{\rho^2}[-dt^2+d\rho^2+n^2 \rho^2dz^2]\right)\,.
\end{align}
where $n$ specifies the ratio of the radius of $S^1$ to that of $AdS_2$.,

Now, since we take $\Delta z=2\pi$, if $n=1$ the three-dimensional boundary metric is related to the flat space metric on $\mathbb{R}^{1,2}$
by a Weyl transformation. Under this transformation, the boundary of the $AdS_2$ space, located at $\rho=0$, gets mapped to the origin
of the spatial $\mathbb{R}^2\subset \mathbb{R}^{1,2}$, the axis of azimuthal symmetry.
If $n<1$ then the spatial $\mathbb{R}_n^2\subset \mathbb{R}^{1,2}$ boundary, has a
co-dimension two conical deficit angle, while if $n>1$ there is a conical excess angle. 
Later we will present various one point functions associated with the conformal boundary $AdS_2\times S^1$; one can translate these to one point functions on flat spacetime via a Weyl transformation.

The gauge fields have an asymptotic expansion as $y\to\infty$ of the form
\begin{align}
a^{\alpha}&=\mu^{\alpha}+\frac{j^{\alpha}}{y}+\dots\,.
\end{align}
The boundary value $g\mu^{\alpha}$ is associated with a background gauge field, $g\mu^{\alpha}dz$, in the dual 
SCFT that cannot, in general, be gauged away.\footnote{We work with gauge fields that are regular in the bulk.}
In the $AdS_2\times S^1$ boundary this corresponds to a non-trivial monodromy for the 
$U(1)^4$ global symmetry along the $S^1$, with the $g\mu^{\alpha}$ being periodic variables.\footnote{The precise periodicity is somewhat subtle.
In the context of free fields, for example, it depends on possible choices of boundary conditions \cite{Alford:1989ie,Bianchi:2021snj}. The issue is also discussed in \cite{Arav:2024exg}.}
After a Weyl transformation to the flat space metric, $g\mu^{\alpha}dz$ corresponds to the insertion of a monodromy defect located at the origin of the spatial $\mathbb{R}^2_n\subset \mathbb{R}^{1,2}$.

For the case of ABJM boundary, as discussed in appendix \ref{appa}, we find that the BPS equations imply  
\begin{align}\label{sussyrcon}
 g\mu_R\equiv g\mu^{0}+g\mu^{1}+g\mu^{2}+g\mu^{3}=-{\kappa}n-s\,,
 \end{align}
 where $s=\pm 1$ appears in the phase of the Killing spinor and $\kappa=\pm 1$. 
 We will see that there are two branches of solutions, one with $s=-\kappa$ which exists for all $0<n$ and
 a second branch with $s=+\kappa$ which can only exist for $0<n<1$. The $s=-\kappa$ branch is of most interest, and will be referred to
 as the ``main branch", since
 it includes the possibility that there is no conical singularity, $n=1$, in which case the supersymmetry condition
is the vanishing of the R-symmetry monodromy, $g\mu_R=0$.
 For ABJM boundary, the solutions with $\varphi=0$ can be constructed analytically in the STU model and are discussed
 in section \ref{sectstu}. 
 
For ABJM boundary we are also interested in solutions with $\varphi\ne 0$; more specifically solutions that
 have boundary conditions on $\varphi$ associated with a source, $\varphi_s$, for a dual $\Delta=2$ mass operator.\footnote{{Further discussion of this mass operator is given at the end of section \ref{abjmcase}.}}
 For $AdS_2\times S^1$ boundary this corresponds to ABJM theory with a constant mass deformation, while for $\mathbb{R}\times \mathbb{R}_n^2$ boundary, after carrying out the Weyl transformation on the boundary,
this corresponds to a spatially dependent and supersymmetric mass source that preserves the superconformal invariance of the defect and are hence
bulk marginal mass deformations. 
For these solutions, we have the additional constraint\footnote{
A way to understand this constraint is to carry out a singular gauge transformation to set $g\mu_B=0$ which then generates a non-zero phase for the complex scalar source. Moving to the flat space Weyl frame this gives rise to a source $(\varphi_s/\rho) e^{i\kappa n z}$, which is a holomorphic function of the spatial coordinates, as required for preservation of supersymmetry (see also \cite{Anderson:2019nlc}). } 
 \begin{align}\label{sussyrconB}
g\mu_B\equiv g\mu^{0}-g\mu^{1}-g\mu^{2}-g\mu^{3}=\kappa n\,.
\end{align}
As noted above, this condition does not need to be imposed in the STU model solutions;
if it is imposed by hand, we call the solutions ``restricted STU solutions" and they then arise as the limiting solutions of those for
ABJM theory with spatially dependent mass deformations when we take $\varphi\to 0$.

For the mABJM case, as discussed in appendix \ref{secappcmabjm}, we must have the constraint
\begin{align}
g\mu_B=0\,.
\end{align}
In addition, the BPS equations imply
that  there is a constraint on the R-symmetry monodromy source 
\begin{align}\label{gmronetime}
&g\mu_R \equiv  g\mu^{0}+g\mu^{1} + g\mu^{2} + g\mu^{3} =-\kappa n - s\,.
\end{align}
This is the same as in \eqref{sussyrcon} and can be understood from the fact that 
when $g\mu_B=0$, we have
\begin{align}\label{gmrtwotime}
g\mu_R^{mABJM}&\equiv\frac{1}{2}(g\mu^0+3g\mu^1+3g\mu^2+3g\mu^3)=2g\mu^0=g\mu^R\,.
\end{align}
where $g\mu_R^{mABJM}$ is the monodromy source for the R-symmetry in the basis for the bulk gauge fields
given in \eqref{mabjmbasis}.
For mABJM boundary there are again two branches of solutions, $s=-\kappa$ which exists for all $n>0$ and $s=+\kappa$ which can exist
for $0<n<1$. In particular, the $s=-\kappa$ ``main branch" includes the case of no conical singularity, $n=1$, and in this case the BPS equations then imply $g\mu_R=0$.
For mABJM boundary, the bulk solutions have $\varphi\ne 0$, but in contrast to the ABJM solutions with $\varphi\ne 0$,
this is \emph{not} associated with the addition
of any additional spatially dependent mass deformations.

In the remainder of this section, we now focus on solutions with 
\begin{align}
\varphi\ne0\,,
\end{align}
covering monodromy defects in ABJM boundary with spatially dependent mass deformations (in flat spacetime) and mABJM theory without such deformations. We analyse various aspects of the BPS equations, leading
to some of our main results which are presented in section \ref{abjmcase} and \ref{secmABJM case}. We then discuss the simpler case of STU solutions with $\varphi=0$ in section \ref{sectstu}. 

\subsection{Integrating the gauge equations of motion}
The equations of motion for the gauge fields can be combined and integrated to give three integrals of motion,
\begin{align} \label{constantmo2}
\mathcal{E}_{R_1}&\equiv e^{2V}\left(e^{-2\lambda_1-2\lambda_2-2\lambda_3}F_{23}^0+e^{-2\lambda_1+2\lambda_2+2\lambda_3}F_{23}^1\right)\,, \notag \\
\mathcal{E}_{R_2}&\equiv e^{2V}\left(e^{-2\lambda_1-2\lambda_2-2\lambda_3}F_{23}^0+e^{2\lambda_1-2\lambda_2+2\lambda_3}F_{23}^2\right)\,, \notag \\
\mathcal{E}_{R_3}&\equiv e^{2V}\left(e^{-2\lambda_1-2\lambda_2-2\lambda_3}F_{23}^0+e^{2\lambda_1+2\lambda_2-2\lambda_3}F_{23}^3\right)\,,
\end{align}
where $\mathcal{E}_{R_i}$ are constant and the field strength components that are appearing are the frame components:
\begin{align}
F^{\alpha}_{23}=f^{-1}h^{-1}(a^{\alpha})'\,.
\end{align}
We can also define the following linear combinations
\begin{align} \label{constantmo1}
\mathcal{E}_{F_1}\equiv\mathcal{E}_{R_1}-\mathcal{E}_{R_2}&=e^{2V}\left(e^{-2\lambda_1+2\lambda_2+2\lambda_3}F_{23}^1-e^{2\lambda_1-2\lambda_2+2\lambda_3}F_{23}^2\right)\,, \notag \\
\mathcal{E}_{F_2}\equiv \mathcal{E}_{R_2}-\mathcal{E}_{R_3}&=
e^{2V}\left(e^{2\lambda_1-2\lambda_2+2\lambda_3}F_{23}^2-e^{2\lambda_1+2\lambda_2-2\lambda_3}F_{23}^3\right)\,, \notag \\
\mathcal{E}_{F_3}\equiv \mathcal{E}_{R_3}-\mathcal{E}_{R_1}&=e^{2V}\left(e^{2\lambda_1+2\lambda_2-2\lambda_3}F_{23}^3-e^{-2\lambda_1+2\lambda_2+2\lambda_3}F_{23}^1\right)\,.
\end{align}
As we will see, the $\mathcal{E}_{R_i}$ are proportional to the expectation values of the three conserved currents in the dual theory (when $\varphi\ne 0$).
In fact, we will see that the three independent combinations $\sum_i\mathcal{E}_{R_i}$, $\mathcal{E}_{F_1}$, $\mathcal{E}_{F_2}$ are rather natural.

After defining
\begin{align}\label{brokeneb1}
\mathcal{E}_B
&\equiv e^{2V}\left(3e^{-2\lambda_1-2\lambda_2-2\lambda_3}F^{0}_{23}
-e^{-2\lambda_1+2\lambda_2+2\lambda_3}F^{1}_{23}
-e^{2\lambda_1-2\lambda_2+2\lambda_3}F^{2}_{23}
-e^{2\lambda_1+2\lambda_2-2\lambda_3}F^{3}_{23}
\right)\,,
\end{align}
the remaining independent gauge equation of motion can be written in the form 
\begin{align}\label{brokenebeom}
\mathcal{E}_B'
&=-3g e^{2V}fh^{-1}\frac{1}{2}\sinh^2\left(2\varphi\right)D_z\theta\,.
\end{align}
In particular, when $\varphi=0$ we see that $\mathcal{E}_B$ is constant, giving another integral of motion, associated
with the fact that there is a fourth conserved current in the boundary theory.

\subsection{BPS equations}\label{bpsbulkdiscsec}

In the obvious orthonormal frame and with a convenient set of gamma matrices, the Killing spinor has the form $\epsilon=\psi\otimes\chi$,
with $\psi$ a two component spinor on $AdS_2$ satisfying $D_m\psi=\frac{1}{2}\kappa\Gamma_m \psi$
with $\kappa=\pm 1$ and\footnote{
Note that this phase differs by a factor of 2 from the analogue in \cite{Arav:2024exg}.}
\begin{align}\label{chiralityspin}
\chi  =e^{V/2}e^{\frac{isz}{2}}\begin{pmatrix}\sin\frac{\xi}{2} \\ \cos\frac{\xi}{2} \end{pmatrix}\,,
\end{align}
where the constant $s$ is the (gauge-dependent) charge of the spinor under the action of $\partial_z$. 
The BPS equations are derived and analysed in \cite{Suh:2022pkg} and discussed in appendix \ref{sec:appa}.
We will assume that $\sin\xi$ is not identically zero.

An integral of the BPS equations implies
\begin{align}\label{hemvszeqtext}
he^{-V}=-n \sin\xi\,,
\end{align}
where $n$ is a constant. In particular this shows that at the core of the solution, where $h$ vanishes and $e^V$ is constant, $\sin\xi\to 0$ and 
the spinor $\chi$ has a definite chirality with respect to $\sigma_3$.
Also recall that $n$ fixes the deficit angle of the boundary in the flat Weyl frame, with no deficit when $|n|=1$.

Using the BPS equations we can write the integrals of motion
\eqref{constantmo2}, \eqref{constantmo1}, \eqref{brokeneb} in the form
\begin{align} \label{er123}
\mathcal{E}_{R_1}\,=&2ge^{2V}\cos\xi-\sqrt{2}\kappa e^V{e}^{-\lambda_1}\cosh\left(\lambda_2+\lambda_3\right)\,, \notag \\
\mathcal{E}_{R_2}\,=&2ge^{2V}\cos\xi-\sqrt{2}\kappa e^V{e}^{-\lambda_2}\cosh\left(\lambda_3+\lambda_1\right)\,, \notag \\
\mathcal{E}_{R_3}\,=&2ge^{2V}\cos\xi-\sqrt{2}\kappa e^V{e}^{-\lambda_3}\cosh\left(\lambda_1+\lambda_2\right)\,,
\end{align}
and hence
\begin{align}\label{eeffs}
\mathcal{E}_{F_1}\,=&\,\sqrt{2}\kappa{e}^Ve^{\lambda_3}\sinh\left(\lambda_1-\lambda_2\right)\,, \notag \\
\mathcal{E}_{F_2}\,=&\,\sqrt{2}\kappa{e}^Ve^{\lambda_1}\sinh\left(\lambda_2-\lambda_3\right)\,, \notag \\
\mathcal{E}_{F_3}\,=&\,\sqrt{2}\kappa{e}^Ve^{\lambda_2}\sinh\left(\lambda_3-\lambda_1\right)\,.
\end{align}
For the BPS STU solutions with $\varphi=0$,  the remaining independent gauge field equation of motion gives another constant of motion with
\begin{align}\label{brokeneb}
\mathcal{E}_B
&=\frac{\kappa}{\sqrt{2}}e^Ve^{-\lambda_1-\lambda_2-\lambda_3}(-3+e^{2\lambda_1+2\lambda_2}
+e^{2\lambda_2+2\lambda_3}
+e^{2\lambda_1+2\lambda_3}
)\,,
\end{align}
with $\mathcal{E}_B$ constant, when $\varphi=0$.

Importantly, the BPS equations can also be used to re-express the field strengths of the gauge fields
in the form
\begin{equation}\label{effaiprime}
F^{\alpha}_{yz} = (a^{\alpha})' = (\mathcal{I}^{\alpha})' \, ,
\end{equation}
where
we have defined 
\begin{align}\label{eq:IntegratedFluxesExpr1text}
\mathcal{I}^0\,\equiv&-\frac{1}{\sqrt{2}}ne^V\cos\xi\,e^{\lambda_1+\lambda_2+\lambda_3}\,, \quad
\mathcal{I}^1\,\equiv-\frac{1}{\sqrt{2}}ne^V\cos\xi\,e^{\lambda_1-\lambda_2-\lambda_3}\,, \notag \\
\mathcal{I}^2\,\equiv&-\frac{1}{\sqrt{2}}ne^V\cos\xi\,e^{-\lambda_1+\lambda_2-\lambda_3}\,, \quad
\mathcal{I}^3\,\equiv-\frac{1}{\sqrt{2}}ne^V\cos\xi\,e^{-\lambda_1-\lambda_2+\lambda_3}\,.
\end{align}
Notice that
\begin{align}\label{prodIs}
\mathcal{I}^0\mathcal{I}^1=\frac{1}{2}n^2\cos^2\xi e^{2V}e^{2\lambda_1},\quad
\mathcal{I}^0\mathcal{I}^2&=\frac{1}{2}n^2\cos^2\xi e^{2V}e^{2\lambda_2},\quad
\mathcal{I}^0\mathcal{I}^3=\frac{1}{2}n^2\cos^2\xi e^{2V}e^{2\lambda_2}\,,\nn
\mathcal{I}^0\mathcal{I}^1\mathcal{I}^2\mathcal{I}^3&=\frac{1}{4}n^4\cos^4\xi e^{4V}\,.
\end{align}

The BPS equations have a discrete symmetry $(h,z)\to -(h,z)$ along
with $Q_z\rightarrow-Q_z$, $s\rightarrow-s$, $a^\alpha\rightarrow-a^\alpha$, $n\rightarrow-n$ and $F_{23}^\alpha\rightarrow+F_{23}^\alpha$. The frame used to analyse the BPS equations is invariant under this transformation. Without loss of generality, away from the core, where $h\to 0$, we can assume $h> 0$.

\subsection{Boundary conditions at the $AdS_4$ boundary}
We are interested in solutions that asymptotically approach $AdS_4$, either the ABJM fixed point or the mABJM fixed point.
The full expansions of solutions to the BPS equations near the
$AdS_4$ boundary are given in the appendices. Here we just make some simple observations.
Near the boundary, using a Fefferman-Graham gauge $f= R/y$, we have $e^{2V}\to e^{2V_0}y^2$, so that $V'\to 1/y$.
Furthermore, for the ABJM case we have $W\to -2$, while for the mABJM case $W\to -3^{3/4}$. In both cases,
we see from the BPS equation for $V$ in \eqref{bpsb1} that we must have $\sin\xi<0$ as we approach the boundary.
Then from \eqref{hemvszeqtext} and $h>0$, as noted above, we deduce that the constant $n$, whose absolute value determines the deficit angle on the boundary, is positive:
\begin{align}\label{klesszero}
n>0\,.
\end{align}
For solutions approaching the ABJM boundary, the leading asymptotic behaviour of $\varphi$ gives a source $\varphi_s$
for a $\Delta=2$ mass operator; for $AdS_2\times S^1$ boundary this is a constant mass deformation, while for $\mathbb{R}\times\mathbb{R}^2_n$ boundary
it is a spatially varying mass deformation; see also the discussion at the end of section \ref{abjmcase}.

\subsection{Boundary conditions on BPS equations for a regular core}\label{sec:bcscore}

We are interested in solutions that are regular at the core of the solution (in the bulk), $y\to y_{core}$.
Working in a gauge which is regular we demand that $a^{\alpha}=0$ at the core. We are focussing on solutions with 
the complex scalar non-vanishing at the core of the solution, $\varphi\ne0$, and so working with regular gauge fields 
implies that the constant appearing in the phase of the complex scalar should be set to zero, 
\begin{align}
\bar\theta=0\,.
\end{align}  We then also have at the core
\begin{align}\label{dthdzcoe}
\text{Core:}\qquad D_z\theta=0\,,\qquad Q_z=0\,.
\end{align}

We next examine regularity of the metric at the core (for related discussion see \cite{Suh:2022pkg}). To do this it is convenient to use\footnote{We only use this gauge to analyse the core of the solution and will use a different gauge when analysing the $AdS$ boundary.} conformal gauge for the radial coordinate:
\begin{align}\label{confgaugetext}
f=e^V\,.
\end{align}
By examining regularity of the metric at the core we deduce that
\begin{align}\label{coreksinsxi}
\text{Core:}\qquad (n\sin\xi)'=-1,\qquad \cos\xi=(-1)^t\,,
\end{align}
where $t=0,1$.
From the BPS equations (the first in \eqref{bpseqsapp}) we then deduce that $(s-Q_z)=(-1)^{t+1}$ at the core and using \eqref{dthdzcoe}
we can also deduce that the constant $s$ appearing in \eqref{chiralityspin} satisfies
\begin{align}\label{essteerel}
s=(-1)^{t+1}\,.
\end{align}
This condition can also be obtained by ensuring\footnote{To check this one should recast the spinor in an orthonormal frame that
is regular at the core.} that the Killing spinor is smooth at the core of the solution. 
We next note that $\partial_{\varphi}Q_z= -\sinh 2\varphi D_z\theta$ and so
$\partial_{\varphi}Q_z=0$ at the core. From the second constraint in \eqref{twoconstraints}
we thus have at the core
\begin{align}\label{betbc}
\text{Core:}\qquad\partial_{\varphi}Q_z =\partial_{\varphi}W=0.
\end{align}
From the expression for $W$ in \eqref{superpottext} we then also have (for $\varphi\ne0$):
\begin{align}\label{betbc2}
\text{Core:}\qquad e^{2\lambda_1}+e^{2\lambda_2}+e^{2\lambda_3}-e^{2\lambda_1+2\lambda_2+2\lambda_3}=0\
\,, \qquad
W=-e^{\lambda_1+\lambda_2+\lambda_3}\,.
\end{align}
This implies that there are two independent core values of the $\lambda_i$.

We pause to highlight that this constraint eliminates any solutions with $\varphi\ne 0$ in the simpler
2+2 truncation for which $\lambda_2=\lambda_3=0$ along with $A^{0}=A^{1}$ and $A^2=A^3$. Also for the 3+1 truncation
with $\lambda_1=\lambda_2=\lambda_3$ and $A^{1}=A^{2}=A^{3}$, the constraint implies that the core value of $\lambda$ is
the same as for the mABJM fixed point. 
In this  3+1 truncation, if $\varphi$ has the mABJM value at the core, then the defect solutions have the scalars taking the mABJM fixed point values everywhere and we have solutions of minimal gauged supergravity (recall sec \ref{secfurthertruncs}). For other values of $\varphi$ at the core in this truncation, we obtain solutions with varying scalars and approaching the ABJM vacuum at the boundary.

\subsection{Evaluating the conserved charges at the core}

We now want to examine the value of the conserved charges $\mathcal{E}_{R_i}$, given in \eqref{er123}, at 
the core (in conformal gauge \eqref{confgaugetext}). 
Similar to \cite{Suh:2022pkg}, it is convenient to first define 
\begin{equation}\label{defm1m2}
M\,\equiv\,\sqrt{2}ge^{\lambda_1+\lambda_2+\lambda_3}e^V\,, \qquad M>0\,.
\end{equation}
In particular we can write
\begin{equation}\label{coreccs}
\mathcal{E}_{R_i}\,=\,\frac{M^2}{g}\cos\xi e^{-2\left(\lambda_1+\lambda_2+\lambda_3\right)}-\frac{\kappa M}{2g}\left(e^{-2\lambda_i}+e^{-2\left(\lambda_1+\lambda_2+\lambda_3\right)}\right)\,,
\end{equation}
and hence e.g.
\begin{align}\label{coreccsf1f2}
\mathcal{E}_{F_1}=-\frac{\kappa M}{2g}\left(e^{-2\lambda_1}-e^{-2\lambda_2}\right)\,,
\quad
\mathcal{E}_{F_2}=-\frac{\kappa M}{2g}\left(e^{-2\lambda_2}-e^{-2\lambda_3}\right)\,.
\end{align}
Using \eqref{betbc2} and
the first constraint in \eqref{twoconstraints} we deduce that at the core
\begin{align}\label{emmbcpoles}
\text{Core:}\qquad M&=-s\kappa+\frac{1}{n}\,.
\end{align}
Note that these conditions have been derived assuming $\varphi\ne 0$.
Taking the $\varphi\to 0$ limit, for the restricted STU solutions we have the extra conserved quantity
\begin{align}\label{eesemmms2}
\varphi\to 0:\qquad \mathcal{E}_B
&=\frac{\kappa M}{{2}g}e^{-2\lambda_1-2\lambda_2-2\lambda_3}(-3+e^{2\lambda_1+2\lambda_2}
+e^{2\lambda_2+2\lambda_3}
+e^{2\lambda_1+2\lambda_3}
)\,.
\end{align}

Now recall from the $AdS_4$ boundary analysis, with $h>0$ we concluded in \eqref{klesszero} that $n>0$.
Since $M>0$ we deduce from \eqref{emmbcpoles} that if $0<n<1$ then $t$ and hence $s$ is unrestricted, while if
$n \ge 1$ then $(-1)^t=\kappa$ and hence from 
\eqref{essteerel} $s=-\kappa$. Thus, we have two branches of solutions:
\begin{align}\label{restrictionsskappn}
\text{Branch 1 (main branch):}\qquad s&=-\kappa,\qquad \text{arbitrary $n>0$}\,,\nn
\text{Branch 2:}\qquad s&=+\kappa,\qquad \text{$0<n<1$}\,,
\end{align}
Notice that the branch 1 has solutions that are continuously connected to having no conical singularity, $n=1$, and hence
are continuously connected to the vacuum, with no monodromy sources; 
thus, we also refer to this as the ``main branch". We continue our analysis with both branches,
for the most part, sometimes focussing on the main branch for simplicity.

For example, when $s=-\kappa$ from \eqref{gmronetime}, \eqref{gmrtwotime}, the R-symmetry sources can be written
for ABJM and mABJM boundary as
\begin{align}
\text{$s=-\kappa$:}\qquad g\mu_R=\kappa(1-n);\qquad g\mu_R^{mABJM}=\kappa(1-n)\,,
\end{align}
and for the case of no conical singularity these read $g\mu_R=0$ and $g\mu_R^{mABJM}=0$. 
In addition, notice that for $n=1$, at the core we have $M=2$ and
$M^2\cos\xi= {4\kappa}$, so
\begin{equation}\label{konecore}
n=1,\qquad \text{Core:}\qquad\mathcal{E}_{R_i}=\frac{\kappa}{g}(3e^{-2\lambda_1-2\lambda_2-2\lambda_3}-e^{-2\lambda_i})\,,
\end{equation}
while
\begin{align}\label{konecore2}
n=1,\quad \text{Core:}\qquad\mathcal{E}_{F_1}=-\frac{\kappa}{g}\left(e^{-2\lambda_1}-e^{-2\lambda_2}\right)\,,
\quad
\mathcal{E}_{F_2}=-\frac{\kappa}{g}\left(e^{-2\lambda_2}-e^{-2\lambda_3}\right)\,.
\end{align}

For future use, recalling \eqref{eq:IntegratedFluxesExpr1text} and using \eqref{betbc2} and \eqref{emmbcpoles}, at the core (for both branches) we have
\begin{align}\label{Icondssatthecore}
(\mathcal{I}^0-\mathcal{I}^1-\mathcal{I}^2-\mathcal{I}^3)|_\text{core}=0,\qquad
(\mathcal{I}^0)|_\text{core}=\frac{1}{2g}(-\kappa n+s)\,.
\end{align}

\subsection{Evaluating the conserved charges at the $AdS_4$ boundary}
By evaluating the conserved quantities $\mathcal{E}_{R_i}$ at the $AdS_4$ boundary we can relate core quantities to boundary quantities.
As we now discuss, the $\mathcal{E}_{R_i}$ are essentially the one point functions for the
conserved currents of the boundary
SCFT in the presence of the monodromy defect.

\subsubsection{ABJM case, $\varphi\ne 0$}
As shown in appendix \ref{appa}, we can use the boundary expansion for the ABJM case
to evaluate $\mathcal{E}_{R_i}$ at the boundary in terms of the conserved currents and we find
\begin{align}  \label{eeejs1}
\mathcal{E}_{R_1}&=c(\langle J^{0}\rangle +\langle J^{1}\rangle )=c(2\langle{J_R^{\varphi}}\rangle+\vev{J_{F_1}})\,,\nn
\mathcal{E}_{R_2}&=c(\langle J^{0}\rangle +\langle J^{2}\rangle )=c(2\langle{J_R^{\varphi}}\rangle-\vev{J_{F_1}}+\vev{J_{F_2}})\,,\nn
\mathcal{E}_{R_3}&=c(\langle J^{0}\rangle +\langle J^{3}\rangle )=c(2\langle{J_R^{\varphi}}\rangle-\vev{J_{F_2}})\,,
\end{align}
where we used \eqref{abjmfnzjs}. We thus have 
\begin{align}\label{expsforefi}
\mathcal{E}_{R_1}+\mathcal{E}_{R_2}+\mathcal{E}_{R_3}&=6c\langle{J_R^{\varphi}}\rangle\,, \nn
\mathcal{E}_{F_1}&=c(\langle J^{1}\rangle -\langle J^{2}\rangle )=
c(2\vev{J_{F_1}}-\vev{J_{F_2}})
\,,\nn
\mathcal{E}_{F_2}&=c(\langle J^{2}\rangle -\langle J^{3}\rangle )
=
c(-\vev{J_{F_1}}+2\vev{J_{F_2}})
\,,
\end{align}
with the constant $c$ given by 
\begin{align}\label{defcee}
c=-3\sqrt{2}\pi e^{V_0}/(ngN^{3/2})\,,
\end{align}
where $N$ is the rank of the gauge groups in ABJM theory which is dual to the vacuum $AdS_4$ solution (recall \eqref{bulkactandnorm}).
The expressions for $\mathcal{E}_{R_i}$ can be equated with the expressions that we obtained at the core
\eqref{coreccs}
and this
relates three independent conserved currents
${J_R^{\varphi}}$, ${J_{F_i}}$, to the value of the $\lambda_i$ at the core. 
Recall also, that due to the constraint \eqref{betbc2} there are two independent core values of the $\lambda_i$.

We also have
\begin{align}
\mathcal{E}_{B}=c(3\langle  J^{0}\rangle -\langle J^{1}\rangle -\langle J^{2}\rangle-\langle J^{}\rangle)
=6c\langle{J_B}\rangle
\,,
\end{align}
where we used \eqref{abjmfnzjs},
and in the limit $\varphi\to0$, associated with restricted STU solutions with an additional conserved current, this is also constant and can be related to the core quantities via \eqref{eesemmms2}.
 
\subsubsection{mABJM case}
For the mABJM case, we have three conserved charges $\mathcal{E}_{R_i}$ and these are again related to three
conserved currents ${J_R^{\varphi}}$, ${J_{F_i}}$.
In fact the expressions in \eqref{eeejs1}-\eqref{defcee} we found for the 
ABJM case with $\varphi\ne 0$ are also valid for the mABJM case.
Since the ABJM case with $\varphi\ne 0$ and the mABJM case have the same symmetries one might have anticipated that the conserved charges are proportional to the same conserved currents, but it is remarkable
that they also have the same proportionality constant.

\subsection{Relating the monodromy sources to core quantities}\label{relatingscesbdysec}
We now explain how we can analytically relate the (constrained) 
monodromy sources for the currents, $g\mu^{\alpha}$, to quantities at the
core of the defect. Then, by using the results of the previous subsections which relate core quantities to boundary currents
we are able to derive relations between the boundary currents and monodromy sources.

To do this, the key step is to integrate the expression \eqref{effaiprime} which was derived from the BPS equations,
\begin{equation}\label{effaiprimet}
F^{\alpha}_{yz} = (A^{\alpha})' = (\mathcal{I}^{\alpha})' \, ,
\end{equation}
where $\mathcal{I}^{\alpha}$ are defined in \eqref{eq:IntegratedFluxesExpr1text}.
Since we are working in a regular gauge (with $\bar\theta=0$) we have $a^{\alpha}=0$ at the core, and at the $AdS$ boundary, as $y\to\infty$, $a^{\alpha}$ defines the monodromy sources, $ga^{\alpha}\to g\mu^{\alpha}$.
We therefore deduce
\begin{equation}
\label{eq:GeneralIntegExpressionForFluxSources}
g\mu^{\alpha}= g\mathcal{I}^{\alpha}|_\text{bdry} - g\mathcal{I}^{\alpha}|_\text{core} \, ,
\end{equation}
with $\mathcal{I}^{\alpha}$ given in \eqref{eq:IntegratedFluxesExpr1text}.
We next evaluate $\mathcal{I}^{\alpha}$ at the boundary using the boundary expansions, given in the appendices,
and at the core using the boundary conditions given in section \ref{sec:bcscore}. Notice that 
the $\mathcal{I}^{\alpha}$ do not depend on the gauge choice for $f$ and hence
can be evaluated in the conformal gauge at the core and a Fefferman-Graham gauge on the boundary.

Proceeding, from \eqref{prodIs} and $\cos\xi=(-1)^t=-s$, as well as using \eqref{eq:GeneralIntegExpressionForFluxSources},
we can obtain an expression for $e^{4V}$ at the core in terms of boundary data:
\begin{align}\label{efvcoreabjm}
e^{4V}|_{\text{core}}&=\frac{4}{n^4}(\mathcal{I}^{0}|_\text{bdry}-\mu^{0})
(\mathcal{I}^{1}|_\text{bdry}-\mu^{1})
(\mathcal{I}^{2}|_\text{bdry}-\mu^{2})
(\mathcal{I}^{3}|_\text{bdry}-\mu^{3})\nn
&=R^4(1-\frac{\mu^{0}}{\mathcal{I}^{0}|_\text{bdry}})
(1-\frac{\mu^{1}}{\mathcal{I}^{1}|_\text{bdry}})
(1-\frac{\mu^{2}}{\mathcal{I}^{2}|_\text{bdry}})
(1-\frac{\mu^{3}}{\mathcal{I}^{3}|_\text{bdry}})\,.
\end{align}
To get the last line we used the fact that $(\mathcal{I}^{0}\mathcal{I}^{1}\mathcal{I}^{2}\mathcal{I}^{3})|_\text{bdry}=
\frac{n^4}{4}(\cos^4\xi e^{4V})|_\text{bdry}=\frac{n^4}{4}R^4$ where $R$ is the radius of the $AdS_4$ fixed point
(recall \eqref{abjmelldef}, \eqref{lsfpsc}). Similarly,
we can also obtain expressions for $e^{2\lambda_i}$ at the core e.g.
\begin{align}\label{lambdascore}
e^{2\lambda_1}|_{\text{core}}=(e^{-2V}|_{\text{core}})R^2e^{2(\lambda_1)_{fp}}
(1-\frac{\mu^{0}}{\mathcal{I}^{0}|_\text{bdry}})
(1-\frac{\mu^{1}}{\mathcal{I}^{1}|_\text{bdry}})\,,
\end{align}
where we used $\frac{2}{n^2}(\mathcal{I}^{0}\mathcal{I}^{1)})|_\text{bdry}=R^2e^{2(\lambda_1)_{fp}}$ and
$(\lambda_1)_{fp}$ is the value of the scalar at the ABJM or mABJM fixed point.

Notice that we must have $(1-\frac{\mu^{\alpha}}{\mathcal{I}^{\alpha}|_\text{bdry}})>0$ or 
$(1-\frac{\mu^{\alpha}}{\mathcal{I}^{\alpha}|_\text{bdry}})<0$ for all $\alpha$. 
Now we have 
$(1-\frac{\mu^{\alpha}}{\mathcal{I}^{\alpha}|_\text{bdry}})=\mathcal{I}^{\alpha}|_\text{core}/\mathcal{I}^{\alpha}|_\text{bdry}$ 
and so from \eqref{eq:IntegratedFluxesExpr1text} we see that the sign of $(1-\frac{\mu^{\alpha}}{\mathcal{I}^{\alpha}|_\text{bdry}})$ is the same as the sign of $(\cos\xi)_\text{core}/(\cos\xi)_\text{bdry}$. Then combining
\eqref{coreksinsxi},\eqref{essteerel} with \eqref{abjmxiexpappb},\eqref{mabjmxiexpappc} we deduce that for either ABJM or mABJM boundary 
we have the following necessary conditions on the monodromy parameters for the two branches of solutions:
\begin{align}\label{branchesbds}
\text{Main branch}:&\quad s=-\kappa,\quad (1-\frac{\mu^{\alpha}}{\mathcal{I}^{\alpha}|_\text{bdry}})>0,\quad n>0\,,\nn
\text{Branch 2}:&\quad s=+\kappa,\quad (1-\frac{\mu^{\alpha}}{\mathcal{I}^{\alpha}|_\text{bdry}})<0,\quad 0<n<1\,.
\end{align}
We have the additional constraints $g\mu_R=-\kappa n-s$ and, when $\varphi\ne 0$, we also have $g\mu_B=\kappa n$ or $0$,
 as given below for ABJM or mABJM asymptotics, respectively. 
We also have the behaviour of $h\to 0 $ at the core from \eqref{hemvszeqtext} and \eqref{coreksinsxi}. The only quantity not specified at the core is the value of $\varphi$. We also note that the above expressions in this subsection, including the conclusion
concerning the signs of $(1-\frac{\mu^{\alpha}}{\mathcal{I}^{\alpha}|_\text{bdry}})$,
 did not assume $\varphi\ne 0$ and also apply to the STU model, as discussed in section \ref{sectstu};
we will also show there that the necessary constraint $0<n<1$ for branch 2 solutions is also valid the STU model.

\subsection{ABJM case, $\varphi\ne 0$}\label{abjmcase}

We calculate $\mathcal{I}^{\alpha}|_\text{bdry}=-\frac{\kappa n}{2g}$ for all $\alpha$ for ABJM boundary with $\varphi\ne 0$
and hence
from \eqref{efvcoreabjm}, \eqref{lambdascore} obtain
expressions for $e^{4V}$ and $e^{2\lambda_i}$ at the core, finding 
\begin{align}\label{efvcoreabjm2}
L^{-2}e^{2V}|_{\text{core}}&=\mathcal{F}^{ABJM}\equiv \left[\Bigl(1+\frac{2g\mu^0}{\kappa n}\Bigr)\Bigl(1+\frac{2g\mu^1}{\kappa n}\Bigr)\Bigl(1+\frac{2g\mu^2}{\kappa n}\Bigr)\Bigl(1+\frac{2g\mu^3}{\kappa n}\Bigr)\right]^{\frac{1}{2}\ }\,,\nn
e^{2\lambda_1}|_{\text{core}}&=\mathcal{F}^{ABJM}(1+\frac{2g\mu^{2}}{\kappa n})^{-1}(1+\frac{2g\mu^{3}}{\kappa n})^{-1}\,,\nn
e^{2\lambda_2}|_{\text{core}}&=\mathcal{F}^{ABJM}(1+\frac{2g\mu^{1}}{\kappa n})^{-1}(1+\frac{2g\mu^{3}}{\kappa n})^{-1}\,,\nn
e^{2\lambda_3}|_{\text{core}}&=\mathcal{F}^{ABJM}(1+\frac{2g\mu^{1}}{\kappa n})^{-1}(1+\frac{2g\mu^{2}}{\kappa n})^{-1}\,,
\end{align}
where we have defined the function $\mathcal{F}^{ABJM}$ which will appear in the sequel
and we recall from \eqref{branchesbds} that $(1+\frac{2g\mu^{\alpha }}{\kappa n})$ and $-\kappa s$
all have the same sign, with $\kappa s=+1$ only possible for branch 2 with  $0<n<1$.

Note that from $\mathcal{I}^{\alpha}|_\text{bdry}=-\frac{\kappa n}{2g}$ and the constraints on 
$\mathcal{I}^{\alpha}|_\text{core}$ in \eqref{Icondssatthecore},
we also have
\begin{align}\label{eq:FluxSourcesFromCore}
&g\mu_R \equiv [g\mu^{0}+ g\mu^{1} + g\mu^{2} + g\mu^{3}] = -\kappa n - s \,,\nn
&g\mu_B \equiv [g\mu^{0}-g\mu^{1} - g\mu^{2} -g\mu^{3}] = \kappa n \, ,
\end{align}
where $g\mu_B$ corresponds to the monodromy source for the broken $U(1)$ symmetry  
and $g \mu_R$ corresponds to the monodromy source for the $U(1)$ R-symmetry.
These constraints can also\footnote{They
can also be deduced by examining the BPS constraints on the boundary ABJM theory coupled to mass sources and monodromies using \cite{Nishimura:2012jh}, analogous to the analysis in section 2 of \cite{Arav:2020obl}.} be obtained from the boundary expansion of the BPS constraints discussed in appendix
\ref{appa}, as already noted in \eqref{sussyrcon} and \eqref{sussyrconB}.
Notice that for $n=1$ case, associated with no conical defect on the flat boundary,
which is on the main branch with $s=-\kappa$, we have
\begin{equation}
g\mu_R = 0\,,\qquad
g\mu_B = \kappa\,.
\end{equation}

We can also define two flavour sources
\begin{align}\label{muflamcore}
g\mu_{F_1}\equiv g\mu^{1} - g\mu^{2} \,,\qquad
g\mu_{F_2}\equiv g\mu^{2} - g\mu^{3} \,.
\end{align}
Inverting these expressions combined with \eqref{eq:FluxSourcesFromCore} we find, for general $n$ and both branches of $s$,
\begin{align}
 g \mu^{0}&=-\frac{s}{2}\,,\qquad\qquad\qquad\qquad\qquad\qquad\,\,
 g \mu^{1}=\frac{1}{6}(4 g\mu_{F_1}+2g \mu_{F_2}-s-2 \kappa n)\,,\nn
g \mu^{2}&=\frac{1}{6}(-2 g \mu_{F_1}+2g \mu_{F_2}-s-2\kappa n)\,,\quad
 g \mu^{3}=\frac{1}{6}(-2 g \mu_{F_1}-4 g\mu_{F_2}-s-2\kappa n)\,.
\end{align}
We then find that the core quantities can be expressed in terms of $g\mu_{F_1}$, $g\mu_{F_2}$ as
\begin{align}\label{efvcoreabjm2again}
e^{2V}|_{\text{core}}&=L^2\mathcal{F}^{ABJM}\equiv \frac{L^2}{3^{3/2}n^2}\Big[(s-\kappa n)(-4 g\mu_{F_1}-2g \mu_{F_2}+s- \kappa n)\nn
&\qquad\qquad\qquad\qquad(2 g \mu_{F_1}-2g \mu_{F_2}+s-\kappa n)(2 g \mu_{F_1}+4 g\mu_{F_2}+s-\kappa n)\Big]^{\frac{1}{2}\ }\,,\nn
e^{2\lambda_1}|_{\text{core}}&=\mathcal{F}^{ABJM}9n^2(2 g \mu_{F_1}-2g \mu_{F_2}+s-\kappa n)^{-1}(2 g \mu_{F_1}+4 g\mu_{F_2}+s-\kappa n)^{-1}\,,\nn
e^{2\lambda_2}|_{\text{core}}&=\mathcal{F}^{ABJM}9n^2(-4 g\mu_{F_1}-2g \mu_{F_2}+s-\kappa n)^{-1}(2 g \mu_{F_1}+4 g\mu_{F_2}+s-\kappa n)^{-1}\,,\nn
e^{2\lambda_3}|_{\text{core}}&=\mathcal{F}^{ABJM}9n^2(-4 g\mu_{F_1}-2g \mu_{F_2}+s- \kappa n)^{-1}(2 g \mu_{F_1}+4 g\mu_{F_2}+s-\kappa n)^{-1}\,.
\end{align}

\begin{figure}[h!]
	\centering
	\includegraphics[scale=0.55]{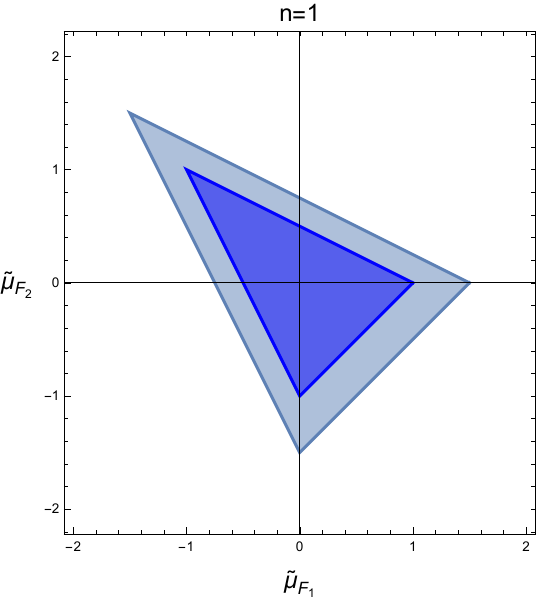}\qquad
		\includegraphics[scale=0.55]{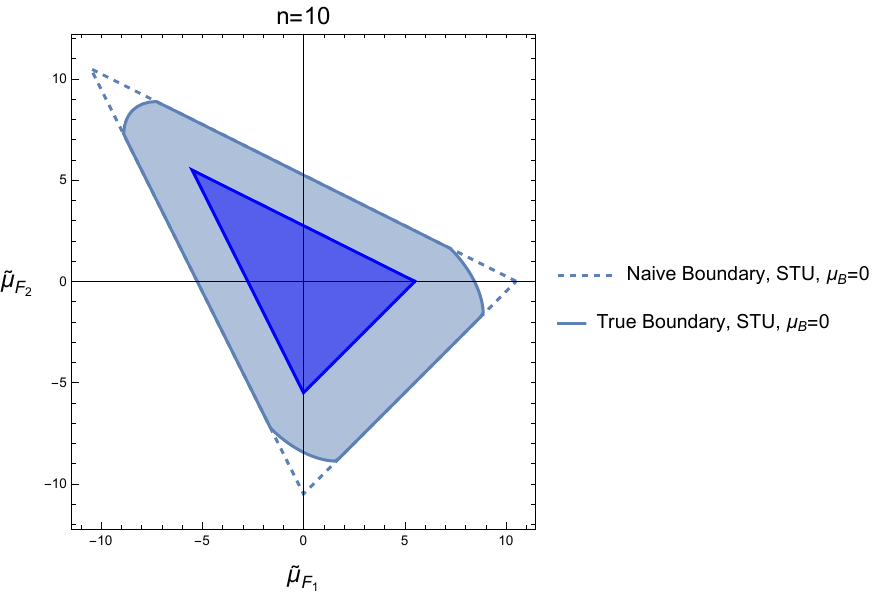}
	\caption{The darker region is the allowed range of $(\kappa g\mu_{F_1},\kappa g\mu_{F_2})\equiv (\tilde\mu_{F_1},\tilde\mu_{F_2})$, 
	for monodromy defects of ABJM theory with spatially dependent mass sources and $g\mu_B=\kappa n$, and also for mABJM theory with $g\mu_B=0$, plotted for $n=1$ in the left panel and $n=10$ in the right panel. 
	The larger lighter region (including the darker interior region) is the allowed range for monodromy defects of ABJM
	theory in the STU model, discussed in section \ref{sectstu}, for the case $g\mu_B=0$, with the dashed line for $n=10$
	encompassing the naive region associated with necessary constraints on the existence of solutions.
	}
	\label{rangesmus}
\end{figure}

The space of possible defect solutions are restricted by the \emph{necessary} conditions given in \eqref{branchesbds} for the two branches.
For the main branch, including $n=1$, we have $s=-\kappa$ and $(1+\frac{2g\mu^{\alpha }}{\kappa n})>0$ for all $\alpha$
which translates into a necessary condition on the allowed
ranges of $g\mu_{F_1}$, $g\mu_{F_2}$, which is plotted as the darker triangular regions in figure \ref{rangesmus} for $n=1$ and $n=10$. 
In section \ref{secmABJM case} we obtain the same result for defects in mABJM theory with $g\mu_B=0$. In section
\ref{numericalsolssec} we construct some numerical solutions and those investigations show, at least for $n=1$ and $n=10$, that these necessary conditions for defect solutions are in fact sufficient for their existence (this is in contrast to the STU solutions with $g\mu_B=0$ that we discuss in
section \ref{sectstu} and also indicated in this figure).

We can now obtain expressions for three independent conserved boundary currents in terms of the monodromy sources, using the three
independent conserved quantities. For example, from \eqref{coreccs}-\eqref{emmbcpoles} and \eqref{expsforefi}
we deduce that the flavour currents satisfy (here for general $n$) 
\begin{align}  \label{eeejs12}
\langle J^1 \rangle -\langle J^2\rangle&=2\langle J_{F_1} \rangle -\langle J_{F_2}\rangle=
\frac{\kappa (1-s \kappa n) N^{3/2}}{6\sqrt{2}\pi e^{V_0}}\left(e^{-2\lambda_1}-e^{-2\lambda_2}\right)|_{\text{core}}
\,,\nn
\langle J^2 \rangle -\langle J^3\rangle&=-\langle J_{F_1} \rangle +2\langle J_{F_2}\rangle=
\frac{\kappa (1-s \kappa n) N^{3/2}}{6\sqrt{2}\pi e^{V_0}}\left(e^{-2\lambda_2}-e^{-2\lambda_3}\right)|_{\text{core}}
\,,
\end{align}
and considering $\mathcal{E}_{R_1}+\mathcal{E}_{R_1}+\mathcal{E}_{R_3}$, using
\eqref{coreccs} and \eqref{eeejs1} we deduce that the
R-symmetry current satisfies
\begin{align}  \label{eeejs123}
&3\langle J^0 \rangle+\langle J^1\rangle+\langle J^2\rangle +\langle J^3\rangle=6\langle J_R^{\varphi} \rangle\nn
&=
\frac{\kappa (1-s \kappa n)N^{3/2}}{6\sqrt{2}\pi e^{V_0}}\left(-(3-\frac{6s\kappa}{n})e^{-2\lambda_1-2\lambda_2-2\lambda_3}+e^{-2\lambda_1}+e^{-2\lambda_2}+e^{-2\lambda_3}\right)|_{\text{core}}\,.
\end{align}
Thus, using \eqref{efvcoreabjm2} or \eqref{efvcoreabjm2again}
the conserved currents $\langle J_{F_1} \rangle$, $\langle J_{F_2} \rangle$ and $\langle J_R^{\varphi} \rangle$
are completely determined by $ g\mu_{F_1}$, $g\mu_{F_2}$ and $n$,
and they are independent of $\varphi_s$, the source for the $\Delta=2$ mass deformation.

We cannot obtain an expression for the remaining fourth independent, non-conserved, current $\vev{J_B}$,
without solving the BPS equations of motion.
This is because when $\varphi\ne 0$, $\mathcal{E}_B$ is not a conserved quantity (recall \eqref{brokenebeom}). However, in
the limit $\varphi\to 0$, associated with a restricted STU solution, we can use a similar approach to deduce
\begin{align}  \label{eeejs123broken}
3\langle J^0 \rangle-&\langle J^1\rangle-\langle J^2\rangle -\langle J^3\rangle=6\langle J_B \rangle\nn
&=\frac{\kappa(1-s \kappa n) N^{3/2}}{6\sqrt{2}\pi e^{V_0}}\left(3e^{-2\lambda_1-2\lambda_2-2\lambda_3}-e^{-2\lambda_1}-e^{-2\lambda_2}-e^{-2\lambda_3}\right)|_{\text{core}}\,.
\end{align}

From the results for the stress tensor in appendix \ref{appa} we obtain for $\varphi\ne 0$ 
\begin{align}
\langle T_{ab}\rangle dx^a dx^b =-\frac{h_D}{2\pi}\left[ds^2(AdS_2)-2 n^2 dz^2\right]\,,
\end{align}
with $h_D$ the conformal weight of the conformal defect (by definition), explicitly given by 
\begin{align}\label{abjmfnzexpccb}
\frac{h_D}{2\pi}=\frac{1 }{4\kappa n }\sum_\alpha \langle J^{\alpha} \rangle&=
\frac{1 }{\kappa n } \langle J_R^{ABJM} \rangle\nn
& =\frac{1}{\kappa n}\left(  \langle J_R^{\varphi} \rangle-\frac{1}{2} \langle J_{B} \rangle \right)\,.
\end{align}
We highlight that for $\varphi\ne 0$, supersymmetry is fixing $h_D$ in terms of the ABJM R-symmetry current $\langle J_R^{ABJM} \rangle$. In particular it is not expressed 
just in terms of the conserved currents $\langle J_R^{\varphi}\rangle$ and $\langle J_{F_i}\rangle$, but also depends
on the non-conserved current $ \langle J_{B} \rangle$. Thus, $h_D$ will vary as a function of $\varphi_s$ and to determine its explicit value requires solving the BPS equations.
It is interesting to also highlight that as we approach the boundary of the allowed values of $g\mu_{F_i}$ in
figure \ref{rangesmus} we find\footnote{This can be contrasted with the 
analogous defect solutions for $d=4$ SCFTs discussed in \cite{Arav:2024exg}.} that there are divergences in the 
conserved currents $\langle J_R^{\varphi}\rangle $ and $\langle J_{F_i}^{\varphi}\rangle $.

Using the results from appendix \ref{appa} we can also obtain the expectation values of the
scalar operators. To be consistent with supersymmetry, the three scalars $\lambda_i$ are dual to operators with scaling dimension $\Delta=1$, 
while $\varphi$ is dual to an operator with $\Delta=2$. To ensure the former, as usual, we need to include additional
boundary terms associated with a suitable Legendre transform of the on-shell action. This gives rise to new 
fields $\Pi_i$. Interestingly,\footnote{This can also be contrasted with the $D=5$ supergravity solutions of
\cite{Arav:2024exg}, where there are separate sources for the fermionic and bosonic mass operators.} 
the BPS equations imply that the associated sources for the
$\Delta=1$ bosonic mass operators necessarily vanish, $\Pi^{(s)}_i=0$. 
Since the deformation parametrised by $\varphi_s$ is preserving supersymmetry, we conclude that
the operator $\mathcal{O}^{\Delta=2}_\varphi$ dual to $\varphi$ 
(when keeping $\Pi^{(s)}_i$ fixed) must be proportional\footnote{The same observation is also applicable in the case of ordinary RG flows from ABJM theory to mABJM theory, driven by a homogeneous mass deformation, a point that seems to have been overlooked in the literature.} to a linear combination
of a fermion mass operator of dimension ${\Delta=2}$, and $\varphi_s$ times a bosonic mass 
operator of dimension ${\Delta=1}$.

The expectation values of the scalars, with an $AdS_2\times S^1$ boundary are given by
\begin{align}\label{scalvevstextabjm}
\langle \mathcal{O}^{\Delta=2}_\varphi \rangle&=0\,,\nn
\langle\mathcal{O}^{\Delta=1}_{\Pi_1} \rangle&=- \frac{g   }{2 \kappa n} 
(\langle J^{0}\rangle +\langle J^{1}\rangle
-\langle J^{2}\rangle-\langle J^{3}\rangle )
= -\frac{g   }{\kappa n}(\vev{J_{B}}+\vev{J_{F_1}}) \,,\nn
\langle\mathcal{O}^{\Delta=1}_{\Pi_2} \rangle&=-\frac{g   }{2 \kappa n}  (
\langle J^{0}\rangle-\langle J^{1}\rangle
+\langle J^{2}\rangle-\langle J^{3}\rangle)=- \frac{g   }{\kappa n}(\vev{J_{B}}-\vev{J_{F_1}}+\vev{J_{F_2}})\,,\nn
\langle\mathcal{O}^{\Delta=1}_{\Pi_3} \rangle&= -\frac{g  }{2 \kappa n} (\langle J^{0}\rangle-\langle J^{1}\rangle
-\langle J^{2}\rangle+\langle J^{3}\rangle)= -\frac{g   }{\kappa n}(\vev{J_{B}}-\vev{J_{F_2}})\,,
\end{align}
where we have used \eqref{abjmfnzjs}. 
Notice that two 
linear combinations of the $\langle\mathcal{O}_{\Pi_i} \rangle$ are
independent of $\vev{J_{B}}$ and hence are fixed by the monodromy sources, for solutions with $\varphi\ne 0$.

\subsection{mABJM case}\label{secmABJM case}

For the mABJM case, the boundary expansion implies that
\begin{align}\label{Isformabjm}
\mathcal{I}^{0}|_\text{bdry}=-\frac{\kappa n}{g}\,,\quad
\mathcal{I}^{i}|_\text{bdry}=-\frac{\kappa n}{3g}\,,\quad \text{for $i=1,2,3$}\,.
\end{align}
We also have 
\begin{align}\label{efvexpMabjmcore}
\tilde L^{-2}e^{2V}|_{\text{core}}=&\mathcal{F}^{mABJM}\equiv \left[\Bigl(1+\frac{g\mu^0}{\kappa n}\Bigr)\Bigl(1+\frac{3g\mu^1}{\kappa n}\Bigr)\Bigl(1+\frac{3g\mu^2}{\kappa n}\Bigr)\Bigl(1+\frac{3g\mu^3}{\kappa n}\Bigr)\right]^{\frac{1}{2}\ }\,.\nn
e^{2\lambda_1}|_{\text{core}}&=\sqrt{3}\mathcal{F}^{mABJM}(1+\frac{3g\mu^{2}}{\kappa n})^{-1}(1+\frac{3g\mu^{3}}{\kappa n})^{-1}\,,\nn
e^{2\lambda_2}|_{\text{core}}&=\sqrt{3}\mathcal{F}^{mABJM}(1+\frac{3g\mu^{1}}{\kappa n})^{-1}(1+\frac{3g\mu^{3}}{\kappa n})^{-1}\,,\nn
e^{2\lambda_3}|_{\text{core}}&=\sqrt{3}\mathcal{F}^{mABJM}(1+\frac{3g\mu^{1}}{\kappa n})^{-1}(1+\frac{3g\mu^{2}}{\kappa n})^{-1}\,,
\end{align}
and we recall the bounds given in \eqref{branchesbds}.
Note that from $\mathcal{I}^{\alpha}|_\text{bdry}$ we also have
\begin{align}\label{eq:FluxSourcesFromCorem}
&g\mu_R \equiv  g\mu^{0}+g\mu^{1} + g\mu^{2} + g\mu^{3} =-\kappa n- s \,,\nn
&g\mu_B\equiv g\mu^{0}-g\mu^{1} - g\mu^{2} - g\mu^{3}=0 \, ,
\end{align}
and in particular, $g\mu_R$ is unchanged from the ABJM case with $\varphi\ne 0$.
For the special case of $n=1$, necessarily on the main branch with $s=-\kappa$, these simplify to
\begin{equation}
g\mu_B= 0, \quad
g\mu_R = 0\,.
\end{equation}

We can again define two flavour sources
\begin{align}\label{muflamcorem}
g\mu_{F_1}\equiv g\mu^{1} - g\mu^{2}\,,\qquad
g\mu_{F_2}\equiv g\mu^{2} - g\mu^{3}\,,
\end{align}
and inverting these and \eqref{eq:FluxSourcesFromCorem} we find 
\begin{align}
g \mu^{0}&=\frac{1}{2}(-s-\kappa n)\,,\qquad\qquad\qquad\qquad\quad
g \mu^{1}=\frac{1}{6}(4 g\mu_{F_1}+2g\mu_{F_2}-s-\kappa n)\,,\nn
g \mu^{2}&=\frac{1}{6}(-2 g\mu_{F_1}+  2g\mu_{F_2}-s-\kappa n)\,,\quad
g \mu^{3}=\frac{1}{6}(- 2g\mu_{F_1}-4g\mu_{F_2}-s-\kappa n)\,.
\end{align}
This can be substituted into \eqref{efvexpMabjmcore} to express the core quantities in terms of the monodromy sources. 
We find that 
\begin{align}
\tilde L^2\mathcal{F}^{mABJM}(g\mu_{F_i},n)=L^2\mathcal{F}^{ABJM}(g\mu_{F_i},n)\,,
\end{align}
with $\mathcal{F}^{ABJM}$ as in \eqref{efvcoreabjm2again}. In fact
the core values of $e^{2V}$ and $e^{2\lambda_i}$ for the mABJM case are \emph{exactly the same}
as in ABJM case with $\varphi\ne 0$, as given in \eqref{efvcoreabjm2again}. Thus, there is a kind of attractor mechanism
at work. The allowed ranges for $\kappa g\mu_{F_1}, \kappa g\mu_{F_2}$  are constrained by exactly the same conditions 
as for the ABJM case, and for $n=1$ and $n=10$ this is plotted in figure \ref{rangesmus}.
Moreover, our numerical investigations in section \ref{numericalsolssec} show, at least for $n=1$ and $n=10$, that
these necessary conditions are also sufficient for the existence of the defect solutions.

We can also determine the conserved currents in terms of the monodromy sources and for mABJM theory
we find \emph{exactly the same} results that we saw for the ABJM case with $\varphi\ne 0$: explicitly expressions
can be obtained from
\eqref{eeejs12}, \eqref{eeejs123} and \eqref{efvexpMabjmcore} or \eqref{efvcoreabjm2again}.
For the stress tensor, from appendix \ref{secappcmabjm}, we obtain for mABJM
\begin{align}
\langle T_{ab}\rangle dx^a dx^b =-\frac{h_D}{2\pi}\left[ds^2(AdS_2)-2 n^2 dz^2\right]\,,
\end{align}
with $h_D$ being expressed in terms of the mABJM conserved R-symmetry current:
\begin{align}\label{hdeemabjm}
\frac{h_D}{2\pi}=
\frac{1 }{\kappa n } \langle J_R^{\varphi} \rangle\,.
\end{align}
For the allowed range of $(\kappa g\mu_{F_1},\kappa g\mu_{F_2})$ in figure \ref{rangesmus}, we have $h_D>0$
and there are divergences in the conserved currents $\langle J_R^{\varphi}\rangle $ and
$\langle J_{F_i}^{\varphi}\rangle $
as we approach the boundary of the solution space, as already noted in section \ref{abjmcase},
and hence for this case we deduce that $h_D$ is also diverging at the boundary.
We emphasise that the value of $h_D$ for the mABJM fixed point is expressed in terms of the conserved R-symmetry current and hence can be explicitly expressed in terms of the monodromy sources
using \eqref{eeejs123} and \eqref{efvexpMabjmcore} or \eqref{efvcoreabjm2again}. Moreover, we note that
$h_D$ for the mABJM fixed point differs from $h_D$ for the restricted STU solutions
(recall the discussion around \eqref{abjmfnzexpccb}), consistent with the fact that $h_D$ varies with $\varphi_s$ along the line of ABJM solutions with $\varphi \ne 0$, as already noted
below \eqref{abjmfnzexpccb}. 
In appendix \ref{positivehdmId} we show that for all solutions on 
the main branch with $n\ge 1$ we have $h_D\ge 0$, but for $n<1$ there are solutions with $h_D<0$. For
branch 2 solutions, which necessarily have conical deficits, we always have $h_D>0$.

Using the results from appendix \ref{secappcmabjm} we can obtain the expectation values of the
scalar operators with an $AdS_2\times S^1$ boundary. There are two scalar operators with dimension $\Delta=1$, one with dimension
$\Delta=\frac{1}{2}(1+\sqrt{17})$ and an irrelevant scalar operator with dimension
$\frac{1}{2}(5+\sqrt{17})$. We find that the expectation values for the relevant operators are 
given, for $AdS_2\times S^1$ boundary and general $n$, by 
\begin{align}
\vev{\mathcal{O}^{\Delta=1}_1}&=\frac{3^{1/4}g}{2\kappa n}\langle J_{F_1}\rangle\,,\qquad
\vev{\mathcal{O}^{\Delta=1}_2}=-\frac{g}{2\times  3^{1/4}\kappa n}(\langle J_{F_1}\rangle-2\langle J_{F_2}\rangle)\,,\nn
\langle{\mathcal{O}^{\Delta=\frac{1}{2}(1+\sqrt{17})}}\rangle&=0\,,
\end{align}
and hence, via \eqref{eeejs12}, \eqref{eeejs123} and  \eqref{efvexpMabjmcore} or \eqref{efvcoreabjm2again}
 are all fixed by the monodromy sources.
To obtain $\langle{\mathcal{O}^{\Delta=\frac{1}{2}(5+\sqrt{17})}}\rangle$ one needs to solve the BPS equations as
explained in appendix \ref{secappcmabjm}.

\subsection{Numerical solutions}\label{numericalsolssec}
We have constructed some solutions numerically corresponding to defect solutions in mABJM theory 
and also in ABJM theory with $\varphi\ne 0$ that are associated with additional mass deformations preserving conformal
invariance. We present some representative plots which capture the essential features. For simplicity
we focus on defect solutions with no conical singularity, $n=1$, and hence vanishing monodromy for
the R-symmetry, $g\mu_R=0$. However, we have considered other values of $n$, including $n=10$ as
in figure \ref{rangesmus}.
 A summary of the overall solution space, including the analytic STU solutions
discussed in the next section, is given in figure \ref{fig:sumsolutions}.

\begin{figure}[htbp]
\begin{center}
\includegraphics[scale=.45]{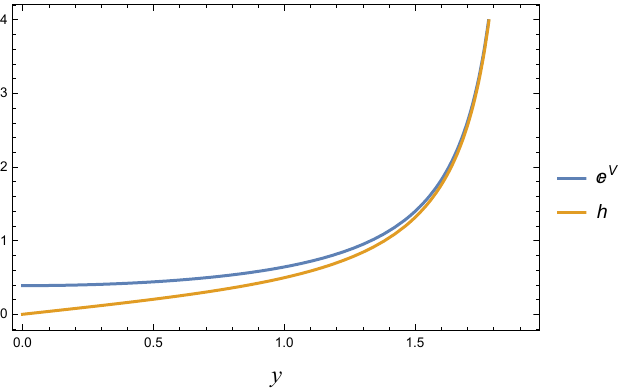}~
\includegraphics[scale=.45]{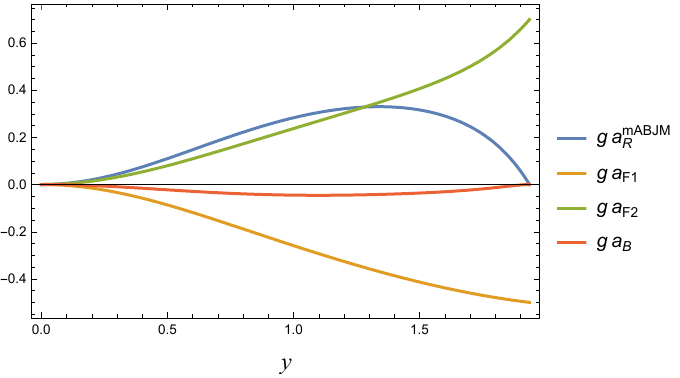}~
\includegraphics[scale=.45]{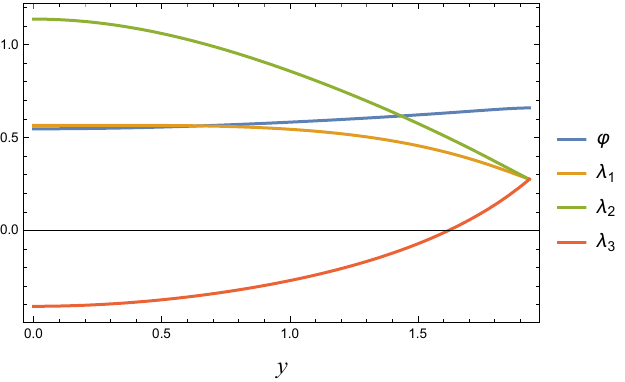}
\includegraphics[scale=.45]{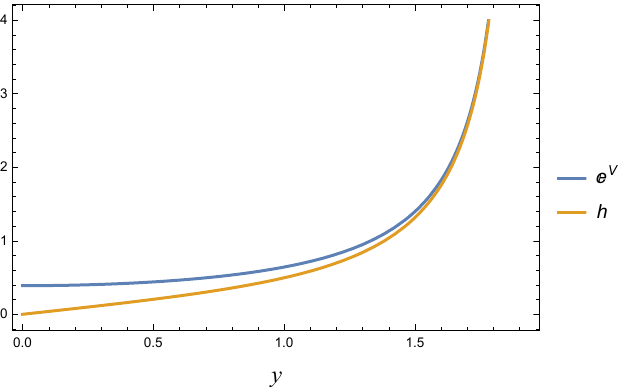}~
\includegraphics[scale=.45]{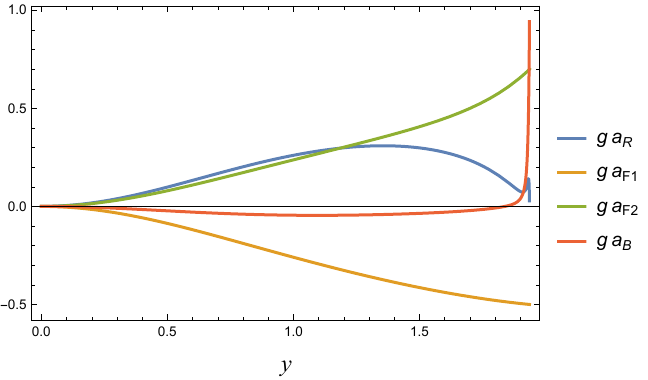}~
\includegraphics[scale=.45]{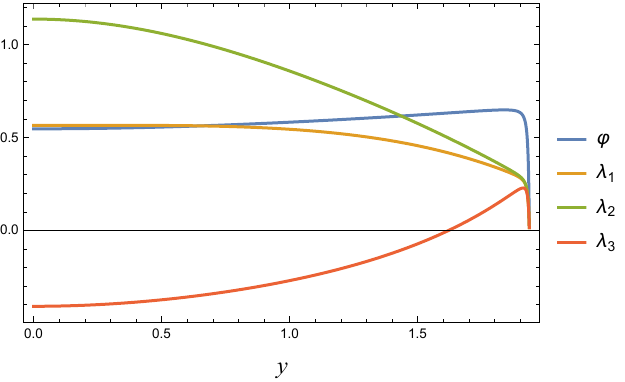}
\caption{
Top plots: monodromy defect solution for mABJM theory with $n=1$, $g\mu_R=0$, $g\mu_B=0$ and $g\mu_{F_1}=\frac{6}{10}$, $g\mu_{F_2} = - \frac{3}{10}$ (with $\kappa=1$).
Bottom plots: monodromy defect solution for ABJM theory with spatially dependent mass deformations preserving conformal invariance, parametrised by $\varphi_s$; the solution is for
$n=1$, $g\mu_R=0$, $g\mu_B=1$ and $g\mu_{F_1}=\frac{6}{10}$,  $g\mu_{F_2} = - \frac{3}{10}$ and $\varphi_s \approx 102$ (with $\kappa=1$).
In the left panels we have plotted the metric functions, in the middle panels the gauge field functions 
(associated with \eqref{mabjmbasis} and \eqref{abjmdefsaphinz}, respectively) and in the right panels the scalar functions.
The ABJM solution with $\varphi_s\ne 0$ in the bottom plots exhibits an intermediate region where the solution is nearly in the mABJM vacuum.
The core values of the scalars $\lambda_i$ in the two right plots are the same, associated with an attractor mechanism.}
\label{fig:mABJM_vortex_num}
\end{center}
\end{figure}

We first consider defect solutions of mABJM theory with $n=1$, $g\mu_R=0$ and $g\mu_B=0$ (as required by supersymmetry).
These solutions can only exist for the range of flavour monodromy sources $g\mu_{F_i}$, restricted by the necessary conditions
imposed by \eqref{branchesbds} and as summarised in figure
\ref{rangesmus}. Our numerical investigations indicate that these necessary conditions are in fact sufficient for $n=1$ and the  whole range is allowed
(in fact the same seems to be true for other values of $n$ on the main branch, including $n=10$).
We also recall that the core values of the scalars
are determined by $g\mu_{F_i}$; those for $\lambda_i$ are given analytically 
in \eqref{efvcoreabjm2again} while the core value of $\varphi$ needs to be determined numerically as a function
of $g\mu_{F_i}$ (and $n$).
In the top panels of figure \ref{fig:mABJM_vortex_num} we have
presented an mABJM defect solution for $g\mu_{F_1}=\frac{6}{10}$, $g\mu_{F_2} = - \frac{3}{10}$ and $n=1$, we have used conformal gauge for the $y$ coordinate as in \eqref{confgaugetext} and chosen $\kappa=1$. For these values we
find $\varphi^{mABJM}_\text{core}=0.54607$.

We next consider defect solutions of ABJM theory which have additional mass deformations that are
spatially dependent (in flat space), which preserve the superconformal symmetry of the defect. There is a one parameter
family of these solutions parametrised by the source $\varphi_s$, as summarised by the line in 
figure \ref{fig:sumsolutions}. With $n=1$ we must have $g\mu_R=0$ and $g\mu_B=\kappa$ to preserve supersymmetry.
The ABJM solutions with $\varphi\ne 0$ are restricted by the necessary conditions imposed by \eqref{branchesbds} and, as for
the mABJM case, our numerical investigations show that these necessary conditions are in fact sufficient for $n=1$ and the  whole range is allowed,
as summarised in figure \ref{rangesmus} (in fact the same seems to be true for other values of $n$ on the main branch, including $n=10$).
The core values of the scalars $\lambda_i$ 
are determined by $g\mu_{F_i}$ as in \eqref{efvcoreabjm2again}, in exactly the same way as for the mABJM defect solutions. 
For $\varphi$, we find the core value ranges from $[0, \varphi^\text{crit}_\text{core})$ with 
$\varphi^\text{crit}_\text{core}=\varphi^{mABJM}_\text{core}$, and over this range the corresponding source $\varphi_s$ monotonically takes the range $[0,\infty)$.
In the bottom panels of figure \ref{fig:mABJM_vortex_num} we have
presented a solution with the same values for $g\mu_{F_1}=\frac{6}{10}$ and $g\mu_{F_2} = - \frac{3}{10}$ that we
considered for the mABJM defect in the top panels of figure \ref{fig:mABJM_vortex_num} (with $\varphi^{mABJM}_\text{core}=0.54607$), and we have also taken
$\varphi_s \approx 102$, with core value $\varphi_\text{core}=0.546$. We have again used conformal gauge
\eqref{confgaugetext} and we have also set $\kappa=1$. Comparing with the top panels we see that the core
values of the scalar fields are indeed identical. Given we have taken a large value of $\varphi_s$ we also notice that the solution closely tracks the mABJM defect solution out from the core up to a value of $y$
where it nearly approaches the mABJM vacuum before sharply returning to the ABJM vacuum. 
Thus, in the limit
$\varphi_s\to \infty$ we end up at the mABJM defect solution (with $g\mu_B=0$) as indicated in figure \ref{fig:sumsolutions}.

\begin{figure}[htbp]
\begin{center}
\includegraphics[scale=.45]{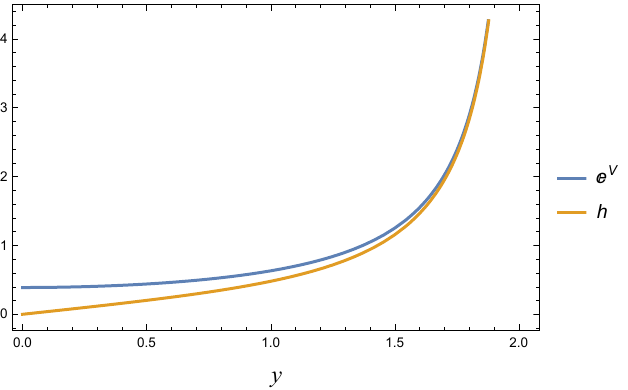}~
\includegraphics[scale=.45]{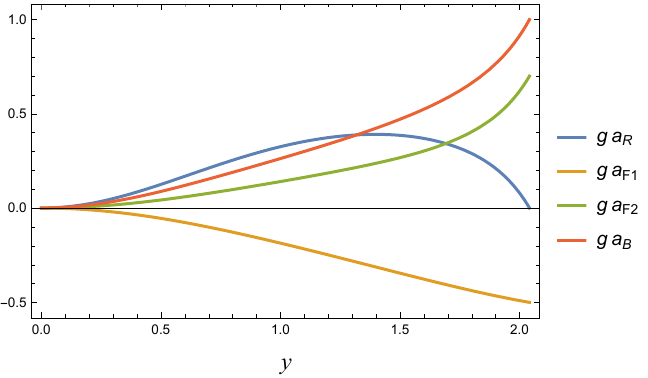}~
\includegraphics[scale=.45]{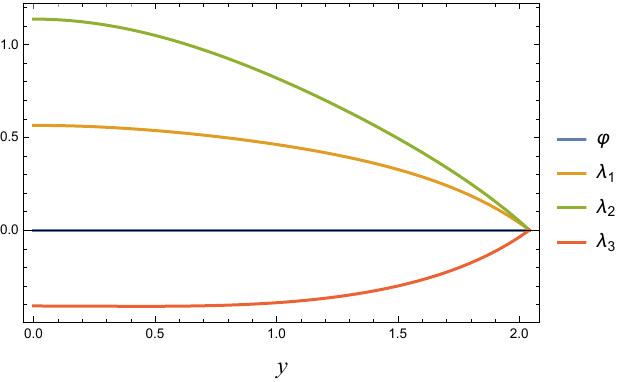}
\caption{
A restricted STU solution with $\varphi =  0$, corresponding to a defect in ABJM theory with $n=1$, $g\mu_R=0$, $g\mu_B=1$ and $g\mu_{F_1}=\frac{6}{10},~ g\mu_{F_2} = - \frac{3}{10}$ (with $\kappa=1$). 
In the left panel we have plotted the metric functions, in the middle panel the gauge field functions (associated with \eqref{abjmdefsaphinz}) and in the right panel the scalar functions. Notice that the core values of the scalar fields $\lambda_i$ are exactly the same as for the solutions in figure \ref{fig:mABJM_vortex_num}.}
\label{fig:N4phi_vortex_num}
\end{center}
\end{figure}

In the next section we will discuss defect solutions in ABJM theory with $\varphi_s=0$ which can be 
constructed analytically in the STU model with $\varphi=0$. We have plotted such a restricted STU solution with $g\mu_B=\kappa n$
and presented in conformal gauge \eqref{confgaugetext}
in figure \ref{fig:N4phi_vortex_num}.

\section{Solutions of the STU model}\label{sectstu}
We now consider solutions of the STU model, associated with defects in ABJM theory, 
by setting the complex scalar $\zeta=\varphi e^{i\theta}=0$. Analytic solutions for
this case are known (see e.g.\cite{Couzens:2021rlk,Ferrero:2021etw}).
However, following our approach in previous sections, 
we do not need to utilise the analytic solutions in order to compute various quantities of interest. 

From the expansion of the BPS equations at the  $AdS_4$ boundary, for the general STU model 
solutions we only have a single constraint
on the monodromy sources $g\mu^{\alpha}$:
\begin{align}\label{n4scebpstextstu}
g\mu_R = g\mu ^{0}+ g\mu^{1} + g\mu^{2} + g\mu^{3} =-\kappa n- s \,.
\end{align}
If we impose, by hand, the additional constraint $g\mu_B=\kappa n$, we
obtain the class of
``restricted STU solutions",  which are the solutions that arise as the $\varphi \to 0$ limit of those
considered in section \ref{abjmcase}.

To proceed, we now consider the following four conserved charges
\begin{align}
\mathcal{E}^0&=e^{2V}e^{-2\lambda_1-2\lambda_2-2\lambda_3}F^0_{23}\,,\qquad
\mathcal{E}^1=e^{2V}e^{-2\lambda_1+2\lambda_2+2\lambda_3}F^1_{23}\,,\nn
\mathcal{E}^2&=e^{2V}e^{2\lambda_1-2\lambda_2+2\lambda_3}F^2_{23}\,,\qquad
\mathcal{E}^3=e^{2V}e^{2\lambda_1+2\lambda_2-2\lambda_3}F^3_{23}\,.
\end{align}
Evaluated at the $AdS_4$ boundary we have 
\begin{align}\label{Ejayrel}
\mathcal{E}^\alpha=c\langle J^\alpha\rangle\,.
\end{align}
where $c=-3\sqrt{2}\pi e^{V_0}/(ngN^{3/2})$ is the same constant appearing in \eqref{defcee}.

Using the BPS equations, these can be rewritten in the form 
\begin{align}\label{abjmconschges}
\mathcal{E}^0&=e^{2V}g\cos\xi-\frac{\kappa}{\sqrt{2}}e^{V-\lambda_1-\lambda_2-\lambda_3}\,,\quad
\mathcal{E}^1=e^{2V}g\cos\xi-\frac{\kappa}{\sqrt{2}}e^{V-\lambda_1+\lambda_2+\lambda_3}\,,\nn
\mathcal{E}^2&=e^{2V}g\cos\xi-\frac{\kappa}{\sqrt{2}}e^{V+\lambda_1-\lambda_2+\lambda_3}\,,\quad
\mathcal{E}^3=e^{2V}g\cos\xi-\frac{\kappa}{\sqrt{2}}e^{V+\lambda_1+\lambda_2-\lambda_3}\,,
\end{align}
and hence we can also write
\begin{align}\label{stucaseei}
\mathcal{E}_\alpha&={g\cos\xi e^{2V}}(1+\frac{\kappa n}{2g \mathcal{I}^\alpha})\,,
\end{align}
where the $\mathcal{I}^\alpha$ are defined in \eqref{eq:IntegratedFluxesExpr1text}.
Notice too that we can write the superpotential $W$, defined in \eqref{superpottext}, in the form
\begin{align}\label{superpotstu}
W=\frac{1}{\sqrt{2}ne^V\cos\xi}\sum_\alpha \mathcal{I}^\alpha\,.
\end{align}

Next, we obtain an expression for $\mathcal{E}_\alpha$ at the core from \eqref{stucaseei} using the same approach as in
section \ref{relatingscesbdysec}.
Specifically, using 
\begin{equation}
\label{eq:GeneralIntegExpressionForFluxSourcesagain}
g\mu^\alpha = g\mathcal{I}^{\alpha}|_\text{bdry} - g\mathcal{I}^{\alpha}|_\text{core} \, ,
\end{equation}
and the fact that for ABJM boundary we have $\mathcal{I}^{\alpha}|_\text{bdry}=-\frac{\kappa n}{2g}$, we obtain
\begin{align}\label{valuesorestukey}
g\mathcal{I}_\alpha|_\text{core}&=-(\frac{\kappa n}{2}+g\mu^\alpha)\,.
\end{align}
Hence, from \eqref{efvcoreabjm} we immediately get 
\begin{align}\label{efvcoreabjm3}
L^{-2}e^{2V}|_{\text{core}}&=\mathcal{F}^{ABJM}\equiv \left[\Bigl(1+\frac{2g\mu^0}{\kappa n}\Bigr)\Bigl(1+\frac{2g\mu^1}{\kappa n}\Bigr)\Bigl(1+\frac{2g\mu^2}{\kappa n}\Bigr)\Bigl(1+\frac{2g\mu^3}{\kappa n}\Bigr)\right]^{\frac{1}{2}\ }\,,\nn
e^{2\lambda_1}|_{\text{core}}&=\mathcal{F}^{ABJM}(1+\frac{2g\mu^{2}}{\kappa n})^{-1}(1+\frac{2g\mu^{3}}{\kappa n})^{-1}\,,\nn
e^{2\lambda_2}|_{\text{core}}&=\mathcal{F}^{ABJM}(1+\frac{2g\mu^{1}}{\kappa n})^{-1}(1+\frac{2g\mu^{3}}{\kappa n})^{-1}\,,\nn
e^{2\lambda_3}|_{\text{core}}&=\mathcal{F}^{ABJM}(1+\frac{2g\mu^{1}}{\kappa n})^{-1}(1+\frac{2g\mu^{2}}{\kappa n})^{-1}\,.
\end{align}

Recall from \eqref{branchesbds} that we have the following bounds on the 
monodromy parameters for the two branches of solutions:\footnote{To obtain the restriction on $n$ for $s=+\kappa$, we 
use \eqref{branchesbds} to deduce $1+2g\mu^\alpha/(\kappa n)<0$ since
$\mathcal{I}^{\alpha}|_\text{bdry}=-\frac{\kappa n}{2g}$. After summing over $\alpha$ and using \eqref{n4scebpstextstu} we deduce $0<n<1$.}
\begin{align}\label{branchesbdsstu}
\text{Main branch}:&\quad s=-\kappa,\quad (1+\frac{2g\mu^{\alpha}}{\kappa n})>0,\quad n>0\,,\nn
\text{Branch 2}:&\quad s=+\kappa,\quad (1+\frac{2g\mu^{\alpha}}{\kappa n})<0,\quad 0<n<1\,,
\end{align}
along with $g\mu_R=-\kappa n-s$.
Notice that the expressions in \eqref{efvcoreabjm3} are exactly the same as in \eqref{efvcoreabjm2} when $\varphi\ne 0$. 
In particular, for the 
restricted STU solutions, with $g\mu_B=\kappa n$ imposed by hand, the core values are given in \eqref{efvcoreabjm2again}, which in turn are the same core values for the mABJM defect solutions (which instead have $g\mu_B=0$). 
Thus, these solutions also can only exist at most in the dark shaded regions of figure \ref{rangesmus}, for $n=1$ and $n=10$, and in fact there are restricted STU solutions for
the whole of these regions. An example of a restricted STU solution for $n=1$ is plotted in figure \ref{fig:N4phi_vortex_num}.
It is also interesting to point out that for STU solutions with $g\mu_B=0$ imposed by hand, the possible range
for $g\mu_{F_i}$, 
with $n=1$, is illustrated by the lighter shaded
region, including the darker region, in the left panel of figure \ref{rangesmus}. For $n=1$ the necessary conditions on the existence of
STU defect solutions with $g\mu_B=0$ in \eqref{branchesbdsstu} are in fact sufficient conditions. By contrast when $n> 1$ this is not the case
as illustrated in the right panel of figure \ref{rangesmus}, where we see that the actual solution space is more restrictive.\footnote{Interestingly this is associated with the appearance of compact spindle type solutions in the solution space; analogous behaviour also occurs for defects of
$D=5$ STU gauged supergravity solutions \cite{Arav:2024exg}.}
We will return to the STU defect solutions with $g\mu_B=0$ 
in section \ref{flows}.

From \eqref{Ejayrel} we can also now express the conserved currents in terms of the monodromy sources
\begin{align}\label{litjistu}
\vev{J^\alpha}&=\frac{N^{3/2}}{6\sqrt{2}\pi}s\kappa e^{-V_0}
\frac{(2g\mu^\alpha)\mathcal{F}^{ABJM}}{(1+\frac{2g\mu^\alpha}{\kappa n})}         
\,.
\end{align}
From the results for the stress tensor in appendix \ref{appa} we obtain
\begin{align}
\langle T_{ab}\rangle dx^a dx^b =-\frac{h_D}{2\pi}\left[ds^2(AdS_2)-2 n^2 dz^2\right]\,,
\end{align}
with
\begin{align}\label{hdforstu}
h_D=\frac{2\pi }{4\kappa n }\sum_\alpha \langle J^{\alpha} \rangle\,.
\end{align}
When $n=1$ this simplifies to 
\begin{align}
h_D=-\frac{N^{3/2}}{12\sqrt{2}}e^{-V_0}
\sum_\alpha \frac{(2\kappa g\mu^\alpha)\mathcal{F}^{ABJM}}{(1+{2g\kappa \mu^\alpha})}\,.       
\end{align}
In appendix \ref{positivehdmId} we show that solutions on 
the main branch ($s=-\kappa$) with $n\ge 1$ always have $h_D\ge 0$, but for $n<1$ there are solutions with $h_D<0$. For
branch 2 solutions ($s=\kappa$), which necessarily have conical deficits, we always have $h_D>0$.

The expectation values of the scalar operators are given in
\eqref{scalvevstextabjm}. For the STU model these can all be 
expressed in terms of the monodromy sources via \eqref{litjistu}.

For minimal gauged supergravity associated with the STU model, we set all of the gauge fields equal and also set
$\lambda_i=0$. The sources are then given by $g\mu^\alpha=\kappa(-\kappa n- s)/4$, and we only have a defect solution for $n\ne 1$.
For these solutions we have
\begin{align}
h_D=\frac{s\kappa(1-n^2)}{n^2}\frac{N^{3/2}}{12\sqrt 2 e^{V_0} }\,.
\end{align}
For the main branch of solutions with $s=-\kappa$, we have explicit solutions for all $n>0$ as we show in appendix \ref{minsugraapp}.
On this branch we see that $h_D>0$ for $n>1$, as noted above, but $h_D<0$ for $n<1$.
For branch 2 solutions with $s=+\kappa$, we show that solutions just exist for $0<n<1/\sqrt{2}$, i.e. smaller than the possible range $0<n<1$, and we have $h_D>0$.

\section{The partition function and supersymmetric Renyi entropy}\label{sec:partfn}
In this section we give the on-shell action and hence partition function for defects of ABJM theory,\footnote{A related
computation of the on-shell action for a sub-class of solutions was also carried out in \cite{Chen:2020mtv}.}
ABJM theory with mass deformations parametrised by $\varphi_s$,
as well as for mABJM theory.
In all cases, we are able to express the result explicitly in terms of the monodromy sources.
After subtracting off the contribution of $n$ times that of the vacuum we obtain a ``defect free energy'' and correspondingly
if we allow the monodromy sources to vary, we will see that this is 
the analogue of the defect central charge $b$ that arises in the context of co-dimension two monodromy defects
in $d=4$ SCFTs  (e.g. see the discussion in section 1 of \cite{Arav:2024exg}).

After a simple analytic continuation we can also use our computation of the on-shell action to 
compute supersymmetric Renyi entropies \cite{Nishioka:2013haa} for the monodromy defects.
We note that for the ABJM case in the STU model a supersymmetric Renyi entropy (SRE) was computed in
\cite{Hosseini:2019and}, but we shall see that more general definitions are possible and we obtain some 
new results. We will also discuss SRE for the mABJM case.

For simplicity (see \eqref{ads3ansbdy}), in this section we have set 
\begin{align}e^{V_0}=1\,.
\end{align}

\subsection{Partition function and defect free energy}

\subsubsection{ABJM case, STU model}

After some computation we find that the on-shell action can be written in the form
\begin{align}
    S&=\frac{L^2}{4G}\mathrm{Vol}(AdS_2)(-s\kappa n)\mathcal{F}^{ABJM}\,,
    \end{align}
where, as before,
\begin{align}
\mathcal{F}^{ABJM}\equiv \left[\Bigl(1+\frac{2g\mu^0}{\kappa n}\Bigr)\Bigl(1+\frac{2g\mu^1}{\kappa n}\Bigr)\Bigl(1+\frac{2g\mu^2}{\kappa n}\Bigr)\Bigl(1+\frac{2g\mu^3}{\kappa n}\Bigr)\right]^{\frac{1}{2}\ }\,,
\end{align}
with the constraint $g\mu_R=\sum_\alpha g\mu^\alpha=-\kappa n- s$.
When $n=1$, associated with no conical singularity, we should set $s=-\kappa$. If we also set $g\mu^\alpha=0$, there is no defect
and we are dual to ABJM theory on $AdS_2\times S^1$. In this case we have the vacuum on-shell action given by
\begin{align}
S_0&={\mathrm{Vol}(AdS_2)}\frac{L^2}{4G}\,.
\end{align}

Note that if we Wick rotate $AdS_2\to H^2$ and regulate the volume of $H^2$ via $\vol(H_2)=-2\pi$ \cite{Casini:2011kv}, we 
obtain the Euclidean action $I=-S$. For no-defect ($n=1$, $g\mu^\alpha=0$) we find
\begin{align}\label{freeabjm}
I_0=F^{ABJM}_{S^3}\equiv\frac{\sqrt{2}\pi}{3}N^{3/2}\,,
\end{align}
where $F^{ABJM}_{S^3}$ is the 
free energy for ABJM theory on $S^3$,
while the full action reads
\begin{align}\label{fullabjmact}
I=(-s\kappa n)\mathcal{F}^{ABJM}F^{ABJM}_{S^3}\,.
\end{align}

We now define  a ``defect free energy" via
\begin{align}\label{Ideeabjmcase}
I_D^{ABJM}\equiv I^{ABJM}-n I^{ABJM}_0= n\left(-s\kappa\mathcal{F}^{ABJM}-1\right)F^{ABJM}_{S^3}\,.
\end{align}
Recall that the branch of solutions with $s=-\kappa$ includes the case of $n=1$ and hence
is continuously connected with the vacuum. Thus, for this branch the defect free energy is the free energy after
subtracting off $n$ times this piece and, in particular,  $I^{ABJM}_D=0$ when we set $g\mu^\alpha=0$ and $n=1$. 
We also note that when $n=1$, the defect free energy is related to the defect $g$-function via $\log g=-I_D$ (e.g. eq (14) of \cite{Cuomo:2021rkm}).
For the $s=+\kappa$ branch of solutions, which are not continuously connected with the vacuum, the meaning of the definition is less clear. In appendix \ref{positivehdmId} we show that for the main branch of solutions $-I^{ABJM}_D\ge 0$ for $n\ge 1$, but it is possible that $-I^{ABJM}_D<0$ for $0<n<1$ (e.g. as it is for minimal gauged supergravity as discussed in appendix \ref{minsugraapp}). For branch 2 solutions we also show there
that we always have $-I^{ABJM}_D> 0$.

We can now consider varying $I^{ABJM}_D$ with respect to the independent monodromy parameters and $n$, subject
to the supersymmetry constraint $\sum_\alpha g\mu^\alpha=-\kappa n- s$. The independent monodromy parameters can 
be taken to be, for example, $(g\mu_{F_1}, g\mu_{F_2}, g\mu_{F'})$ as in \eqref{abjmdefsa}
or $(g\mu_{F_1}, g\mu_{F_2}, g\mu_{B})$ as in \eqref{abjmdefsaphinz}. Interestingly, we find
\begin{align}\label{varyIdee}
\frac{1}{(2\pi)^2}dI^{ABJM}_D&=\frac{1}{n}\Big(  \sum_\alpha\vev{J^\alpha}d[g\mu^\alpha] \Big)   +(2\frac{h_D}{2\pi}{}-\frac{F^{ABJM}_{S^3}}{(2\pi)^2}) dn       \,,\nn
&=\frac{1}{n}\Big(  \sum_\alpha\vev{J^\alpha}d[g\mu^\alpha]+\frac{\kappa}{2}\sum_\alpha\langle J^{\alpha}\rangle dn \Big)   -\frac{F^{ABJM}_{S^3}}{(2\pi)^2}dn       \,,
\end{align}
and we also recall that 
\begin{align}
 \sum_\alpha\vev{J^\alpha}d[g\mu^\alpha] &=\vev{J_{F_1}}d[g\mu^{F_1}]+\vev{J_{F_2}}d[g\mu^{F_2}]+\vev{J_{F'}}d[g\mu_{F'}]
 +\langle{J^{ABJM}_{R}}\rangle d[g\mu_{R}]\,,
 \end{align}
 with $d[g\mu_{R}]=-\kappa dn$.
 The result in \eqref{varyIdee} is the direct analogue of the variation of the $b$ central charge for monodromy defects
 in $d=4$ SCFTs discussed in \cite{Arav:2024exg}.

\subsubsection{mABJM case}
For the mABJM case things take a similar form. We find the on-shell action is given by
\begin{align}
    S  &=-\frac{\tilde L^2}{4G}(s\kappa n)\mathrm{Vol}(AdS_2)\mathcal{F}^{mABJM}\,,
      \end{align}
where $\tilde L$ is the radius of the $AdS_4$ vacuum dual to the mABJM $AdS_4$ vacuum \eqref{lsfpsc} and, as before,
\begin{align}\label{mfabjmexp}
\mathcal{F}^{mABJM}\equiv \left[\Bigl(1+\frac{g\mu^0}{\kappa n}\Bigr)\Bigl(1+\frac{3g\mu^1}{\kappa n}\Bigr)\Bigl(1+\frac{3g\mu^2}{\kappa n}\Bigr)\Bigl(1+\frac{3g\mu^3}{\kappa n}\Bigr)\right]^{\frac{1}{2}\ }\,.
\end{align}
For mABJM we have
$g\mu_R \equiv  \sum g\mu^{\alpha} =-\kappa n- s $ and 
$g\mu_B\equiv g\mu^{0}-g\mu^{1} - g\mu^{2} - g\mu^{3}=0$.

When $n=1$, associated with no conical singularity, we should set $s=-\kappa$. If we also set $g\mu^\alpha=0$, there is no defect
and we are dual to mABJM theory on $AdS_2\times S^1$. In this case we have the vacuum on-shell action given by
\begin{align}
S_0&={\mathrm{Vol}(AdS_2)}\frac{\tilde L^2}{4G}\,.
\end{align}
Wick rotating $AdS_2\to H^2$ and regulating the volume via $\vol(H_2)=-2\pi$, we find a Euclidean action 
\begin{align}
I_0=F^{mABJM}_{S^3}\,,
\end{align}
which is precisely the free energy for mABJM theory on $S^3$,
\begin{align}\label{freemabjm}
F^{mABJM}_{S^3}\equiv\frac{4\sqrt{2}\pi}{9\sqrt{3}}N^{3/2}=\frac{4}{3\sqrt{3}}F^{ABJM}_{S^3}\,.
\end{align}

We can again define a defect free energy via
\begin{equation}\label{ideemabjm}
I^{mABJM}_D=n\left(-s\kappa\mathcal{F}^{mABJM}-1\right)F^{mABJM}_{S^3}\,.
\end{equation}
Here $I^{mABJM}_D$ is a function of the two independent monodromy sources
$I^{mABJM}_D=I^{mABJM}_D(g\mu_{F_1},g\mu_{F_2})$. For the $s=-\kappa$ branch, we are continuously connected 
with the vacuum and setting $n=1$  we have $I^{mABJM}_D(0,0)=0$. We also note that
for $n=1$, within the allowed range of $g\mu_{F_i}$ illustrated in figure \ref{rangesmus}, we find that $I^{mABJM}_D<0$.
More generally, in appendix \ref{positivehdmId} we show that for the main branch of solutions $-I^{mABJM}_D\ge 0$ for $n\ge 1$, but it is possible that $-I^{mABJM}_D<0$ for $0<n<1$ (e.g. as it is for minimal gauged supergravity associated with the mABJM vacuum). For branch 2 solutions we show that we always have $-I^{mABJM}_D> 0$.

We now vary $I^{mABJM}_D$ with respect to $g\mu_{F_i}$ and $n$, subject to the supersymmetry constraints
to find
\begin{align}
 \frac{1}{(2\pi)^2}dI^{mABJM}_D       &=\frac{1}{n}\vev{J_{F_1}}d[g\mu^{F_1}]+\frac{1}{n}\vev{J_{F_2}}d[g\mu^{F_2}]
    +\frac{1}{(2\pi)^2}(2\pi h_D-F^{mABJM}_{S^3})dn  \,,\nn
     &=\frac{1}{n}\Bigl(\vev{J_{F_1}}d[g\mu^{F_1}]+\vev{J_{F_2}}d[g\mu^{F_2}]
    +\kappa\langle J^\varphi_R\rangle dn   \Bigr)
    -\frac{F^{mABJM}_{S^3}}{(2\pi)^2}dn\,.
\end{align}

\subsubsection{Partition function for ABJM with $\varphi\ne0$.}\label{actgenfnzflows}
We can also consider the on-shell action for the one-parameter family of solutions parametrised by $\varphi_s$,
and recall that as $\varphi_s\to  0$ this family smoothly connects to the STU solutions with 
$g\mu_B=\kappa n$ set by hand (see figure \ref{fig:sumsolutions}).
Computing the Euclidean on-shell action we find the remarkable result that the answer is independent of $\varphi_s$. 
Explicitly, again regulating with $\vol(H_2)=-2\pi$, we find
\begin{align}\label{abjmfnzacthere}
I^{ABJM,\varphi\ne 0}&=(-s\kappa n)\mathcal{F}^{ABJM}F^{ABJM}_{S^3}|_{g\mu_B=\kappa n}\nn
&=\frac{(-s\kappa)}{4n}\Big[  (s-n\kappa)(s-n\kappa-4g\mu_{F_1}-2g\mu_{F_2})\nn
&\qquad(s-n\kappa+2g\mu_{F_1}-2g\mu_{F_2}) (s-n\kappa+2g\mu_{F_1}+4g\mu_{F_2})         \Big]^{1/2}\frac{4F^{ABJM}_{S^3}}{3\sqrt 3}\,.
\end{align}
Moreover, this is exactly the same as the action for the restricted STU solutions:
\begin{align}
I^{ABJM,\varphi\ne 0}=I^{ABJM}|_{g\mu_B=\kappa n}\,.
\end{align}
The fact that the on-shell action is independent of $\varphi_s$ aligns with the fact that
$\langle \mathcal{O}^{\Delta=2}_\varphi \rangle=0$, as we saw in \eqref{scalvevstextabjm}. Correspondingly we can view
the $\varphi_s$ deformation as a Q-exact deformation of ABJM defects with $g\mu_B=\kappa n$
 by a linear combination of a fermionic mass deformation with $\varphi_s$ times a bosonic mass deformation.

Recall that as $\varphi_s\to \infty$ this one-parameter family of solutions closely approaches the defect solution in mABJM
theory, as we saw in the numerical solutions discussed in section \ref{numericalsolssec} and illustrated in
figure \ref{fig:sumsolutions}. 
Therefore, one might anticipate that
\eqref{abjmfnzacthere} is also the same as the on-shell action for the solutions dual to defects in mABJM theory with $g\mu_B=0$ and this is exactly
what we find, 
\begin{align}
I^{ABJM,\varphi\ne 0}=I^{mABJM}\,.
\end{align}
Interestingly, this means that from a knowledge of the
on-shell action for the STU solutions as a function of $n$ and the monodromy sources,
just associated with ABJM theory, one can infer $F^{mABJM}_{S^3}$ and $I_D^{mABJM}$.

\subsubsection{RG flows from ABJM to mABJM}\label{flows}

Consider ABJM theory with a defect with monodromy sources $g\mu^\alpha$ subject
to the constraint $\sum_\alpha g\mu^\alpha=-\kappa n- s$, as demanded by supersymmetry, and further 
set $g\mu_B=0$ by hand (note, by contrast, we defined the restricted STU solutions to have $g\mu_B=\kappa n$). This
line defect is then parametrised by two independent monodromy sources $g\mu_{F_1}$, $g\mu_{F_2}$ in addition to $n$.
We now imagine switching on a homogeneous (in flat space), supersymmetric mass deformation (i.e. not localised on the defect) which will induce a bulk RG flow. 
The IR end point of this RG flow is expected to be a defect in mABJM 
theory parametrised by the same values of $g\mu_{F_1}$, $g\mu_{F_2}$ and $n$, $s$, 
\emph{provided that such a solution exists}, as illustrated in figure \ref{fig:sumsolutions}. 
A subtlety is that the mABJM defect solution in the IR may not exist. For example, recall
figure \ref{rangesmus} associated with $n=1$: the dark region is where the ABJM defect solution with $g\mu_B=0$ 
and the mABJM defect solution both exist and so for these values of $g\mu_{F_i}$
we expect an RG flow as in figure \ref{fig:sumsolutions}.
However, for the lighter region outside of the dark region in figure \ref{rangesmus}, an ABJM solution 
with $g\mu_B=0$ exists but not an mABJM solution. It would be interesting to determine
the IR fate of the associated RG flows for these values of the monodromy sources.
We also highlight that for general $n$ the space of solutions can be more constrained than
the bounds coming from \eqref{branchesbds}. In figure \ref{rangesmus} we see that for $n=10$ the
actual space of STU solutions with $g\mu_B=0$ is more constrained than \eqref{branchesbds} but nevertheless
the darker region is still strictly contained within it, so the preceding comments are still valid. 
Something similar was also seen in the context of defect solutions of $D=5$ gauged supergravity in \cite{Arav:2024exg}, and
we shall leave further exploration to future work.

\begin{figure}[h!]
	\centering
	\includegraphics[scale=0.4]{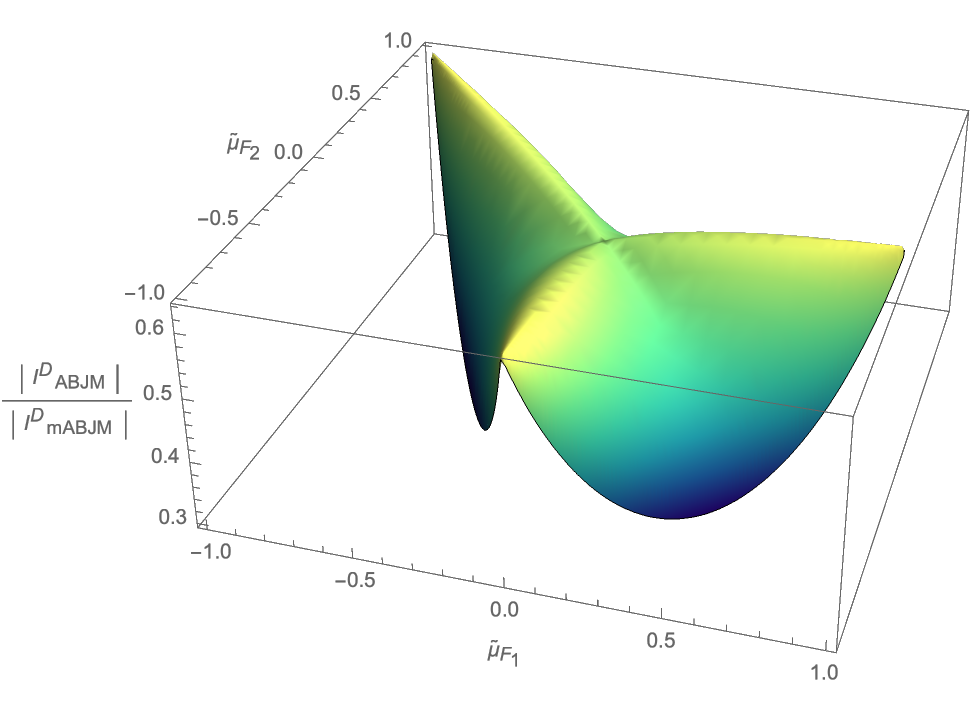}
	\caption{Ratio of the absolute value of the defect free energy for ABJM theory to that of mABJM theory, $|I_D^{ABJM}|/|I_D^{mABJM}|$, with the same monodromy sources
	$g\mu_{F_1}$ and $g\mu_{F_2}$. The plot is for the case of no conical singularity, $n=1$, so that $g\mu_R=0$ and for ABJM 
	we also impose $g\mu_B=0$ by hand. The plotted range of 
	$(\kappa g\mu_{F_1},\kappa g\mu_{F_2})\equiv (\tilde\mu_{F_1},\tilde\mu_{F_2})$, 
	is the same as the darker region of figure \ref{rangesmus}. Observe that $-I_D^{ABJM}<-I_D^{mABJM}$.
	}
	\label{idsratio}
\end{figure}

It is also interesting to compare the values of $I_D$ at the two
end points of such a flow.\footnote{Note that the RG flows we are considering are not the same as those considered in \cite{Cuomo:2021rkm}
which are driven by deformations localised on the defect itself and for which a
$g$-theorem was proven.} We first consider the case of no conical singularity, $n=1$.
Recall that associated with the allowed dark region in figure \ref{rangesmus}, we have $I_D^{mABJM}<0$. 
For the same region we also find that $I_D^{ABJM}(g\mu_B=0)<0$ and in addition that
$-I_D^{ABJM}<-I_D^{mABJM}$, as illustrated in
figure \ref{idsratio}. Thus, $-I_D$ would necessarily increase under such an RG flow.\footnote{That $-I_D$ is increasing is 
possibly not too surprising since it has been shown in the context of boundaries in $d=2$ CFTs that the boundary entropy $\log g$ can both increase and decrease 
under bulk RG flows \cite{Green:2007wr}.}
This also implies that the total action $I=I_0-(-I_D)$ will necessarily decrease under the RG 
flow.
This is directly analogous to what was seen in the analogous setting of monodromy defects in $d=4$ SCFTs
discussed in \cite{Arav:2024exg}, with $-I_D$ the analogue of the defect central charge $b$. It is also suggestive
that this occurs more generally for bulk RG flows in the presence of defects when $n=1$.

\begin{figure}[h!]
	\centering
			\includegraphics[scale=0.2]{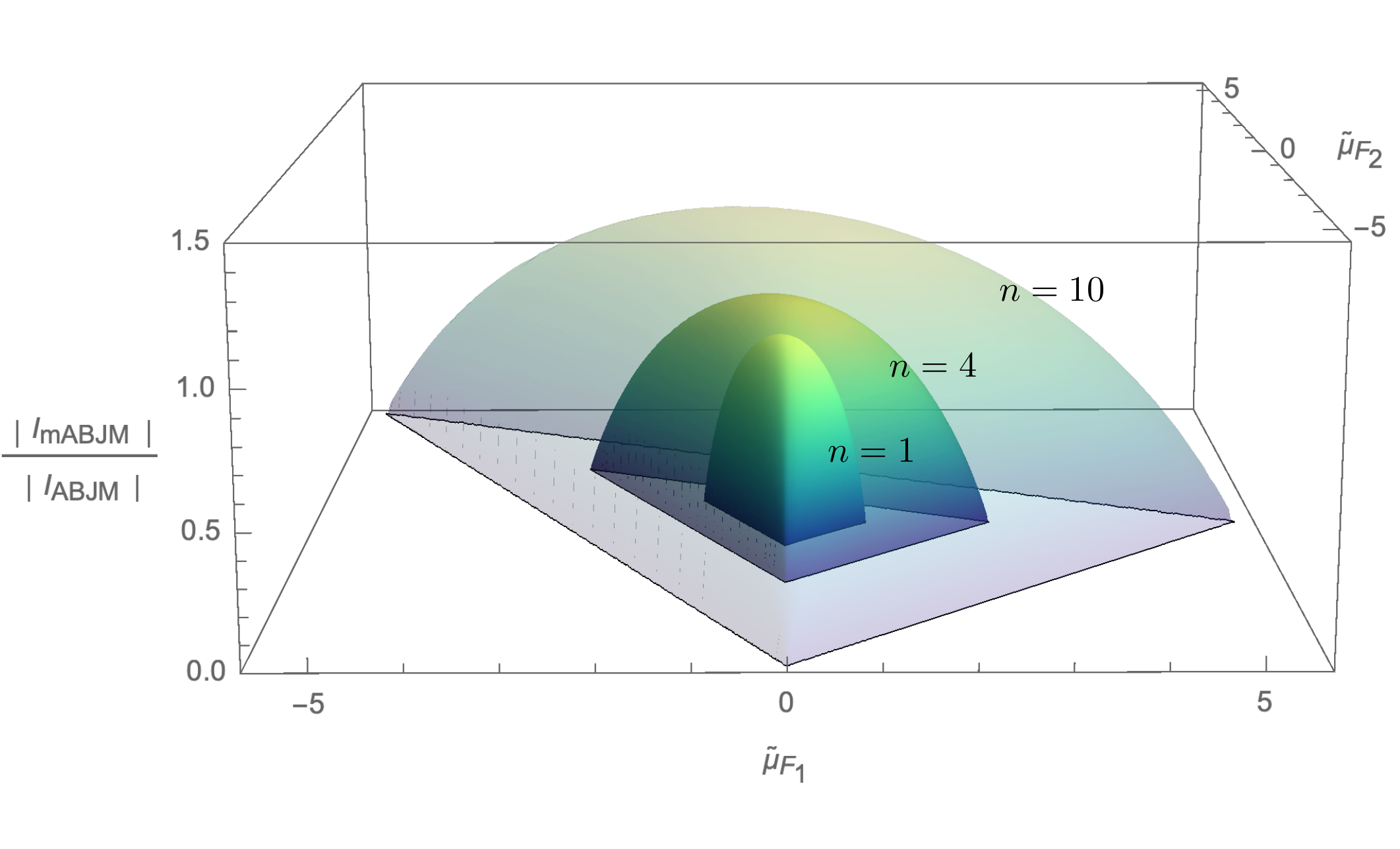}
		\includegraphics[scale=0.21]{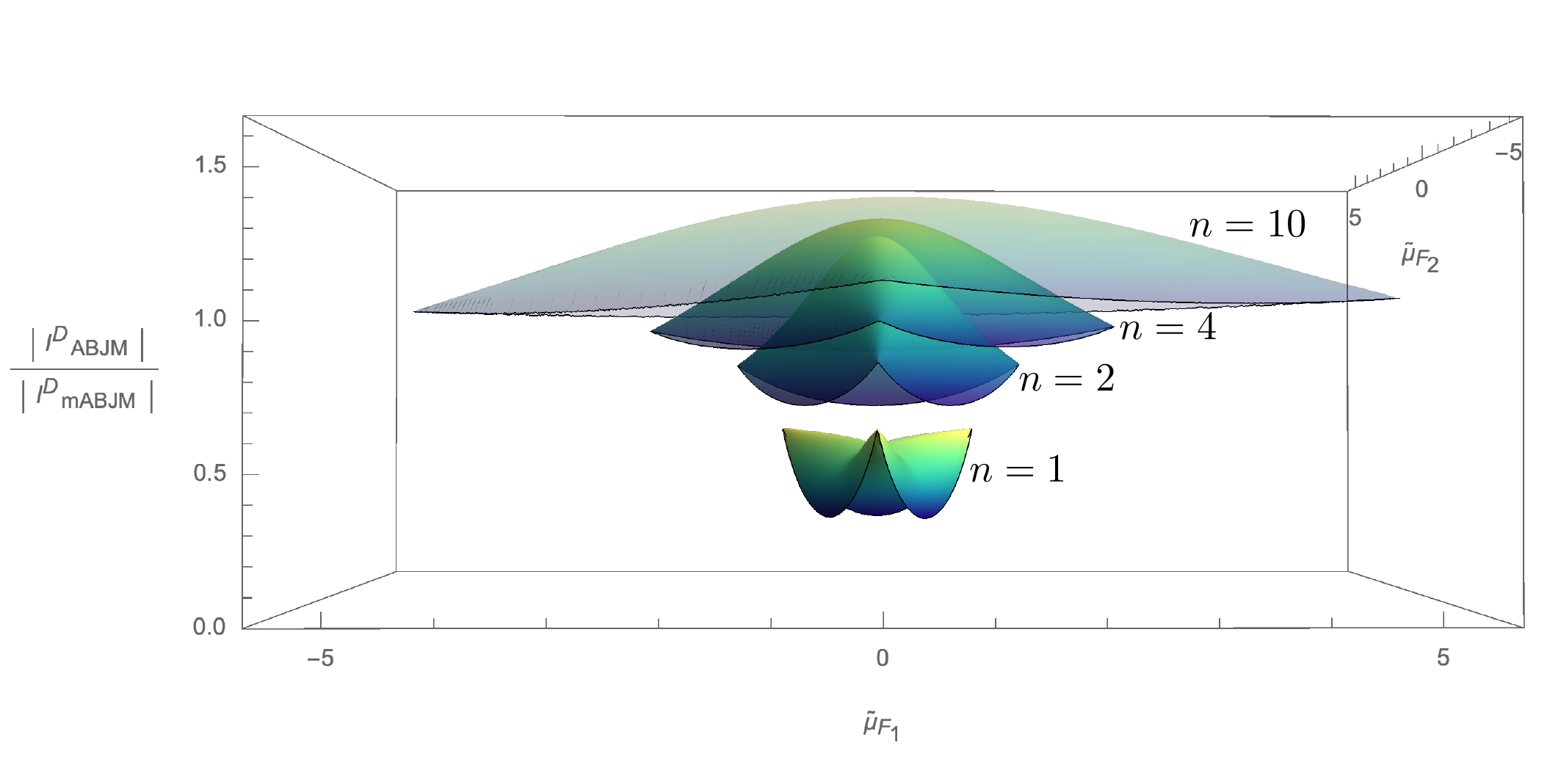}
	\caption{Ratios of free energies for defects of ABJM theory to those of mABJM theory with the same monodromy sources
	$g\mu_{F_1}$ and $g\mu_{F_2}$ for $n\ge 1$. The left panel plots the ratio of the total free energies for $n=1,4,10$.
	For $n\le 4$ we have 
		$I^{mABJM}<I^{ABJM}$ and the free energy must decrease in an RG flow. However, for $n>4$ the ratio $I^{mABJM}/I^{ABJM}$ can be both bigger than or smaller than 1. The right panel plots the ratio of the absolute value of the defect free energies, $|I_D^{ABJM}|/|I_D^{mABJM}|$ (as in figure 
		\ref{idsratio}), for $n=1,2,4,10$.
	The plots have $g\mu_R=\kappa(1-n)$ and for ABJM 
	we also impose $g\mu_B=0$ by hand. Note that $(\kappa g\mu_{F_1},\kappa g\mu_{F_2})\equiv (\tilde\mu_{F_1},\tilde\mu_{F_2})$. 
	}
	\label{idsratioplot2}
\end{figure}

We now briefly discuss $n>1$ (with $s=-\kappa$). By analogy with what was seen for defects in $d=4$ SCFTs in \cite{Arav:2024exg}, 
one might wonder if the total action $I=n I_0-(-I_D)$ always decreases along the RG flows (when they exist). 
By plotting this as a function of the monodromy sources, as shown in figure \ref{idsratioplot2},
we find that for $n\le 4$ this is indeed the case with $I^{mABJM}<I^{ABJM}$ and so the free energy would decrease in an RG flow. However, for 
$n> 4$ we find that ratio can be both bigger than or smaller than one,
depending on the monodromy, in contrast to the what was seen for the defects of $d=4$ SCFTs in
\cite{Arav:2024exg}.
Notice that we have plotted the range for $g\mu_{F_1}$ and $g\mu_{F_2}$ in figure \ref{idsratioplot2} consistent with
\eqref{branchesbds}, \eqref{Isformabjm} with $g\mu_B=0$ and $g\mu_R=\kappa(1-n)$ 
for the mABJM theory (the analogue of the darker region in figure \ref{rangesmus}); as discussed earlier we expect for all $n>1$
this region corresponds to mABJM defect solutions that actually exist and we also expect this region corresponds to
STU solutions with $g\mu_B=0$ that actually exist (which are the analogue of the lighter region in figure \ref{rangesmus}).
In figure \ref{idsratioplot2} we also see for certain values of $n>1$ that $-I_D$ is also not monotonic and can potentially decrease or increase, depending on the values of the monodromy. In fact closer examination reveals that $-I_D$ is not monotonic for any $n>1$.

\subsection{Supersymmetric Renyi entropy}
We briefly recall some aspects of (supersymmetric) Renyi entropy. We consider a 3d SCFT on Euclidean
$\mathbb{R}_{t_E}\times \mathbb{R}^2$, with metric $ds^2=dt_E^2+d\rho^2+\rho^2 d\theta^2$. We are interested in an entangling region consisting of a solid disc in $\mathbb{R}^2$ with surface located at $\rho=1$. To compute the Renyi entropy one uses the replica method by introducing an $n$-fold covering along a cut $0\le\rho\le 1$ to compute the partition function $Z_n$. 
The Renyi entropy for the disc is then $(1-n)^{-1}\log Z_n/(Z_1)^n$. Using the change of coordinates $t_E=\cos\alpha\sin\tau(1+\cos\alpha\cos\tau)^{-1}$, $\rho=\sin\alpha(1+\cos\alpha\cos\tau)^{-1}$, the starting metric is conformal to that on $S^3$: $ds^2=\cos^2\alpha d\tau^2+d\alpha^2+\sin^2\alpha d\theta^2$, with $0\le\alpha\le\pi/2$ and $\Delta \tau=2\pi$. 
To compute the Renyi entropy, we observe that in these coordinates the branch cut is at $\tau=0$ and the branch locus
is at $\alpha=\pi/2$, so that we should evaluate the partition on $S^3_n$, the $n$-branched cover of $S^3$, by taking $\Delta\tau=2\pi n$.
Finally, we notice that $S^3_n$ is conformal to $S^1\times H_2$ with metric 
$ds^2=d\tau^2+(1+q^2)^{-1}dq^2+q^2 d\theta^2$, where $q=\tan\alpha$ so that the branch locus has been pushed off to infinity.
Computing the partition function on\footnote{Note that the starting metric can also be written
$ds^2=\rho^2(\rho^{-2}dt_E^2+\rho^{-2}d\rho^2+d\theta^2)$, i.e conformal to $S^1\times H_2$, but notice that the circle parametrised
by $\theta$ is not the same circle parametrised by $\tau$.}
 $S^1\times H_2$, with the $S^1$ having period $2\pi n$, then provides
a convenient way to compute the Renyi entropy \cite{Casini:2011kv}.
In \cite{Nishioka:2013haa} it was observed that for $\mathcal{N}=2$ SCFTs one can switch on additional R-symmetry gauge fields
on $S^3_n$ or $S^1\times H_2$ of the form $A\propto d\tau$, so that supersymmetry is preserved. A corresponding
supersymmetric Renyi entropy was then defined which was also shown to reduce to the ordinary entanglement entropy in the limit
$n\to 1$. 

After Wick rotating our holographic $D=4$ solutions by taking $ds^2(AdS_2)\to ds^2(H_2)$, we obtain solutions with conformal boundary 
$H_2\times S^1$ and the period of the $S^1$ is given by $2\pi n$, where $n\equiv -k$, and preserving supersymmetry
by construction. Focusing on $n\ge 1$, in the rest of this section we now set 
\begin{align}
s=-\kappa\,.
\end{align}
We then define the supersymmetric Renyi entropy
via
\begin{equation}\label{sreexp}
    S_n^\mathrm{SRE}=-\frac{I_n(g\mu^\alpha)-nI_{n=1}(g\mu^\alpha)}{1-n}\,.
\end{equation}
Here $I_n(g\mu^\alpha)=-S$ is the on-shell, Euclidean free energy with the constraints of supersymmetry imposed
and is a function of $n$ and the independent flavour monodromy sources; we will be more explicit below.
Also, $I_{n=1}(g\mu^\alpha)$ is the same quantity in the limit $n\to 1$, but we need to be careful how this limit is taken.
In the STU model context,
one possible limit, and hence one possible definition for SRE, was given in \cite{Hosseini:2019and}, but we emphasise
that there are other possibilities too, as we highlight below.

The quantity $S_n^\mathrm{SRE}$ is UV divergent but one can extract a universal part free of ambiguities, by regulating the volume
of $H^2$ via $\vol(H^2)=-2\pi$ \cite{Casini:2011kv}.
The quantity $ S_n^\mathrm{SRE}$
has a finite $n\to 1$ limit 
which is then a kind of charged
entanglement
entropy expressed in terms of the independent flavour monodromy sources

\subsubsection{ABJM theory, STU solutions}

For the ABJM case, we have
\begin{align}
I_n(g\mu^i)=n\mathcal{F}^{ABJM}F^{ABJM}_{S^3}\,,
\end{align}
with
\begin{align}
\mathcal{F}^{ABJM}= \left[\Bigl(1+\frac{2g\mu^0}{\kappa n}\Bigr)\Bigl(1+\frac{2g\mu^1}{\kappa n}\Bigr)\Bigl(1+\frac{2g\mu^2}{\kappa n}\Bigr)\Bigl(1+\frac{2g\mu^3}{\kappa n}\Bigr)\right]^{\frac{1}{2}\ }\,,
\end{align}
and $F^{ABJM}_{S^3}$ given in \eqref{freeabjm}.
The supersymmetry constraint
is $g\mu_R=  \sum_\alpha g\mu^{\alpha}=\kappa(1-n)$.

We can view $I_n(g\mu^\alpha)$ and $S_n^\mathrm{SRE}$ as a function of $n$ and 
three independent flavour monodromy sources which can be taken to be, for example,
$(g\mu_{F_1}, g\mu_{F_2}, g\mu_{F'})$ as in \eqref{abjmdefsa}. 
We can then take the limit $I_{n=1}(g\mu^\alpha)$ in \eqref{sreexp} holding $(g\mu_{F_1}, g\mu_{F_2}, g\mu_{F'})$ fixed.
This gives us one possible definition of SRE expressed in terms of $(g\mu_{F_1}, g\mu_{F_2}, g\mu_{F'})$, but as the explicit expression for general $n$ is rather long, we skip it.

We can also consider the class of STU model solutions that arise in the context of minimal gauged supergravity.  
For this case we impose $g\mu^{0}=g\mu^{1} = g\mu^{2} = g\mu^{3}=\frac{\kappa}{4}(1-n)$ and now $I_{n}(g\mu^\alpha)=\frac{(1+n)^2}{4n}F^{ABJM}_{S^3}$ is just a function 
of $n$. Computing $S_n^\mathrm{SRE}$
we obtain
\begin{align}\label{mingssre}
    S_n^\mathrm{SRE}=-\frac{1+3n}{4n}F^{ABJM}_{S^3}
    \,,
    \end{align}
   in agreement with (2.8) of \cite{Nishioka:2014mwa} (who also used a regulated volume $\mathrm{vol}(H_2)=-2\pi$). Note that this expression
   can be obtained from the expression for  $S_n^\mathrm{SRE}$ in the previous paragraph by setting 
   $g\mu_{F_1}= g\mu_{F_2}= g\mu_{F'}=0$

We could also consider defining the SRE using the basis \eqref{abjmdefsaphinz} and instead of holding
$(g\mu_{F_1}, g\mu_{F_2}, g\mu_{F'})$ fixed in \eqref{sreexp} as above, we hold
$(g\mu_{F_1}, g\mu_{F_2}, g\mu_{B})$ fixed. We then get a lengthy result which we don't record.
 In the special case of $g\mu_{F_1}=g\mu_{F_2}= g\mu_{B}=0$ or equivalently
    $\frac{1}{3}g\mu^{0}=g\mu^{1} = g\mu^{2} =g\mu^3=\frac{\kappa}{6}(1-n)$ (which gives an action just depending on $n$)
we find
\begin{align}\label{abjmspcase}
    S_n^\mathrm{SRE}=-\frac{9 n^{2}-\sqrt{3}(1+2n)^{3/2}}{9n(n-1)}F^{ABJM}_{S^3}\,.
\end{align}

We next make contact with the definition of SRE given in
\cite{Hosseini:2019and}. We parametrise the monodromy sources via
\begin{align}
g\mu^\alpha=\kappa(\Delta^\alpha-\frac{1}{2})+\frac{\kappa}{2}(\Delta^\alpha-1)(n-1)\,, \qquad \sum_\alpha\Delta^\alpha=2\,,
\end{align}
where the constraint on the $\Delta^\alpha$ ensures that the supersymmetry constraint is satisfied.
Clearly the monodromy sources are non-vanishing, in general, when $n=1$.
In this case we find the Euclidean action has the form
$I_n(\Delta^\alpha)= \frac{(1+n)^2}{n}\sqrt{\Delta^0 \Delta^1\Delta^2\Delta^3}F^{ABJM}_{S^3}$.
Then, holding $\Delta^\alpha$ fixed when we take $n\to 1$ in \eqref{sreexp}, we obtain an SRE given by
\begin{align}
S^{SRE}_n=-\frac{1+3n}{n}\sqrt{\Delta^0 \Delta^1\Delta^2\Delta^3}F^{ABJM}_{S^3}\,,
\end{align}
 in agreement with \cite{Hosseini:2019and}. Observe that when we set $\Delta^\alpha=\frac{1}{2}$
 we obtain the result for minimal gauged supergravity \eqref{mingssre}.
 
 Finally, we can consider a parametrisation of the monodromy sources that uniformly vanishes as $n\to 1$, similar to what was consider
 in \cite{Huang:2014pda} in a different setting:
 \begin{align}\label{uniformvanihs}
 g\mu^\alpha=\frac{\kappa}{2}(1-n)\Delta^\alpha\,, \qquad \sum_\alpha\Delta^\alpha=2\,.
 \end{align}
 If we now hold the $\Delta^\alpha$ fixed when we take $n\to 1$ in \eqref{sreexp}, we obtain an SRE given by
 \begin{equation}
{S}^{SRE}_n = -\frac{\sqrt{\prod_{\alpha=0}^3\left(n(\Delta^\alpha-1)-\Delta^\alpha\right)}-n^2}{n(1-n)}F^{ABJM}_{S^3}\,.
\end{equation}
 Observe that taking the limit $n\to 1$ we find a result that is independent of $\Delta^\alpha$:
 $\lim_{n\to 1}{S}^{SRE}_n=-F^{ABJM}_{S^3}$. 
 
 \subsubsection{mABJM}
 For the mABJM case, associated with solutions with $\varphi\ne 0$, we have
\begin{align}
I_n=n\mathcal{F}^{mABJM}F^{mABJM}_{S^3}\,,
\end{align}
with
\begin{align}
\mathcal{F}^{mABJM}\equiv \left[\Bigl(1+\frac{g\mu^0}{\kappa n }\Bigr)\Bigl(1+\frac{3g\mu^1}{\kappa n}\Bigr)\Bigl(1+\frac{3g\mu^2}{\kappa n}\Bigr)\Bigl(1+\frac{3g\mu^3}{\kappa n}\Bigr)\right]^{\frac{1}{2}\ }\,,
\end{align}
where $F^{mABJM}_{S^3}$ is given \eqref{freemabjm}.
The supersymmetry constraints are
now $g\mu_R=  \sum_\alpha g\mu^{\alpha}=\kappa(1-n)$ and $g\mu_B=g\mu^{0}-(g\mu^{1} + g\mu^{2} + g\mu^{3})=0$.

In this case we can view $I_n(g\mu^\alpha)$ and $S_n^\mathrm{SRE}$ as being functions of $n$ and 
two independent flavour monodromy sources which can be taken to be
$(g\mu_{F_1}, g\mu_{F_2})$.
The general expression for $S_n^\mathrm{SRE}$ is still rather long so we skip it. We just note that when $g\mu_{F_1}=g\mu_{F_2}=0$
then we have solutions with $\varphi$ taking the constant value as in the mABJM vacuum \eqref{lsfpsc} and we have solutions
of minimal gauged supergravity (see the discussion in section \ref{secfurthertruncs}). For this case we find
\begin{align}\label{mabjmsspcase}
    S_n^\mathrm{SRE}=-\frac{1+3n}{4n}F^{mABJM}_{S^3}
    \,.\end{align}
    as expected.

 At this point we pause to make a comment regarding the RG flows, discussed in section \ref{flows}, that start in the UV from a monodromy defect in ABJM theory with
$ g\mu_{B}=0$ and parametrised by $g\mu_{F_1},g\mu_{F_2}$. 
For values of $g\mu_{F_1},g\mu_{F_2}$ in the dark region of figure \ref{rangesmus},
a homogeneous mass deformation away from the line defect would then
induce an RG flow that is expected to flow to the monodromy defect in mABJM theory parametrised by the same values of $g\mu_{F_1},g\mu_{F_2}$, as illustrated in figure \ref{fig:sumsolutions}. It is interesting
 to compare the SRE entropy. Focussing on the simple case of $g\mu_{F_1}=g\mu_{F_2}=0$, we 
 point out that the ratio of the supersymmetric Renyi entropy in the UV, given in \eqref{abjmspcase}, to that in the IR, given in
 \eqref{mabjmsspcase} is a monotonically increasing function of $n$.

We can instead parametrise the monodromy sources similar to 
\cite{Hosseini:2019and} in the STU case. Specifically, we now take
\begin{align}
g\mu^0&=\kappa(\Delta^0-{1})+\frac{\kappa}{2}(\Delta^0-2)(n-1)\,, \nn
g\mu^a&=\kappa(\Delta^a-\frac{1}{3})+\frac{\kappa}{2}(\Delta^a-\frac{2}{3})(n-1)\,, \quad a=1,2,3\,,
\end{align}
with 
\begin{align}
\sum_\alpha \Delta^\alpha=2\,,\qquad \Delta^0-\Delta^1-\Delta^2-\Delta^3=0\,,
\end{align}
to satisfy the supersymmetry constraints. Clearly the monodromy sources are non-vanishing, in general, when $n=1$.
In this case we find the Euclidean action has the form
$I_n(\Delta^\alpha)=3\sqrt{3}\frac{(1+n)^2}{4n}\sqrt{\Delta^0 \Delta^1\Delta^2\Delta^3}F^{mABJM}_{S^3}$.
Then, holding $\Delta^\alpha$ fixed when we take $n\to 1$ in \eqref{sreexp}, we obtain a SRE given by 
\begin{align}
S^{SRE}_n=-3\sqrt{3}\frac{1+3n}{4n}\sqrt{\Delta^0 \Delta^1\Delta^2\Delta^3}F^{mABJM}_{S^3}\,.
\end{align}
Observe that when we set $\Delta^0=1$, $\Delta^1=\Delta^2=\Delta^3=\frac{1}{3}$
 we obtain the result for minimal gauged supergravity \eqref{mabjmsspcase}.   
    
  An alternative definition of SRE will arise from parametrising the sources as in
 \eqref{uniformvanihs}, which uniformly vanishes in the limit $n\to 1$:
  \begin{align}
 g\mu^\alpha=\frac{\kappa}{2}(1-n)\Delta^\alpha\,, \qquad \sum_\alpha\Delta^\alpha=2\,,
 \quad \Delta^0-\Delta^1-\Delta^2-\Delta^3=0\,.
 \end{align}
 If we now hold the $\Delta^\alpha$ fixed when we take $n\to 1$ in \eqref{sreexp}, we obtain an SRE given by
 \begin{equation}
{S}^{SRE}_n = -\frac{\sqrt{(\Delta^0(n-1)-2n)\prod_{\alpha=1}^3\left(3\Delta^\alpha(n-1)-2n\right)}-4n^2}{4n(1-n)}F^{mABJM}_{S^3}\,.
\end{equation}

\section{Discussion}\label{sec:discussion}
Within holography we have computed various observables for superconformal line defects with non-trivial monodromy for the global symmetry in two $d=3$ SCFTs, ABJM theory and mABJM theory. We derived
expressions for the conserved currents and the stress tensor, and hence the conformal weight of the defect, $h_D$, 
in terms of the monodromy parameters and a parameter $n$; if the SCFT is in flat spacetime there is a conical singularity with deficit angle fixed by $n$, while
if the SCFT is on $AdS_2\times S^1$, the parameter $n$ fixes the ratio of the radius of the $S^1$ to the radius of the $AdS_2$. 
We also obtained analogous formula for the partition function as well as the supersymmetric Renyi entropy (SRE) for circular entangling surfaces. 
We emphasised that  there are different supersymmetric Renyi entropies that one can define, depending on which monodromy sources
one holds fixed
as one adjusts the conical deficit. We computed various SREs for both ABJM theory and mABJM theory and recovered the results of \cite{Hosseini:2019and} for the ABJM case as a particular example.

Many of the results we have obtained did not require explicitly solving the BPS equations. For ABJM theory the solutions exist in the STU model and
are known in explicit form \cite{Hosseini:2019and,Ferrero:2021etw}. For solutions of mABJM theory, we obtained some representative solutions by solving the BPS equations numerically. The fact that we could obtain certain results explicitly, without an analytic solution is related to recent
general results concerning supersymmetric solutions and localization \cite{BenettiGenolini:2023kxp,BenettiGenolini:2024kyy} and it would be interesting to make this connection more precise.
We showed there are two branches of solutions, in general, with necessary bounds on the value of $n$ and
the monodromy parameters as given in \eqref{branchesbds}. By analogy with what was 
seen in the context of defect solutions of $D=5$ gauged supergravity in \cite{Arav:2024exg}, these are not expected to
be sufficient condition; we illustrated this in figure \ref{rangesmus} and it would be of interest to explore this in more detail.

We also considered gravitational solutions that are dual to monodromy line defects in ABJM theory with additional mass sources.
For ABJM theory in flat spacetime these correspond to spatially dependent mass sources that preserve the superconformal invariance of the defect. For ABJM theory on $AdS_2\times S^1$ these correspond to constant mass source deformations.
There is a one parameter family of such solutions, labelled by the sources $\varphi_s$ as illustrated in figure
\ref{fig:sumsolutions}, and they can be viewed as a line of exactly marginal deformations in the sense of \cite{Herzog:2019bom}.  When $\varphi_s=0$ we have ABJM defects corresponding to restricted STU solutions
with $g\mu_B=\kappa n$ imposed by hand. In the limit as $\varphi_s\to\infty$ the solutions become arbitrarily close to the
mABJM defect solutions. We showed that there is an attractor mechanism at work with the behaviour of the
metric function $V$ and the scalars $\lambda_i$ having exactly the same value along the line of solutions parametrised by $\varphi_s$, for fixed values of $g_{\mu_{F_i}}$ and $n$. Additionally, the conserved currents
and also the on-shell action take the same value along the line of solutions, for fixed values of $g_{\mu_{F_i}}$ and $n$.
By contrast, the conformal weight of the defect, $h_D$, is not constant along this line.

For all of the defect solutions we have discussed, we showed that $h_D$ 
is proportional to the one-point function of the R-symmetry current. This is very suggestive that this is a general result
for co-dimension two defects in $d=3$ SCFTs with the amount of supersymmetry we have been considering,
and it would be interesting to try and prove this in general. This would be a direct analogue of the result for 
co-dimension two defects of $d=4$ SCFTs preserving $\mathcal{N}=(0,2)$ supersymmetry, that was derived in
\cite{Bianchi:2019sxz}.

We also discussed novel RG flows that start in the UV from ABJM defects that have been deformed
by a homogeneous (in flat spacetime) mass deformation. Take $n=1$ for simplicity and consider UV defects with
$g\mu_R=0$, as demanded by supersymmetry, and $g\mu_B=0$ imposed by hand. The defects are then parametrised by
$g\mu_{F_i}$ as in the larger shaded region in figure \ref{rangesmus}. If we restrict to values of 
$g\mu_{F_i}$ as in the smaller shaded region in figure \ref{rangesmus}, we also have
an mABJM defect solution
parametrised by the same $g\mu_{F_i}$, and this solution is then expected to be the IR limit of the RG flow. It remains
an interesting open question to determine the IR fate of the RG flows for other values of $g\mu_{F_i}$.
For such RG flows with $n=1$, we showed that the defect free energy $-I_D$ would necessarily increase 
along such an RG flow. On the other hand the total free energy $I$ would decrease and this may also be true more generally for bulk RG flows with $n=1$. These results are analogous to what was seen for defects in $d=4$ SCFTS in \cite{Arav:2024exg}.
We also investigated the free energy for $n>1$: we find $-I_D$ is not monotonic in general and can 
 potentially decrease or increase, depending on the values of the monodromy. For 
$1<n\le4$ we find that the total free energy $I$ would decrease 
but for $n>4$ we find that $I$ could potentially decrease or increase, depending on the values of the monodromy.

We highlighted that at the boundary of the space of defect solutions, as illustrated in figure \ref{rangesmus},
there is a divergence in the conformal weight $h_D$ and the currents. This can be contrasted with what happens for
the analogous defect solutions of $\mathcal{N}=4$ SYM theory and the LS $\mathcal{N}=1$ SCFT in $d=4$. It would be interesting
to understand the physical reason for this differing behaviour.

We showed that varying the defect free energy, $I_D$, with respect to the flavour monodromy sources not constrained
by supersymmetry we obtained the associated dual currents. The $b$ central charge for codimension two defect in $d=4$ SCFTs
shares the same property \cite{Bianchi:2021snj}. It would be interesting to make a precise connection with
\cite{Kobayashi:2018lil}, but we leave that to future work. 

Finally, we note that the monodromy parameters are periodic variables, but the details are somewhat subtle. 
In \cite{Arav:2024exg} a discussion in the context of defects in $\mathcal{N}=4$ SYM theory was made and many of the observations have a direct analogue in the present setting. It would be interesting to analyse this in more detail.

\section*{Acknowledgements}

\noindent 
We thank Lorenzo Bianchi and Minwoo Suh for helpful discussions.
This work was supported in part by FWO project G003523N,
STFC grants ST/T000791/1, ST/X000575/1,
the National Research Foundation of Korea (NRF) grant funded by the Korea government (MSIT) (No. 2023R1A2C1006975), an appointment to the JRG Program at the APCTP through the Science and Technology Promotion Fund and Lottery Fund of the Korean Government,
the European  MSCA grant HORIZON-MSCA-2022-PF-01-01 and by the H.F.R.I call ``Basic research
Financing (Horizontal support of all Sciences)" under the National Recovery and Resilience Plan ``Greece 2.0" funded by the
European Union - NextGenerationEU (H.F.R.I. Project Number: 15384.).
JPG is supported as a Visiting Fellow at the Perimeter Institute.

\appendix

\section{BPS equations}\label{sec:appa}

We summarise the BPS equations for the $AdS_2$ ansatz
\eqref{ads2ans}
derived in \cite{Suh:2022pkg}, with some small notational changes.

We use the orthonormal frame $e^a=e^V\bar e^a$, $e^2=f dy$, $e^3=h dz$, where $\bar e^a$
is an orthonormal frame for $ds^2(AdS_2)$. Associated with the spinors discussed in section \ref{bpsbulkdiscsec},
with $\sin\xi\ne0$, the complete BPS equations are given by 
\begin{align}\label{bpsb1}
f^{-1}\xi'\,=&\,\sqrt{2}gW\cos\xi+\kappa{e}^{-V}\,, \notag \\
f^{-1}V'\,=&\,\frac{g}{\sqrt{2}}W\sin\xi\,, \notag \\
f^{-1}\lambda_i'\,=&\,-\frac{g}{\sqrt{2}}\partial_{\lambda_i}W\sin\xi\,, \notag \\
f^{-1}\varphi'\,=&\,-\frac{g}{\sqrt{2}}\frac{\partial_\varphi{W}}{\sin\xi}\,, \notag \\
f^{-1}\frac{h'}{h}\sin\xi\,=&\,\kappa{e}^{-V}\cos\xi+\frac{gW}{\sqrt{2}}\left(1+\cos^2\xi\right)\,, 
\end{align}
with two constraints, 
\begin{align}\label{twobpsconsts}
\left(s-Q_z\right)\sin\xi&=-\sqrt{2}gWh\cos\xi-\kappa{h}e^{-V}\,, \notag \\
\sqrt{2}g\partial_\varphi{W}\cos\xi&=\partial_\varphi{Q}_z\sin\xi{h}^{-1}\,.
\end{align}
The frame components of the field strengths that are dressed with scalar fields (see \eqref{effrels})
 are given by
\begin{align}\label{bpseffexps}
\overline{F}_{23}^{12}\,=&\,-g\partial_{\lambda_1}W\cos\xi\,, \notag \\
\overline{F}_{23}^{34}\,=&\,-g\partial_{\lambda_2}W\cos\xi\,, \notag \\
\overline{F}_{23}^{56}\,=&\,-g\partial_{\lambda_3}W\cos\xi\,, \notag \\
H_{23}\,=&\,-gW\cos\xi-\sqrt{2}\kappa{e}^{-V}\,.
\end{align}
We get expressions for the field strengths $F^{\alpha}$ from the inverse of \eqref{effrels}:
\begin{align}
F^{0}\,=&\,\frac{1}{2}e^{\lambda_1+\lambda_2+\lambda_3}\left(\overline{F}^{12}+\overline{F}^{34}+\overline{F}^{56}+\overline{F}^{78}\right)\,, \notag \\
F^{1}\,=&\,\frac{1}{2}e^{\lambda_1-\lambda_2-\lambda_3}\left(\overline{F}^{12}-\overline{F}^{34}-\overline{F}^{56}+\overline{F}^{78}\right)\,, \notag \\
F^{2}\,=&\,-\frac{1}{2}e^{-\lambda_1+\lambda_2-\lambda_3}\left(\overline{F}^{12}-\overline{F}^{34}+\overline{F}^{56}-\overline{F}^{78}\right)\,, \notag \\
F^{3}\,=&\,-\frac{1}{2}e^{-\lambda_1-\lambda_2+\lambda_3}\left(\overline{F}^{12}+\overline{F}^{34}-\overline{F}^{56}-\overline{F}^{78}\right)\,,
\end{align}
with $F^\alpha_{yz}\,=\,\left(a^\alpha\right)'$ and $F^{\alpha}_{23}=f^{-1}h^{-1}(a^{\alpha})'$.

An integral of the BPS equations is given by
\begin{equation}
he^{-V}\,=-n\sin\xi\,,
\end{equation}
where $n$ is a constant. Employing this to eliminate $h$, we find the BPS equations are given by
\begin{align}\label{bpseqsapp}
f^{-1}\xi'\,=&\,n^{-1}\left(s-Q_z\right)e^{-V}\,, \notag \\
f^{-1}V'\,=&\,\frac{g}{\sqrt{2}}W\sin\xi\,, \notag \\
f^{-1}\lambda_i'\,=&\,-\frac{g}{\sqrt{2}}\partial_{\lambda_i}W\sin\xi\,, \notag \\
f^{-1}\varphi'\,=&\,-\frac{g}{\sqrt{2}}\frac{\partial_\varphi{W}}{\sin\xi}\,,
\end{align}
with two constraints
\begin{align} \label{twoconstraints}
\left(s-Q_z\right)&=n\left(\sqrt{2}gWe^V\cos\xi+\kappa\right)\,, \notag \\
\sqrt{2}g\partial_\varphi{W}\cos\xi&= -n^{-1}e^{-V}\partial_\varphi{Q}_z\,.
\end{align}

From the definition of $Q_z$ in \eqref{hhbbdef}, we find
\begin{equation}
\partial_\varphi{Q}_z=
- \sinh 2\varphi D_z\theta\,.
\end{equation}
If $\varphi\ne0$, from the second constraint in \eqref{twoconstraints}, we find
\begin{equation} \label{dspsidzpsi}
D_z\theta=\frac{\sqrt{2}gne^V\partial_\varphi{W}\cos\xi}{\sinh2\varphi}\,,
\end{equation}
and the right hand side is independent of $\varphi$. 

We have checked that the BPS equations for $\lambda_i$, $V$, $h$ and $\varphi$ are consistent with
the full equations of motion. We can also differentiate the first constraint
in \eqref{twobpsconsts} (after dividing both sides by $\sin\xi$) and find a consistent result with
the signs as in \eqref{dthetastext}.

The BPS equations \eqref{bpsb1} can be used to express the field strengths in the form
\begin{equation}
F^\alpha_{yz}\,=\,\left(a^\alpha\right)'\,=\,\left(\mathcal{I}^\alpha\right)'\,,
\end{equation}
where
\begin{align}
\mathcal{I}^0\,\equiv&-\frac{1}{\sqrt{2}}ne^V\cos\xi\,e^{\lambda_1+\lambda_2+\lambda_3}\,, \qquad
\mathcal{I}^1\,\equiv-\frac{1}{\sqrt{2}}ne^V\cos\xi\,e^{\lambda_1-\lambda_2-\lambda_3}\,, \notag \\
\mathcal{I}^2\,\equiv&-\frac{1}{\sqrt{2}}ne^V\cos\xi\,e^{-\lambda_1+\lambda_2-\lambda_3}\,, \qquad
\mathcal{I}^3\,\equiv-\frac{1}{\sqrt{2}}ne^V\cos\xi\,e^{-\lambda_1-\lambda_2+\lambda_3}\,.
\end{align}
These expressions are consistent with taking the derivative of \eqref{dspsidzpsi}.

There is a symmetry of the BPS equations,
\begin{equation} \label{hsymm}
h\,\rightarrow\,-h\,, \qquad z\,\rightarrow\,-z\,,
\end{equation}
when we have $Q_z\rightarrow-Q_z$, $s\rightarrow-s$, $a^\alpha\rightarrow-a^\alpha$, $n\rightarrow-n$ and $F_{23}^\alpha\rightarrow+F_{23}^\alpha$. The frame is invariant under this transformation. We fix $h\ge0$ by this symmetry in the main text.

\section{ABJM $AdS_4$ boundary }\label{appa}

\subsection{Expansion for the equations of motion}\label{Eomnotbps}
In the gauge
\begin{align}
f=\frac{L}{y}\,,
\end{align}
where $L^2\equiv 1/(2g^2)$,
we find the following expansion as $y\to \infty$ for the scalars and gauge fields
\begin{align}
\lambda_i&=\frac{\lambda_i^{(1)}}{y}+\frac{\lambda_i^{(2)}}{y^2}+\dots \,,\nn
\varphi&=\frac{\varphi^{(1)}}{y}+\frac{\varphi^{(2)}}{y^2}+\dots\,, \nn
a^{\alpha}&=\mu^{\alpha}+\frac{j^{\alpha}}{y}+\dots\,.
\end{align}
Here $g\mu^\alpha$ are the monodromy sources for the global symmetry currents and
$\varphi^{(1)}$ is the source for the operator dual to $\varphi$ with scaling dimension $\Delta=2$.
In the main text we have written
\begin{align}
\varphi_s\equiv\varphi^{(1)}\,.
\end{align}
Since we want the operator dual to $\lambda_i$ to have scaling dimension $\Delta=1$,
$\lambda_i^{(1)}$ is \emph{not} the associated source; we return to this below.

For the remaining metric functions we have the expansion
\begin{align}
e^{2V}&=e^{2V_0}y^2+\frac{1}{2}(L^2-e^{2V_0}[\sum_i (\lambda_i^{(1)})^2+(\varphi^{(1)})^2])+\frac{V_{(2)}}{y}+\dots\,, \nn
\frac{h^2}{h_0^2}&=y^2
-\frac{1}{2}(\frac{L^2}{e^{2V_0}}+\sum_i (\lambda_i^{(1)})^2+(\varphi^{(1)})^2)\nn
&\quad-2\left(\frac{V_{(2)} }{e^{2V_0}}+\frac{4}{3}[ \sum_i \lambda_i^{(1)}\lambda_i^{(2)}
+\varphi^{(1)}\varphi^{(2)}      ] \right)\frac{1}{y}+\dots \,.
\end{align}
Note that the bulk metric approaches the boundary at $y\to \infty$ as
\begin{align}
ds^2=\gamma_{ab}dx^a dx^b+\frac{L^2}{y^2}dy^2+\dots\,,
\qquad
\gamma_{ab}=y^2 h_{ab}\,,
\end{align}
and the boundary metric can be written
\begin{align}
h_{ab} dx^a dx^b= e^{2V_0}\left(ds^2(AdS_2)+\frac{h_0^2}{e^{2V_0} }dz^2\right)\,.
\end{align}
\subsection{Expansion for the BPS equations}
Using the BPS equations in \eqref{bpsb1}, \eqref{twobpsconsts}
we can develop the expansion, when $\varphi\ne0$, given by
\begin{align}\label{abjmxiexpappb}
\xi=-\frac{\pi}{2}+\frac{\kappa L}{e^{V_0}y}+\dots\,,
\end{align}
with
\begin{align}\label{somebps2}
n=\frac{h_0}{e^{V_0}}\,,\qquad
\sum_\alpha g\mu^{\alpha}&=-{\kappa}n-s\,,\nn
\bar\theta- (g\mu^{0}-g\mu^{1}-g\mu^{2}-g\mu^{3})&=-\frac{\kappa h_0}{ e^{V_0}}\,,
\end{align}
where $\bar\theta z$ is the phase of the complex scalar (in the text we took $\bar\theta=0$),
and
\begin{align}
\lambda_1^{(2)}&=\lambda_2^{(1)}\lambda_3^{(1)}-\frac{1}{2}(\varphi^{(1)})^2\,,\nn
\lambda_2^{(2)}&=\lambda_1^{(1)}\lambda_3^{(1)}-\frac{1}{2}(\varphi^{(1)})^2\,,\nn
\lambda_3^{(2)}&=\lambda_1^{(1)}\lambda_2^{(1)}-\frac{1}{2}(\varphi^{(1)})^2\,,\nn
\varphi^{(2)}&=-(\lambda_1^{(1)}+\lambda_2^{(1)}+\lambda_3^{(1)}  )\varphi^{(1)}\,,
\end{align}
as well as
\begin{align}
V_{(2)}=
 \frac{4}{3} e^{2V_0} \left((\lambda_1^{(1)}+\lambda_2^{(1)}+\lambda_3^{(1)}) (\varphi^{(1)})^2
 -2 \lambda_1^{(1)} \lambda_2^{(1)} \lambda_3^{(1)}\right)
 -\frac{\kappa e^{V_0}}{6 g h_0} (j^{0}+j^{1}+j^{2}+j^{3})\,.
\end{align}

From the first three in \eqref{bpseffexps}, and using \eqref{effrels}, we also obtain 
\begin{align}
\lambda_1^{(1)}&= -\frac{g  }{2 \kappa n} (j^{0}+j^{1}-j^{2}-j^{3})\,,\nn
\lambda_2^{(1)}&= -\frac{g  }{2 \kappa n} (j^{0}-j^{1}+j^{2}-j^{3})\,,\nn
\lambda_3^{(1)}&= -\frac{g  }{2 \kappa n} (j^{0}-j^{1}-j^{2}+j^{3})\,,
\end{align}

If we evaluate the conserved quantities in \eqref{constantmo2} we obtain
\begin{align} 
\mathcal{E}_{R_1}&=-\frac{e^{2V_0}}{h_0 L} (j^{0}+j^{1})\,,\nn
\mathcal{E}_{R_2}&=-\frac{e^{2V_0}}{h_0 L} (j^{0}+j^{2})\,,\nn
\mathcal{E}_{R_3}&=-\frac{e^{2V_0}}{h_0 L} (j^{0}+j^{3})\,,
\end{align}
and we also get the same result from \eqref{er123}. Shortly, we will relate these to the conserved currents.
For $\mathcal{E}_{B}$ in \eqref{brokeneb}, which is not constant unless $\varphi=0$, we have
\begin{align}
\mathcal{E}_{B}&=-\frac{e^{2V_0}}{h_0 L} (3j^{0}-j^{1}-j^2-j^3)\,.
\end{align}

Note that in the gauge $\bar\theta=0$, \eqref{somebps2} includes the constraints on the monodromy sources:
\begin{align}\label{somebps23}
g\mu_R&\equiv g\mu^0+g\mu^1+g\mu^2+g\mu^3=-s-{\kappa}n\,,\nn
g\mu_B&\equiv g\mu^{0}-g\mu^{1}-g\mu^{2}-g\mu^{3}=\kappa n\,,
\end{align}
which must be imposed on solutions with $\varphi\ne 0$. For solutions of the STU model with $\varphi=0$, we only
need to impose the first condition.

\subsection{Holographic renormalisation}
We first consider the on-shell action.
Since\footnote{
Consider the theory
\begin{align}\label{emax}
\mathcal{L}=\gamma R-V(\varphi)-\frac{Z(\varphi)}{4}F^2-\frac{1}{2}\nabla_\mu\varphi\nabla^\mu\varphi-X(\varphi)(D_\mu \theta D^\mu\theta)\,,
\end{align}
with $D\theta=d\theta+A$ and $\gamma$ a constant. Define the two form
$q_{\mu\nu}=2\gamma\nabla_{\mu}k_{\nu}+Z(\varphi)(A_\rho k^\rho)F_{\mu\nu}$
where $k^\mu$ is a Killing vector. Then, assuming $\mathcal{L}_kA_\mu=\mathcal{L}_k\varphi=0$,
on-shell 
$ \nabla_\nu q^{\mu\nu}=k^\mu \mathcal{L}+2XD^\mu\theta (k^\rho \partial_\rho\theta)$.
} $\partial_t$ is a Killing vector that preserves the whole solution, we can rewrite the on-shell bulk
action as a total derivative:
\begin{align}
\sqrt{-g}\mathcal{L}=\partial_y\left(-\frac{e^{2V}h V'}{\rho^2 f}\right)+\partial_\rho\left(\frac{f h}{\rho}\right)\,.
\end{align}
For the solutions of interest with $h\to 0$ at the core, the first term will only have a boundary contribution at $y\to\infty$.
When $\varphi=0$ we also have another Killing vector $\partial_z$ and we can write the on-shell action as a total derivative in
another way. In fact, for $\varphi\ne 0$, we find that we can write
\begin{align}
\sqrt{-g}\mathcal{L}=&-\partial_y\Bigg(\frac{e^{2V}h'}{\rho^2 f}
+\frac{e^{2V}}{\rho^2f h}\Big[
e^{-2\left(\lambda_1+\lambda_2+\lambda_3\right)}a_0 a_0'
+e^{2\left(-\lambda_1+\lambda_2+\lambda_3\right)}a_1 a_1' \nn
&\qquad\qquad\qquad\qquad+e^{2\left(\lambda_1-\lambda_2+\lambda_3\right)}a_2 a_2'
+e^{2\left(\lambda_1+\lambda_2-\lambda_3\right)}a_3 a_3'
\Big]
\Bigg)\nn
&-\frac{e^{2V} f}{2 \rho^2 h}\sinh^22\varphi\bar\theta D_z\theta\,.
\end{align}
Interestingly, notice that in a gauge with $\bar\theta=0$ the last term vanishes; this arises because the ansatz is invariant under a combination of gauge transformations and translations in the $z$ direction. 
The first term will get contributions from
both $y\to\infty$ and $y\to core$ for solutions to our BPS equations (both $h\to 0$ and $a_\alpha\to 0$).

In order to carry out holographic renormalisation that is sufficient for the BPS solutions of interest,
we start by considering an action given by
\begin{align}
S=S_{bulk}+S_{bdy}\,,
\end{align}
where the boundary action $S_{bdy}$ is given by (e.g. \cite{Freedman:2013oja})
\begin{align}\label{s1abjmcase}
S_{bdy}=\frac{1}{8\pi G}\int d^3x \sqrt{-\gamma}\left(Tr K+\frac{1}{L}W+ L R(\gamma)[-\frac{1}{2}+\sum_i b_i\lambda_i+b_4\varphi] \right)\,.
\end{align}
Here $K$ is the trace of the extrinsic curvature, $R(\gamma)$ is the Ricci scalar of the boundary metric.
We have temporarily included the constants $b_i$ and $b_4$, which are associated with finite counter-terms;
however, shortly we will use a scheme in which we set them to zero.
Varying the 
total action $S_{bulk}+S_{bdy}$ and using our $AdS_2$ ansatz
we deduce that it depends on the following source terms:
\begin{align}
[S+S_{bdy}] =[S+S_{bdy}](g\mu^{\alpha},\lambda_i^{(1)},\varphi^{(1)},h_0,e^{2V_0})\,.
\end{align}

To preserve supersymmetry, we want to do alternative quantisation for the scalar fields $\lambda_i$, so that they are dual to operators with scaling dimension $\Delta=1$. 
This requires adding additional boundary terms via a suitable Legendre transform.
We first define 
\begin{align}\label{pisdefabjm}
\Pi_i\equiv \left(\frac{\delta (\sqrt{-g}\mathcal{L})}{\delta \partial_y \lambda_i}
+\frac{\delta ( \sqrt{-\gamma}[ \frac{1}{L}W+ L R(\gamma)(\sum_i b_i\lambda_i+b_4\varphi)])}{\delta \lambda_i}\right)\,.
\end{align}
Notice that as $y\to\infty$ we have 
\begin{align}\label{pidef}
\Pi_i&=
\frac{h_0 e^{2V_0} }{\rho^2}\Pi_i^{(s)}\frac{y}{L}+\dots \,,
\end{align}
where, for example, the first component of $\Pi_i^{(s)}$ is given by 
\begin{align}\label{pidefs}
 \Pi_1^{(s)}\equiv 
2 \lambda_1^{(2)}-2 \lambda_2^{(1)}\lambda_3^{(1)}+(\varphi^{(1)})^2 -2e^{-2V_0} b_1L^2\,.
\end{align}
To change the quantisation for the $\lambda_i$, we now define 
\begin{align}
S_{Tot}=S_{bulk}+S_{bdy}-\frac{1}{8\pi G}\int d^3 x\sum_i  \Pi_i\lambda_i\,,
\end{align}
with $S_{Tot}=S_{Tot}(g\mu^{\alpha},\Pi_i^{(s)},\varphi^{(1)},h_0,e^{2V_0})$ 
and $-\Pi_i^{(s)}$ the source of the 
$\Delta=1$ operators.

We can now define the one-point functions for various operators. For the scalar operators, recalling that $\sqrt{-h}=e^{2V_0} h_0/u^2$, 
we define 
\begin{align}
\delta S_{Tot}=\int d^3x \sqrt{-h}\left[ \langle \mathcal{O}_\varphi \rangle \delta \varphi^{(1)}
-\langle\mathcal{O}_{\Pi_i} \rangle \delta \Pi^{(s)}_i
 \right]\,,
\end{align}
with
\begin{align}\label{appbvevsoetc}
\langle \mathcal{O}_\varphi \rangle&=
\frac{1}{8\pi G}\frac{1}{L}\left(2\sum_i\lambda_i^{(1)}\varphi^{(1)}+2\varphi^{(2)}
- 2b_4 L^2e^{-2V_0}\right)\,,\nn
\langle\mathcal{O}_{\Pi_i} \rangle&=\frac{1}{8\pi G}\frac{1}{L}\lambda_i^{(1)}\,.
\end{align}
For the symmetry currents, the sources are $g\mu^{\alpha}$ and we have
\begin{align}\label{abjmcrt1}
\delta S_{Tot}=\int d^3x \sqrt{-h} \langle J^{\alpha z} \rangle \delta (g\mu^{\alpha})\,,
\end{align}
with
\begin{align}\label{abjmcrt2}
\langle J^{\alpha z} \rangle&=\frac{1}{8\pi G}\frac{1 }{gL h_0^2 }{j^{\alpha}},
\quad
\Rightarrow \quad 
\langle J^{\alpha}_ z \rangle=\frac{1}{8\pi G}\frac{1 }{gL }{j^{\alpha}}\equiv \langle J^\alpha\rangle,
\end{align}
where we lowered the index using the boundary metric $h_{ab}$.
Finally, we consider the boundary stress tensor, which has the form $T^{ab}=diag(-T^{uu},T^{uu},T^{zz})$.
Using $\delta S_{Tot}=\frac{1}{2}d^3x\sqrt{-h}T^{ab}\delta h_{ab}$ we have
\begin{align}
\delta S_{Tot}=\int d^3x \sqrt{-h}\left[ \langle T^{uu} \rangle \delta \left(\frac{e^{2V_0}}{\rho^2}\right)
+ \frac{1}{2}\langle T^{zz} \rangle \delta h_0^2
\right]\,,
\end{align}
where
\begin{align}
\langle T^{uu}\rangle&=\frac{1}{8\pi G}\rho^2 
\frac{e^{-2V_0}}{L}\left(
4\lambda_1^{(1)}\lambda_2^{(1)}\lambda_3^{(1)}
+2\varphi^{(1)}\varphi^{(2)}+\frac{3}{2}e^{-2V_0}V_{(2)}
+\sum_i2 b_i L^2e^{-2V_0}\lambda_i^{(1)}\right)\,,\nn
\langle T^{zz}\rangle&=\frac{1}{8\pi G}\frac{h_0^{-2}}{L}\left(-4\sum_i\lambda_i^{(1)}\lambda_i^{(2)}+4\lambda_1^{(1)}\lambda_2^{(1)}\lambda_3^{(1)}
-2\varphi^{(1)}\varphi^{(2)}-3e^{-2V_0}V_{(2)}
-2 b_4 L^2e^{-2V_0}\varphi^{(1)}\right)\,.
\end{align}
With these results, we can verify that the following Ward identity holds:
\begin{align}
\langle T^{a}{}_a\rangle=(3-\Delta_{\mathcal{O}_\varphi })\langle \mathcal{O}_\varphi \rangle \varphi^{(1)} -\sum_i (3-\Delta_{\mathcal{O}_{\lambda_i }})\langle\mathcal{O}_{\Pi_i} \rangle \Pi_i^{(s)}=0\,,
\end{align}
with $\mathcal{O}_\varphi =2$ and $\Delta_{\mathcal{O}_{\lambda_i }}=1$.

If we further impose the BPS equations we find 
\begin{align}\label{abjmstressapp1}
\langle T^u{}_u\rangle&=
-\frac{\kappa n}{4 h_0^2}\sum_\alpha \langle J^{(\alpha)} \rangle+
\frac{1}{8\pi G}
\frac{1}{L}\left(
\sum_i2 b_i L^2e^{-2V_0}\lambda_i^{(1)}\right)\,,\nn
\langle T^z{}_z\rangle&=\frac{\kappa n}{2h_0^2  }\sum_\alpha\langle J^{(\alpha)} \rangle+
\frac{1}{8\pi G}\frac{1}{L}\left(
-2 b_4 L^2e^{-2V_0}\varphi^{(1)}\right)\,,\nn
\langle \mathcal{O}_\varphi \rangle&=
\frac{1}{8\pi G}\frac{1}{L}\left(
- 2b_4 L^2e^{-2V_0}\right)\,,\nn
\langle\mathcal{O}_{\Pi_i} \rangle&=\frac{1}{8\pi G}\frac{1}{L}\lambda_i^{(1)}\,.
\end{align}
It is also interesting to note that the BPS equations imply 
$\Pi^{(s)}_i=-2e^{-2V_0}b_i L^2$ and also
\begin{align}\label{rewrtielamvev}
\frac{1}{8\pi G L}\lambda_1^{(1)}&= -\frac{g   }{2 \kappa n} 
(\langle J^{0}\rangle +\langle J^{1}\rangle
-\langle J^{2}\rangle-\langle J^{3}\rangle )\,,\nn
\frac{1}{8\pi G L}\lambda_2^{(1)}&=-\frac{g   }{2 \kappa n}  (
\langle J^{0}\rangle-\langle J^{1}\rangle
+\langle J^{2}\rangle-\langle J^{3}\rangle)\,,\nn
\frac{1}{8\pi G L}\lambda_3^{(1)}&= -\frac{g  }{2 \kappa n} (\langle J^{0}\rangle-\langle J^{1}\rangle
-\langle J^{2}\rangle+\langle J^{3}\rangle)\,.
\end{align}

Given the expressions in \eqref{abjmstressapp1}, 
we see that the conformal weight $h_D$ will be determined by the R-symmetry current one-point function provided that
we impose
\begin{align}\label{appbscheme}
b_i=b_4=0\,.
\end{align}
Thus, we strongly suspect that this scheme \eqref{appbscheme} is demanded by supersymmetry 
and is the one we have utilised in the text. It is interesting to highlight that this scheme also implies that 
$\langle \mathcal{O}_\varphi \rangle=0$
and correspondingly we can view the $\varphi_s$ deformation as a Q-exact mass deformation 
as discussed in section \ref{actgenfnzflows}.

In the scheme \eqref{appbscheme}, the BPS configurations have 
\begin{align}
\langle T_{ab} \rangle dx^a dx^b &=-\frac{1}{4\kappa n}\sum_\alpha \langle J^{(\alpha)} \rangle\,[ds^2(AdS_2)-2n^2dz^2]\,,\nn
\langle \mathcal{O}_\varphi \rangle&=0\,,\nn
\langle\mathcal{O}_{\Pi_1} \rangle&= -\frac{g   }{2 \kappa n} 
(\langle J^{0}\rangle +\langle J^{1}\rangle
-\langle J^{2}\rangle-\langle J^{3}\rangle )\,,\nn
\langle\mathcal{O}_{\Pi_2} \rangle&=-\frac{g   }{2 \kappa n}  (
\langle J^{0}\rangle-\langle J^{1}\rangle
+\langle J^{2}\rangle-\langle J^{3}\rangle)\,,\nn
\langle\mathcal{O}_{\Pi_3} \rangle&=- \frac{g  }{2 \kappa n} (\langle J^{0}\rangle-\langle J^{1}\rangle
-\langle J^{2}\rangle+\langle J^{3}\rangle)\,,
\end{align}
and $\Pi^{(s)}_i=0$.

Since the deformation parametrised by $\varphi_s$ preserves supersymmetry, combined with the fact that
supersymmetry implies that the source for a bosonic mass operator must vanish, $\Pi^{(s)}_i=0$, 
we conclude that
the operator $\mathcal{O}^{\Delta=2}_\varphi$ dual to $\varphi_s$ must be proportional\footnote{The same observation is also applicable in the case of ordinary RG flows from ABJM theory to mABJM theory, driven by a homogeneous, Poincar\'e invariant mass deformation.} to a linear combination
of a fermion mass operator of dimension ${\Delta=2}$, and $\varphi_s$ times a bosonic mass 
operator of dimension ${\Delta=1}$.
Notice also from \eqref{appbvevsoetc} that the expansion coefficient $\varphi_2$ directly determines 
the expectation value of the following operator:
$\langle \mathcal{O}_\varphi    -\varphi_s  \sum_i \mathcal{O}_{\Pi_i} \rangle=
\frac{1}{4\pi G}\frac{1}{L}\varphi^{(2)}$. 
Since $\varphi_2$ is associated with an expansion 
coefficient of an operator that has a definite charge under the broken $U(1)$ symmetry, we 
deduce that $\varphi_2$ determines the vev of the fermionic mass operator.

\section{mABJM $AdS_4$ boundary}\label{secappcmabjm}

\subsection{Expansion for the equations of motion}
The mABJM $AdS_4$ vacuum has radius ${\tilde L}=\frac{\sqrt{2}}{3^{3/4}g}$ with scalars given by
\begin{equation}\label{mabjmcssc}
\lambda_i  = \frac{1}{4}\ln 3 ,\qquad
\varphi  = \frac{1}{2}\mathrm{arccosh} 2\,.
\end{equation}
In this background $\lambda_i,\varphi$ mix and are dual
to relevant operators with scaling dimension $\Delta =1, 1, \frac{1}{2}+\frac{1}{2}\sqrt{17}$ 
as well as an irrelevant operator of dimension $\frac{5}{2}+\frac{1}{2}\sqrt{17}$. There is
a massive vector in the bulk, $A_B\equiv A^{0}-A^{1} -A^{2}-A^{3}$, arising from a mixing with the complex scalar
which is dual to an irrelevant vector operator with scaling dimension $\Delta=\frac{3}{2}+\frac{1}{2}\sqrt{17}$.
The mixing and the irrational scaling dimensions give rise to a rather complicated near boundary expansion.

We work in the gauge
\begin{align}
f=\frac{{\tilde L}}{y}\,,\qquad {\tilde L}=\frac{\sqrt{2}}{3^{3/4}g}\,,
\end{align}
where ${\tilde L}$ is the radius of the $AdS_4$ mABJM vacuum. We then develop the schematic expansion 
\begin{align}\label{expofmabjmappcfg}
V & =V_0 + \ln y + y^{-2\delta}\left(\upsilon_s + \ldots \right) + \frac{v_1}{y} + \frac{v_2}{y^2}+\frac{v_3}{y^3}+y^{-\delta}\left(\frac{\upsilon_m}{y}+\ldots \right)+\ldots\,,\nn
h & = h_0y\left[1+y^{-2\delta}\left(\eta_s + \ldots \right) + \frac{h_1}{y} + \frac{h_2}{y^2}+\frac{h_3}{y^3}+y^{-\delta}\left(\frac{\eta_m}{y}+\ldots \right)+\ldots\right]\,,\nn
A^\alpha & = \mu^\alpha +\frac{j^\alpha}{y} +\ldots + y^{-3+\delta}\left(m_v^\alpha +\ldots \right) +\ldots\,,\nn
\lambda_i & =\frac{1}{4}\ln 3+ y^{-\delta}\left(\zeta_i^s +\ldots \right) + \frac{l^1_i}{y} +  \frac{l^2_i}{y^2} + \ldots +y^{-3+\delta}\left(\zeta_i^v +\ldots \right) + y^{-5+\delta}\left(\zeta_i^{Iv} +\ldots \right)+\ldots\,,\nn
\varphi & =\frac{1}{2}\mathrm{arccosh} 2+ y^{-\delta}\left(Z^s +\ldots \right) + \frac{f^1}{y} +  \frac{f^2}{y^2} + \ldots +y^{-3+\delta}\left(Z^v +\ldots \right) + y^{-5+\delta}\left(Z^{Iv} +\ldots \right)+\ldots\,,
\end{align}
where $\delta = (5-\sqrt{17})/2$.
We have set $\bar\theta=0$ and also set to zero any terms in the expansion that would be dual to sources for irrelevant operators. 
In particular we have the condition
\begin{align}\label{appcgbzero}
g\mu_B\equiv g\mu^0-g\mu^1-g\mu^2-g\mu^3=0\,.
\end{align}
We highlight that with $g\mu_B=0$ we have 
\begin{align}
g\mu_R^{mABJM}&\equiv\frac{1}{2}(g\mu^0+3g\mu^1+3g\mu^2+3g\mu^3)=2g\mu^0=g\mu^R\,,
\end{align}
where $g\mu_R^{mABJM}$ is the monodromy associated with the bulk gauge field basis given in 
\eqref{mabjmbasis}.
The equations of motion enforce algebraic relationships between the coefficients appearing in this expansion and to find them in practice, it is efficient to define the quantity $\sigma \equiv y^\delta$ and perform a double series expansion of the equations of motion around both $y\to \infty$ and $\sigma\to\infty$. 
In particular we find the constraint
\begin{align}\label{jappcthelast}
0&=j^0-j^1-j^2-j^3\,.
\end{align}

\subsection{Expansion for the BPS equations}
We can consider a similar expansion of the BPS equations. We expand
\begin{equation}\label{mabjmxiexpappc}
\xi = -\frac{\pi}{2} + \frac{x_1}{y}+\frac{x_2}{y^2} + \ldots\,,
\end{equation}
with
\begin{equation}
x_1 = \frac{\sqrt{2}}{3^{3/4}}\frac{\kappa}{g}e^{-V_0}\,, \qquad x_2 = -\frac{\sqrt{2}}{3^{3/4}}\frac{e^{-V_0}}{n}j^0\,,
\end{equation}
and a number of other constraints on the expansion parameters including
\begin{equation}
\zeta_i^s  = Z^s = \upsilon_s = \eta_s = f_1 = v_1 = 0\,,
\end{equation}
as well as 
\begin{align}
h_0&=ne^{V_0}\,,\qquad f_2   = \frac{2}{\sqrt{3}}\left(l_1^1\,^2 + l_1^1 l_2^1+(l_2^1)^2 \right)\,, \nn
l_1^2 & = \frac{1}{3}\left(4 l_1^1\,^2 + l_1^1 l_2^1+(l_2^1)^2 \right)\,,\qquad
l_2^2  =  \frac{1}{3}\left( l_1^1\,^2 + l_1^1 l_2^1+4(l_2^1)^2 \right)\,,\nn
l_3^2  &=  \frac{1}{3}\left(4 (l_1^1)^2 +7 l_1^1 l_2^1+4(l_2^1)^2 \right)\,,\qquad
\zeta_i^v  = -\frac{1+\sqrt{17}}{4\sqrt{3}}Z^v\,,\nn
v_2 & = \frac{1}{18}\Big(-9\left((l_1^1)^2 + l_1^1 l_2^1+(l_2^1)^2 \right)+\frac{\sqrt{3}}{g^2}e^{-2V_0}\Big)\,,\nn
v_3 &= \frac{1}{9}\Big(12\, l_1^1l_2^1\left(l_1^1+l_2^1\right)+3^{1/4}\sqrt{2}\,x_2\frac{\kappa}{g}e^{-V_0} \Big)\,,
\end{align}
and
\begin{align}\label{eq:BPSfos}
l^1_1 & = \frac{g  \kappa}{2n}\left(j^0-3j^1 \right),\quad
l^1_2  = \frac{g  \kappa}{2n}\left(j^0-3j^2 \right),\quad
l^1_3  = -\frac{g  \kappa}{2n}\left(2j^0-3(j^1+j^2) \right).
\end{align}
Furthermore, we also have the constraint on the R-symmetry monodromy source
\begin{align}
&g\mu_R \equiv  g\mu^{0}+g\mu^{1} + g\mu^{2} + g\mu^{3} =-\kappa n- s \,,
\end{align}
in addition to \eqref{appcgbzero}.

\subsection{Holographic renormalisation}
As in the previous appendix we start by considering an action given by
\begin{align}\label{sbdybulkmabjm}
S=S_{bulk}+S_{bdy}\,,
\end{align}
where the boundary action $S_{bdy}$ is given by\footnote{
The boundary actions \eqref{s1abjmcase} and \eqref{sbdymabjm1}
can be viewed as part of a more general boundary action that has been expanded about the ABJM and mABJM vacua, respectively. 
In particular, it should be possible to replace the Ricci scalar term with $f(\lambda_i,\varphi) R(\gamma)$ consistent with supersymmetry.
For both cases, with the addition of the further boundary terms we consider below for the mABJM case, this scheme
is sufficient for renormalising the BPS solutions of interest. However, for non-BPS solutions that asymptote to the mABJM vacua
further modifications are required.}
\begin{align}\label{sbdymabjm1}
S_{bdy}=\frac{1}{8\pi G}\int d^3x \sqrt{-\gamma}\left(Tr K+\frac{1}{L}W-\frac{\tilde L}{2} R(\gamma) \right)\,,
\end{align}
with some possible finite counter terms that we can consider set to zero.

We first consider the one point functions associated with the gauge fields. We need to separate out the massless and massive vectors in the variation. 
Thus, we consider the variation 
\begin{align}
\delta[S_{bulk}+S_{bdy}]&=\frac{1}{8\pi G}\int d^3 x (\Pi^{(A)})_\alpha^\mu\delta A_\mu^\alpha\nn
&=\frac{1}{8\pi G}\int d^3 x(\Pi^{(A)})_\alpha^\mu M^\alpha{}_\beta \delta \tilde A_\mu^\beta\,,
\end{align}
where $\tilde A_\mu^\beta$ are defined as in \eqref{mabjmbasis} with the matrix $M^\alpha{}_\beta$ given by
\begin{equation}
{M}^\alpha\,_\beta
= \frac{1}{12}
\begin{pmatrix}
6 & 0 & 0 & 9\\
2 & 8 & 4 & -1\\
2 & -4 & 4 & -1\\
2 & -4 & -8 & -1
\end{pmatrix}\,.
\end{equation}
Associated with the variations $\delta A_R^{mABJM}$ and $\delta A_{F_i}$ we obtain the one point functions for the
conserved current one point functions. 
After lowering an index with the boundary metric (see \eqref{abjmcrt1}, \eqref{abjmcrt2}), recalling \eqref{jappcthelast},
we obtain
\begin{align}
\vev{J_R^{\varphi}}&=
\frac{1}{8\pi G}\frac{3^{1/4}}{\sqrt{2}}\frac{1}{6}(3j^0+j^1+j^2+j^3)
\,,\nn
\vev{J_{F_1}}&=
\frac{1}{8\pi G}\frac{3^{1/4}}{\sqrt{2}}(2j^1-j^2-j^3)
\,,\nn
\vev{J_{F_2}}&=
\frac{1}{8\pi G}\frac{3^{1/4}}{\sqrt{2}}(j^1+j^2-2j^3)
\,.
\end{align}
Note the similarity to \eqref{abjmfnzjs}, but we highlight that they are not in exactly the same ratio.

Considering variations $\delta A_B$ associated with the massive vector, we find that there is an additional counter term that we must add, of the form 
\begin{align}\label{massvecct}
S^{mv}_{ct}=-\frac{1}{8\pi G}\frac{g3^{1/4}}{2\sqrt{2}}\int d^3 x\sqrt{-\gamma}A_B^2\,,
\end{align}
Then, carrying out the variation, and again lowering the $z$ index, we get the associated expectation value
\begin{align}
\vev{J^{mABJM}_B}=\frac{1}{8\pi G}\frac{\kappa n}{g2^{5/2} 3^{1/4}}(5-3\sqrt{17})Z^v\,.
\end{align}

We next consider the stress tensor. With the expansion given in 
\eqref{expofmabjmappcfg}, with in particular $g\mu_B=0$ in \eqref{appcgbzero},
it is clear that the counterterm \eqref{massvecct} will not give any contribution.
While additional boundary terms are required to suitably renormalise the scalar sector, discussed below, it turns out that they do not give rise to any additional contributions to the stress tensor. So the stress tensor for the BPS configurations of interest can be obtained from
\eqref{sbdybulkmabjm}. After some computation we find 
\begin{align}
\langle T_{ab}\rangle dx^a dx^b =-\frac{h_D}{2\pi}\left[ds^2(AdS_2)-2 n^2 dz^2\right]\,,
\end{align}
with 
\begin{align}
h_D=\frac{2\pi}{\kappa n}  J_R^{mABJM}\,.
\end{align}

Finally, we consider the scalar sector. We need to consider both the mixing of modes in the scalar sector as well
as additional boundary terms to ensure suitable alternative quantisation for some of the modes. For variations of
the scalar fields we write $\delta\Phi^a\equiv (\delta\lambda_i,\delta\varphi)$ and we also define $\delta\Phi^a=\mathbb{S}^a{}_b\delta\tilde \Phi^b$,
where $\tilde \Phi^b$ are a basis for mass eigenmodes for the mABJM vacuum; we give an expression for $\mathbb{S}^a{}_b$ below.
We next consider the on-shell variation with respect to the scalars of the total action so far (c.f. \eqref{pisdefabjm})
\begin{align}
\delta[S_{bulk}+S_{bdy}+S^{mv}_{ct}]=\frac{1}{8\pi G}\int d^3x( \Pi_a\delta\Phi^a+\dots)=\frac{1}{8\pi G}\int d^3x( \Pi_a \mathbb{S}^a{}_b\delta\tilde \Phi^b+\dots)\,.
\end{align}
Now recall that these mass eigenmodes modes are dual to operators with conformal dimension
 $\Delta =1, 1, \frac{1}{2}+\frac{1}{2}\sqrt{17}$ and $\frac{5}{2}+\frac{1}{2}\sqrt{17}$. To ensure the appropriate quantisation for
 the first two, we need to
 add in an extra boundary term to the action  
\begin{align}\label{extrabctmabjm}
S^{\delta\tilde\Phi^{1,2}}=-\frac{1}{8\pi G}\int d^3 x\sum_{a=1}^4\sum_{b=1}^2  \Pi_a \mathbb{S}^a{}_b\tilde \Phi^b\,,
\end{align}
The total on-shell action can be considered to be a function of the scalar sources 
$( -\Pi_a \mathbb{S}^a{}_1, -\Pi_a \mathbb{S}^a{}_2,\tilde \Phi^3,\tilde \Phi^4)$ dual to operators with the scaling dimensions given above, respectively.

Interestingly, we find that the BPS equations imply that the scalar sources for the $\Delta=1$ operators vanish
$ \Pi_a \mathbb{S}^a{}_1=\Pi_a \mathbb{S}^a{}_2=0$ (similar to what we saw in the ABJM case). For this reason, the extra counterterm
action \eqref{extrabctmabjm} does not give rise to any additional contribution to the boundary stress tensor for BPS configurations, as noted above. 

A convenient choice for $\mathbb{S}$ is given by the orthogonal matrix
\begin{equation}
\begin{pmatrix}
-\sqrt{\frac{2}{3}} & 0 & \frac{1}{\sqrt{6}}\sqrt{1-\frac{1}{\sqrt{17}}} & 2 \frac{\sqrt{17}-3}{\sqrt{17}-5}\sqrt{\frac{2}{3(17+\sqrt{17})}}\\
\frac{1}{\sqrt{6}} & -\frac{1}{\sqrt{2}} & \frac{1}{\sqrt{6}}\sqrt{1-\frac{1}{\sqrt{17}}} & 2 \frac{\sqrt{17}-3}{\sqrt{17}-5}\sqrt{\frac{2}{3(17+\sqrt{17})}}\\ 
\frac{1}{\sqrt{6}} & \frac{1}{\sqrt{2}} & \frac{1}{\sqrt{6}}\sqrt{1-\frac{1}{\sqrt{17}}} & 2 \frac{\sqrt{17}-3}{\sqrt{17}-5}\sqrt{\frac{2}{3(17+\sqrt{17})}}\\ 
0 & 0 & \sqrt{\frac{1}{34}(17+\sqrt{17})} &2 \sqrt{\frac{2}{17+\sqrt{17}}}
\end{pmatrix}.
\end{equation}
In particular, for the two $\Delta=1$ modes we can write\footnote{Notice that the constant terms in \eqref{mabjmcssc} drop
out of these linear combinations.}
\begin{align}
\tilde\Phi^1 & = \frac{1}{\sqrt{6}}\left(\lambda^3+\lambda^2-2\lambda^1 \right) \,,\nn
\tilde\Phi^2 & = \frac{1}{\sqrt{2}}\left(\lambda^3-\lambda^2\right) \,.
\end{align}
Using \eqref{eq:BPSfos}, the associated expectation values for the scalar operators is given by 
\begin{align}
\vev{\mathcal{O}^{\Delta=1}_1}=\frac{3^{1/4}g}{2\kappa n}\langle J_{F_1}\rangle,\qquad
\vev{\mathcal{O}^{\Delta=1}_2}=-\frac{g}{2\times  3^{1/4}\kappa n}(\langle J_{F_1}\rangle-2\langle J_{F_2}\rangle).
\end{align}
Varying the on-shell action with respect to $\tilde \Phi^3$ gives rise to a vanishing expectation value 
\begin{align}
\langle{\mathcal{O}^{\Delta=\frac{1}{2}(1+\sqrt{17})}}\rangle=0\,.
\end{align}
Last, we consider varying the on-shell action with respect to $\tilde \Phi^4$, dual to the irrelevant scalar operator.
An additional divergence appears that can be cancelled by having an additional boundary term with variation
given by
\begin{align}\label{extrabctmabjm2}
\delta S^{\delta\tilde\Phi^{4}}=\frac{1}{8\pi G}\frac{\sqrt{13+\frac{43}{\sqrt{17}}}}{2\times 3^{3/4} g}\int d^3 x
\sqrt{-\gamma}(\mathbb{S}^{-1})^4{}_a\delta\Phi^a (\varphi-\frac{1}{2}\text{arccosh 2})R(\gamma)\,,
\end{align}
and this leads to an expectation value
\begin{align}
\vev{\mathcal{O}^{\Delta=\frac{1}{2}(5+\sqrt{17})}}=\frac{2}{3^{3/4}g}\sqrt{5+\frac{13}{\sqrt {17}}}e^{-2V_0}Z^v\,.
\end{align}
This variation can be integrated to find the contribution to the counterterm action but we will not do so here. 

Finally, it is worth highlighting that with vanishing sources for the massive vector,
the $\Delta=1$
scalars ($ \Pi_a \mathbb{S}^a{}_1=\Pi_a \mathbb{S}^a{}_2=0$), and the irrelevant scalar operator, the on-shell action for BPS configurations
is given by $S_{bulk}+S_{bdy}$ in \eqref{sbdybulkmabjm}. Furthermore, this is sufficient to get the expectation values for the
conserved currents and the stress tensor for BPS equations.
To get the one-point function for the massive vector, the $\Delta=1$ scalar operators and the irrelevant scalar operator, we need to consider the additional counterterms $S^{mv}_{ct}$ in \eqref{massvecct}, 
$S^{\delta\tilde\Phi^{1,2}}$ in \eqref{extrabctmabjm} and $S^{\delta\tilde\Phi^{4}}$ in \eqref{extrabctmabjm2}.

\section{Solutions of minimal gauged supergravity}\label{minsugraapp}
For minimal gauged supergravity, associated with the STU model, we take $\varphi = \lambda_i = 0$, as well as $a^0 = a^1=a^2= a^3  \equiv a_R/4$. The BPS equations then take the form
\begin{align}
\label{eq:ads4_minsugra_genl}
f^{-1} \xi' + 2^{3/2} g \cos\xi - \kappa e^{-V} = 0\,, \nonumber \\
f^{-1} V' + 2^{1/2} g \sin\xi = 0\,, \nonumber \\
h = - n e^{V} \sin\xi\,, \nonumber \\
a_R = -2^{3/2}n e^{V} \cos\xi + c_0\,,
\end{align}
where the constant $xc_0$ is given by
\begin{align}\label{constappD}
g c_0 - \kappa n + s = 0\,,
\end{align}
as follows from \eqref{twoconstraints}.
We want to examine the known solution (e.g. \cite{Ferrero:2020twa}) in our conventions, so we consider
the ansatz
\begin{align}
e^V = e^{V_0} y\,, \qquad f = \frac{h_0}{h}\,.
\end{align}
The regularity condition for the metric at the core then becomes
\begin{align}
\label{eq:ads4_core_reg}
(h^2)'|_{y=y_*} = 2 h_0\,.
\end{align}
The second and third equation of \eqref{eq:ads4_minsugra_genl} then fix $h_0$ to be
\begin{align}
\sqrt{2}g h_0 = n e^{V_0}\,.
\end{align}
Substituting into the first equation of \eqref{eq:ads4_minsugra_genl} we obtain
\begin{align}
\xi'= \frac{2}{y\tan\xi}-\frac{\kappa e^{-V_0}}{\sqrt{2} g y^2 \sin\xi}\,.
\end{align}
It is convenient to introduce the rescaled coordinate
\begin{align}
\tilde y=2^{3/2}g e^{V_0}\kappa y\,,
\end{align}
to obtain the solution
redone from here
\begin{align}
\xi = \pm \arccos\left[\frac{2\tilde y+a}{\tilde y^2} \right]\,,
\end{align}
where $a$ is a constant.
This then gives rise to the full solution in the form
\begin{align}
ds^2&=\frac{1}{2g^2}\Big[\frac{\tilde y^2}{4}ds^2(AdS_2)+\frac{\tilde y^2}{q}d\tilde y^2+\frac{q}{4\tilde y^2}n^2 dz^2\Big]\,,\nn
ga_R&=-s-n\kappa(1+\frac{a}{\tilde y})\,.
\end{align}
Here $q$ is the quartic
\begin{align}
q=(\tilde y^2-[2 \tilde y+a])(\tilde y^2+[2\tilde y+a])\,,
\end{align}
with
four roots given by
\begin{align}
\tilde y_*(\sigma_i) =\sigma_2 +\sigma_1\sqrt{1+\sigma_2a}\,,
\end{align}
with $\sigma_1^2 = \sigma_2^2 = 1$.
When $a=0$ or $-1$ we have a double root.

As in the text, we take $y$ to range to positive infinity. For simplicity we take 
\begin{align}
\kappa=+1\,,\end{align}
 so that the same is true for $\tilde y$.
We then want to consider solutions with $y_*\le \tilde y<\infty$ where $y_*>0$ the largest positive root and $y\to\infty$ associated with the 
$AdS_4$ boundary.
We first consider the case where $a>-1$ and $a\ne 0$ so the outermost single root is given by $\sigma_1=\sigma_2=+1$. From \eqref{eq:ads4_core_reg} we find
$(1+a)^{-1/2}=n$, or
\begin{align}
a = \frac{1-n^2}{n^2}\,,
\end{align}
and notice that this is consistent with $a>-1$, $a\ne 0$ for all $n>0$, $n\ne 1$. This branch of solutions corresponds to ``branch 1" solutions in the main text \eqref{restrictionsskappn}. To see this,
demanding that $a_R$ vanishes at the root we must have
$s = -1$ and hence  $s= - \kappa$. Expanding at the $AdS_4$ boundary $\tilde y\to \infty$ we find that
$g \mu_R = (1-n)$ as expected. Notice that when $n\to 1$ we have no conical singularity and $g\mu_R\to 0$ so we have no defect; this case corresponds to the $a=0$ solutions; after making the coordinate transformation $\tilde y^2=4\cosh^2\rho$ we find that that the $a=0$ solution is in fact just the vacuum $AdS_4$ solution.
This is expected as there is no defect in minimal gauged supergravity when $n=1$.
We comment on the $n\to \infty$ or $a\to -1$ limiting solution below.
We also record expressions for $h_D$ and $-I_D$ for these main branch solutions (with $e^{V_0}=1$):
\begin{align}
h_D=\frac{n^2-1}{ n^2} \frac{N^{3/2}}{12 \sqrt{2} }
\,,\qquad 
-I_D=\frac{ (n-1) (3 n+1)}{4 n}{F^{ABJM}_{S^3}}\,.
\end{align}
In particular, we see that for $n>1$ we have $h_D>0$ and $-I_D>0$, while for $0<n<1$ we have
 $h_D<0$ and $-I_D<0$.

To find ``branch 2" solutions we should assume that $a<-1$, and then there is only two real roots. In this case 
the outer root corresponds to $\sigma_1=+1$ and $\sigma_2=-1$.
From \eqref{eq:ads4_core_reg} we now find
$(1-a)^{-1/2}=n$, or
\begin{align}
a = \frac{n^2-1}{n^2}.
\end{align}
Consistency with $a<-1$ requires that for such defect solutions $n$ lies in the range
\begin{align}
\label{eq:ineq_branch2_ads4}
0<n<1/\sqrt{2}\,,
\end{align}
which we note is more restricted than the condition $0<n<1$ discussed below \eqref{efvcoreabjm3}.
One can explicitly check that regularity of $a_R$ at the root implies $s = + 1$ and hence $s= + \kappa$, so we have a branch 2 solution.
We can also check $g \mu_R = - n - 1$.
Notice that if we consider the range $1/\sqrt{2}<n<1$ where there is the potential for additional branch 2 solutions from \eqref{restrictionsskappn}, 
we have $-1<a<0$ and we find that $\sigma_1=+1$ and $\sigma_2=-1$ is not the outermost root and instead there are associated spindle solutions, with $\tilde y$ running between two roots.
We also record expressions for $h_D$ and $-I_D$ for these branch 2 solutions (with $e^{V_0}=1$):
\begin{align}
h_D=\frac{1-n^2}{n^2}\frac{ N^{3/2}}{12 \sqrt{2}  }
\,,\qquad
-I_D=\frac{ \left(5 n^2-2 n+1\right)}{4 n}F^{ABJM}_{S^3}\,,
\end{align}
and $h_D>0$, $-I_D>0$ for the allowed range $0<n<1/\sqrt{2}$.

Finally, we briefly comment on the solution associated with $a=-1$. 
The metric and field strength are given by (dropping the tilde on $y$):
\begin{align}\label{qeqminone}
ds^2&=\frac{1}{2g^2}\Big[\frac{y^2}{4} ds^2(AdS_2)+\frac{ y^2}{( y-1)^2( y^2+2 y-1)} dy^2+\frac{( y-1)^2( y^2+2 y-1)}{ 4y^2}n^2dz^2\Big]\,,\nn
F^R&=-\frac{n}{g  y^2}d y\wedge dz\,.
\end{align}
Here we can take $ nz$ to be non-compact
(as is natural in thinking of this solution
as arising as the $n\to\infty$ limit of the main branch solutions above),
and $1< y<\infty$. As $ y\to 1$, the solution approaches the standard $AdS_2\times H_2$ solution, with $y,z$ parametrising $H_2$ written as the upper-half plane and
supported by a magnetic flux on the $H_2$. Interestingly, as $ y\to \infty $ the solution approaches $AdS_4$ with an $AdS_2\times \mathbb{R}$ boundary.
This solution can be compared with the standard magnetically charged black hole solution \cite{Caldarelli:1998hg} (e.g. see (3.25) of \cite{Gauntlett:2001qs})
that interpolates between $AdS_4$ with $\mathbb{R}_t\times H^2$ boundary and $AdS_2\times H_2$. 

\section{Positivity of $h_D$ and $-I_D$ }\label{positivehdmId}

\subsection{The main branch (branch 1 solutions)}
In this subsection we show that for $n\ge 1$, which necessarily means for solutions on the main branch with $s=-\kappa$, we have $h_D\ge 0$ and $-I_D\geq 0$.
This applies to solutions of both ABJM and mABJM theory.
These conditions are not necessarily true on the main branch when $n<1$ and in particular for solutions of minimal gauged supergravity 
we have $h_D<0$ and $-I_D<0$, as we saw in appendix \ref{minsugraapp}.

We begin by recalling that if $f(x)$ is a concave function (i.e.\ if $f''(x)\leq 0$) then
\begin{equation}\label{concavecond}
\frac{1}{k}\sum_{i=1}^k f(x_i) \leq f\left(\frac{1}{k}\sum_{i=1}^k x_i\right) \,,
\end{equation}
and similarly if $f(x)$ is convex ($f''(x)\geq 0$) then
\begin{equation}
\frac{1}{k}\sum_{i=1}^k f(x_i) \geq f\left(\frac{1}{k}\sum_{i=1}^k x_i\right) \,.
\end{equation}

To show that $I^{ABJM}_D\leq 0$ when $n\geq 1$, we first note that from the definition in \eqref{efvcoreabjm2}
\begin{equation}
\log\mathcal{F}^{ABJM} = \frac{1}{2} \sum_{\alpha=0}^3 f\left( \frac{2g\mu^\alpha}{\kappa n} \right) \,,
\end{equation}
where $f(x) \equiv \log(1+x)$. Since $f(x)$ is concave, we can use \eqref{concavecond} to obtain:
\begin{equation}
\log\mathcal{F}^{ABJM} \leq 
2 \log\left( 1 + \frac{1}{4} \sum_{\alpha=0}^3 \frac{2g\mu^\alpha}{\kappa n} \right) = 2\log \left( 1 - \frac{n-1}{2n} \right) \,,
\end{equation}
where in the last equality we used the supersymmetry restriction $ \sum_{\alpha=0}^3 g\mu^\alpha = \kappa(1-n)$ for the main branch of solutions. We therefore have
\begin{equation}
\mathcal{F}^{ABJM} \leq \left( 1 - \frac{n-1}{2n} \right)^2 = \mathcal{F}^{ABJM} |_{\mu_0=\mu_1=\mu_2=\mu_3}\,,
\end{equation}
and when $n \geq 1$ we have $\mathcal{F}^{ABJM} \leq 1$ and hence from \eqref{Ideeabjmcase}
\begin{equation}
I^{ABJM}_D = n (\mathcal{F}^{ABJM} - 1) F_{S^3}^{ABJM} \leq 0 \,.
\end{equation}
Note that the maximal value of $I_D$ is obtained for the minimal gauged supergravity case ($\mu_0=\mu_1=\mu_2=\mu_3$).

Next, to show that in ABJM theory $h_D \geq 0$ for $n\geq 1$, we note from equations \eqref{litjistu}-\eqref{hdforstu}
that
\begin{equation}
- c \frac{h_D}{\mathcal{F}^{ABJM}} = \frac{1}{4} \sum_{\alpha=0}^3 f\left( \frac{2g\mu^\alpha}{\kappa n} \right),
\end{equation}
where here we take $f(x)\equiv \frac{x}{1+x}$, and $c$ is some positive constant. We again notice that $f(x)$ is concave for $1+x \geq 0$, which corresponds to the restriction on $g\mu^\alpha$ for the main branch (recall the comment below \eqref{efvcoreabjm3}). Thus, we conclude
\begin{equation}
- c \frac{h_D}{\mathcal{F}^{ABJM}} \leq f\left( \frac{1}{4} \sum_{\alpha=0}^3 \frac{2g\mu^\alpha}{\kappa n} \right) = \frac{1-n}{1+n} \leq 0\,,
\end{equation}
for $n\geq 1$, and we again used the supersymmetry restriction for the monodromy sources. Since
we always have $\mathcal{F}^{ABJM} \geq 0 $ we conclude that $h_D \geq 0$ for $n\geq 1$.

The analysis for the mABJM case is similar but slightly more involved. We
start with $I^{mABJM}_D$ given in \eqref{ideemabjm}. Recalling
 \eqref{mfabjmexp} we have
\begin{align}
\log \mathcal{F}^{mABJM} &= \frac{1}{2} f\left( \frac{g\mu^0}{\kappa n} \right) + \frac{1}{2} \sum_{i=1}^3 f\left( \frac{3g\mu^i}{\kappa n} \right) \leq
\frac{1}{2} f\left( \frac{g\mu^0}{\kappa n} \right) +
\frac{3}{2} f\left( \sum_{i=1}^3 \frac{g \mu^i}{\kappa n}\right)  \nn
& = 2 f\left( \frac{g\mu^0}{\kappa n} \right) = 
2 \log \left( 1- \frac{n-1}{2n} \right) \,,
\end{align}
where we again took $f(x) \equiv \log(1+x)$, and used 
$ \sum_{i=1}^3 g\mu^i = g\mu^0 =\frac{\kappa}{2}(1-n)$ following \eqref{eq:FluxSourcesFromCorem}. Therefore
\begin{equation}
\mathcal{F}^{mABJM} \leq \left( 1 - \frac{n-1}{2n} \right)^2 = \mathcal{F}^{mABJM} |_{\mu_0/3=\mu_1=\mu_2=\mu_3} \leq 1 \,,
\end{equation}
for $n \geq 1$, leading to $I^{mABJM}_D \leq 0$. Again we see that $I_D$ is maximal for the minimal gauged supergravity case
associated with mABJM, here defined via $\frac{1}{3} g\mu^0 = g\mu^1 = g\mu^2 = g\mu^3$.

Finally, we consider $h_D$ for the mABJM case. We now use equations
\eqref{eeejs123} along with \eqref{efvexpMabjmcore} and \eqref{hdeemabjm} to express $h_D$ in terms of the sources. After some algebra, one can show that
\begin{align}
c\, h_D\, & \mathcal{F}^{mABJM}  \prod_{i=1}^3 \left( 1 + \frac{3g\mu_i}{\kappa n} \right)^{-1} = 
\sum_{i=1}^3 f\left( \frac{3g\mu^i}{\kappa n} \right) - \left( 1 + \frac{2}{n}\right) f\left( \frac{g\mu^0}{\kappa n} \right) \nn
&\geq 3 f\left( \sum_{i=1}^3 \frac{g\mu^i}{\kappa n} \right) - \left( 1 + \frac{2}{n}\right) f\left( \frac{g\mu^0}{\kappa n} \right) = f\left( \frac{g\mu^0}{\kappa n} \right) \frac{2(n-1)}{n} \geq 0\,,
\end{align}
for $n \geq 1$, where $c$ is some positive constant (that depends on $n$) and here $f(x) \equiv \frac{1}{1+x}$, which is a convex function in the relevant range $1+x>0$. We again used 
$ \sum_{i=1}^3 g\mu^i = g\mu^0 =\frac{\kappa}{2}(1-n)$ following \eqref{eq:FluxSourcesFromCorem}.
Since the factors on the LHS multiplying $h_D$ are all positive for the main branch solutions, we deduce that $h_D \geq 0$ for $n\geq 1$.
 
\subsection{Branch 2 solutions}
For branch 2 solutions with $s=+\kappa$, which can only exist for $0<n<1$, it is straightforward to show $h_D>0$ and $-I_D> 0$.

For $I_D$ we argue as follows.
Since $\mathcal{F}^{ABJM} > 0$ and $\mathcal{F}^{mABJM} > 0$, it is easy to see from equations 
\eqref{fullabjmact}, \eqref{Ideeabjmcase} and \eqref{ideemabjm} that  $I_D < 0 $ (and also $ I < 0 $)
for both ABJM and mABJM branch 2 solutions. 

We next turn to $h_D$.
For ABJM solutions, we consider \eqref{litjistu}-\eqref{hdforstu} and use the fact that 
for branch 2 solutions we have $ 1 + \frac{2g\mu^\alpha}{\kappa n} < 0$ (recall the comment below \eqref{efvcoreabjm3}).
Since $f(x) \equiv \frac{x}{1+x}>0$ for $1+x<0$ we conclude that $h_D > 0$.

Finally, for mABJM solutions, since branch 2 solutions can only exist for $n<1$, with $s=\kappa$ we have $1-s\kappa n >0$ and $ 3- \frac{6s\kappa}{n} < 0 $. Then using equation \eqref{eeejs123} and \eqref{hdeemabjm}, we see that once again $h_D > 0$ for branch 2 solutions.


\providecommand{\href}[2]{#2}\begingroup\raggedright\endgroup

\end{document}